\documentclass[epj,nopacs,final]{svjour}
\usepackage{graphics}
\usepackage{ifpdf,latexsym,overpic,rotating}
\usepackage{subfigure,color,showkeys}
\usepackage{epsfig,color,rotating,amsmath,delarray,array}
\usepackage{makeidx,pifont,float,amssymb}
\newcommand{\be}{\begin{eqnarray}}
\newcommand{\ee}{\end{eqnarray}}
\newcommand{\bc}{\begin{center}}
\newcommand{\ec}{\end{center}}\definecolor{gray01}{gray}{0.9}
\definecolor{gray02}{gray}{0.8}
\definecolor{gray03}{gray}{0.7}
\definecolor{gray04}{gray}{0.6}
\definecolor{gray05}{gray}{0.5}
\definecolor{gray06}{gray}{0.4}
\definecolor{gray07}{gray}{0.3}
\definecolor{gray08}{gray}{0.2}
\definecolor{gray09}{gray}{0.1}

\newcommand{\er}{$\pm$}
\newcommand{\oo}{$^o$}
\title{High statistics study of the reaction \boldmath$\gamma p\to p\pi^0\eta$
}
\titlerunning{High statistics study of the reaction $\gamma p\to p\pi^0\eta$}
\authorrunning{The CBELSA/TAPS Collaboration}
\author{The CBELSA/TAPS Collaboration \medskip \\
 E.~Gutz\inst{1,2},
 V.~Crede\inst{3},
 V.~Sokhoyan\inst{1}$^,$\thanks{{\em Present address:} Institut f\"{u}r Kernphysik, Universit\"{a}t Mainz, Germany},
 H.~van Pee\inst{1},
 A.V.~Anisovich\inst{1,4},
 J.C.S.~Bacelar\inst{5},
  B.~Bantes\inst{6},
 O.~Bartholomy\inst{1},
 D.~Bayadilov\inst{1,4},
  R.~Beck\inst{1},
 Y.A.~Beloglazov\inst{4},
 R.~Castelijns\inst{5},
 H.~Dutz\inst{6},
 D.~Elsner\inst{6},
 R.~Ewald\inst{6},
 F.~Frommberger\inst{6},
  M.~Fuchs\inst{1},
  Ch.~Funke\inst{1},
 R.~Gregor\inst{2},
 A.B.~Gridnev\inst{4},
 W.~Hillert\inst{6},
 Ph.~Hoffmeister\inst{1},
 I.~Horn\inst{1},
 I.~Jaegle\inst{7},
 J.~Junkersfeld\inst{1},
 H.~Kalinowsky\inst{1},
 S.~Kammer\inst{6},
 V.~Kleber\inst{6}$^,$\thanks{{\em Present address:} German Research School for Simulation Sciences, J\"{u}lich, Germany},
 Frank~Klein\inst{6},
 Friedrich~Klein\inst{6},
 E.~Klempt\inst{1},
 M.~Kotulla\inst{2,7},
 B.~Krusche\inst{7},
 M.~Lang\inst{1},
 H.~L\"ohner\inst{5},
 I.V.~Lopatin\inst{4},
 S.~Lugert\inst{2},
 T.~Mertens\inst{7},
  J.G.~Messchendorp\inst{5},
  V.~Metag\inst{3},
  M.~Nanova\inst{3},
  V.A.~Nikonov\inst{1,4},
  D.~Novinsky\inst{1,4},
  R.~Novotny\inst{2},
  M.~Ostrick\inst{6}$^,$\footnotemark[1],
  L.~Pant\inst{2}$^,$\thanks{{\em On leave from:} Nucl. Phys. Div., BARC, Mumbai, India},
  M.~Pfeiffer\inst{2},
  D.~Piontek\inst{1},
  A.~Roy\inst{2}$^,$\thanks{{\em On leave from:} Department of Physics, IIT, Mumbai, India},
  A.V.~Sarantsev\inst{1,4},
  Ch.~Schmidt\inst{1},
  H.~Schmieden\inst{6},
  S.~Shende\inst{5},
  A.~S\"ule\inst{6},
  V.V.~Sumachev\inst{4},
  T.~Szczepanek\inst{1},
  A.~Thiel\inst{1},
  U.~Thoma\inst{1},
  D.~Trnka\inst{2},
  R.~Varma\inst{2}$^,$\footnotemark[4],
  D.~Walther\inst{1,6},
  Ch.~Wendel\inst{1},
  A.~Wilson\inst{1,3}\\
}                     
\institute{\inst{1}Helmholtz-Institut f\"ur Strahlen- und Kernphysik, Universit\"at Bonn, Germany\\
\inst{2}II. Physikalisches Institut, Universit\"at Gie{\ss}en, Germany\\
\inst{3}Department of Physics, Florida State University, Tallahassee, USA\\
\inst{4}Petersburg Nuclear Physics Institute, Gatchina, Russia\\
\inst{5}Kernfysisch Versneller Instituut, Groningen, The Netherlands\\
\inst{6}Physikalisches Institut, Universit\"at Bonn, Germany\\
\inst{7}Institut f\"ur Physik, Universit\"at Basel, Switzerland}
\date{Received: date / Revised version: date}

\abstract{Photoproduction off protons of the $p\pi^0\eta$ three-body
final state was studied with the Crystal Barrel/TAPS detector at the
electron stretcher accelerator ELSA at Bonn for incident energies
from the $\pi^0\eta$ production threshold up to 2.5\,GeV.
Differential cross sections and the total cross section are presented. The use of
linearly polarized photons gives access to the polarization
observables $\Sigma$, $I^{s}$ and $I^{c}$, the latter two
characterize beam asymmetries in case of three-body final states.
$\Delta(1232)\eta$, $N(1535){1/2^-}\pi$, and $p a_0(980)$ are the
dominant isobars contributing to the reaction. The partial wave
analysis confirms the existence of some nucleon and $\Delta$
resonances for which so far only fair evidence was reported. A large
number of decay modes of known nucleon and $\Delta$ resonances is
presented. It is shown that detailed investigations of decay
branching ratios may provide a key to unravelling the structure of
nucleon and $\Delta$ resonances.}
\begin{document}
\maketitle
%
\section{Introduction}
At medium energies, our present understanding of QCD is limited. In
the energy regime of meson and baryon resonances, the strong
coupling constant is large and perturbative methods cannot be
applied. Lattice QCD has made great progress in the calculation of
properties of ground-state baryons, and even excited states are
being explored \cite{Edwards:2011jj}. But there is still a long path
ahead before the baryon resonance spectrum from the lattice
can be considered understood. One of the key issues in this
energy regime is therefore to identify the relevant
degrees of freedom and the effective forces between them. A
necessary step towards this aim is undoubtedly a precise knowledge
of the experimental spectrum of hadron resonances and of their
properties.

Quark models are in general amazingly successful in describing the
spectrum of known hadronic states. However, in meson spectroscopy,
there seems to be an overpopulation: more states are found
experimentally than are expected from a $q\bar q$ scheme. Intrusion
of glueballs, hybrids, multiquark states, and of molecules have been
suggested to explain the proliferation of states
\cite{Anisovich:2008zz}. However, a conservative approach does not
confirm the need for additional resonances \cite{Klempt:2007cp}. In
baryon spectroscopy, the situation is reverse \cite{Klempt:2009pi}.
Due to the three-body nature of the system - which is characterized
by two independent oscillators - quark models predict many more
resonances than have been observed so far, especially at higher
energies
\cite{Capstick:1986bm,Glozman:1997ag,Loring:2001kx,Loring:2001ky}.
This is the problem of the so-called \emph{missing resonances}. The
problem is aggravated by the prediction of additional states, hybrid
baryons, in which the gluonic string mediating the interaction
between the quarks itself can be excited. Hybrid baryons, if they exist,
would increase the density of states at high masses even further;
their mass spectrum is predicted to start at 1.8 or 2\,GeV
\cite{Capstick:2002wm} or at 2.5\,GeV \cite{Dudek:2012ag}. The
properties of some baryon resonances are also difficult to reconcile
in quark models. The best known example is the $N(1440){1/2}^+$
Roper resonance which is predicted in quark models to have a mass
above the negative-parity resonance $N(1535){1/2}^-$. The
interpretation of the Roper resonance as a dynamically generated
object \cite{Krehl:1999km} is, however, not supported by recent
electro-production experiments at JLab \cite{Aznauryan:2008pe}.

High-mass positive and negative parity resonances are often (nearly)
degenerate in mass while quark models most\-ly predict alternating
groups of resonances of opposite parity (see, however, also
\cite{Glozman:1997ag,Santopinto:2012nq}). The mass-degeneracy of
positive- and nega\-tive-parity resonances in both the meson
\cite{Glozman:2003bt} and the baryon \cite{Glozman:1999tk} spectrum
has attracted considerable interest. It has been suggested that in
high-mass hadrons, a phase transition may possibly occur causing
chiral symmetry to be restored. The constituent-quark mass could
evolve in the direction of the current-quark mass
\cite{Bicudo:2009cr}, and chiral multiplets could be formed:
mass-degenerate spin doublets and/or spin and isospin quartets of
resonances with identical spin, opposite parity, and similar masses
\cite{Jaffe:2004ph,Glozman:2007ek,Glozman:2008vg}.

In ``gravitational'' theories (AdS/QCD), baryon mass spectra can be
calculated analytically \cite{Berenstein:2002ke}, and give rather
simple results. When confinement is parameterized by a soft wall,
the squared masses are just proportional to the sum of the intrinsic
orbital angular momentum $L$ and a radial quantum number $N$
\cite{Brodsky:2006uqa}. Diquarks which are antisymmetric in spin and flavor
 - called `good diquarks' by Wilczek
\cite{Wilczek:2004im} - reduce the mass \cite{Forkel:2008un}. A
two-parameter fit to all baryon masses emerges which reproduces the
baryon excitation spectrum with remarkable accuracy
\cite{Forkel:2008un}.

Quarks and gluons may not be the most suitable degrees of freedom to
describe hadron resonances. Instead, baryon resonances can be
generated dynamically from the interaction of pseudoscalar or vector
mesons and ground-state octet or decuplet baryons. In case of
pseudoscalar mesons, the meson-baryon interaction is extracted from
chiral Lagrangians, from an effective theory of QCD at low energies
\cite{Weinberg:1978kz,Meissner,Lutz}. For vector mesons, transition
amplitudes provided by hidden-gauge Lagrangians can be used to
derive an effective interaction \cite{Bando:1984ej}. A recent survey
of the field can be found in \cite{Oset:2010ab}. One might speculate that
high-mass resonances could possibly be generated from the
interactions between pions and nucleon resonances, or nucleons and
meson resonances. To pursue such scenarios, knowledge on cascading
decays of baryon resonances are mandatory.

Experimentally, most information on baryon resonances stems from
partial wave analyses of $\pi N$ elastic scattering data performed
in the early 80s of the last century
\cite{Hohler:1979yr,Cutkosky:1980rh,Arndt:2006bf}, even though now a
number of new resonances has been reported recently which were
deduced from photoproduction data \cite{Anisovich:2011fc}. The
resonances predicted but not observed experimentally seem not to be
randomly distributed; instead, complete baryon multiplets have
remained unobserved
\cite{Anisovich:2011sv,Klempt:2012fy,Crede:2013kia}. A popular
possibility to reduce the number of predicted resonances is to
assume that resonances are formed from a quasi-stable `good diquark'
and a third quark \cite{Anselmino:1992vg}. There are two paths 
to identify the additional degrees of freedom of the three-body 
dynamics of a full quark model compared to a quark-diquark picture. 

i) The number of expected resonances is 
considerably reduced in quark-diquark models: This was the reason 
why they were introduced. However, the formation of
quasi-stable diquarks seems to be excluded by the existence of a
spin-quartet ($S=3/2$) of positive parity ($L=2$) nucleon $I=1/2$
resonances with $J^P=1/2^+\cdots 7/2^+$ at about 2\,GeV
\cite{Klempt:2012fy,Anisovich:2011su}: A $L=2$, $S=3/2$, $I=1/2$ wave
function is only symmetric when two oscillators -- with orbital
excitations $l_i$ -- are excited, with coherent contributions from
excitations with $|L=2; l_1=2,l_2=0\rangle$ and $|L=2;
l_1=0,l_2=2\rangle$, {\it and} from simultaneous excitations of both
oscillators with $|L=2; l_1=1,l_2=1\rangle$. If only one oscillator
was excited, the orbital and spin wave functions would be
symmetric, the isospin wave function of mixed symmetry, the color
wave function is fully antisymmetric, and the full wave function
would be incompatible with the Pauli principle. 

ii) The excitation of two independent oscillators may lead to  
changes in the decay pattern. In nuclear physics, the excitation of two
nucleons often leads to a step-wise de-excitation of the first, then the second nucleon. Anticipating a
result of this paper we may speculate that the $|L=2;
l_1=1,l_2=1\rangle$ component of the nucleon wave function  could
prefer decay modes via a cascade, where the intermediate state has
negative parity and a wave function with one unit of orbital angular
momentum (i.e. a superposition of $|L=1; l_1=1,l_2=0\rangle$ and
$|L=1; l_1=0,l_2=1\rangle$).
\begin{figure*}[pt]
\centering
\includegraphics[width=0.95\textwidth]{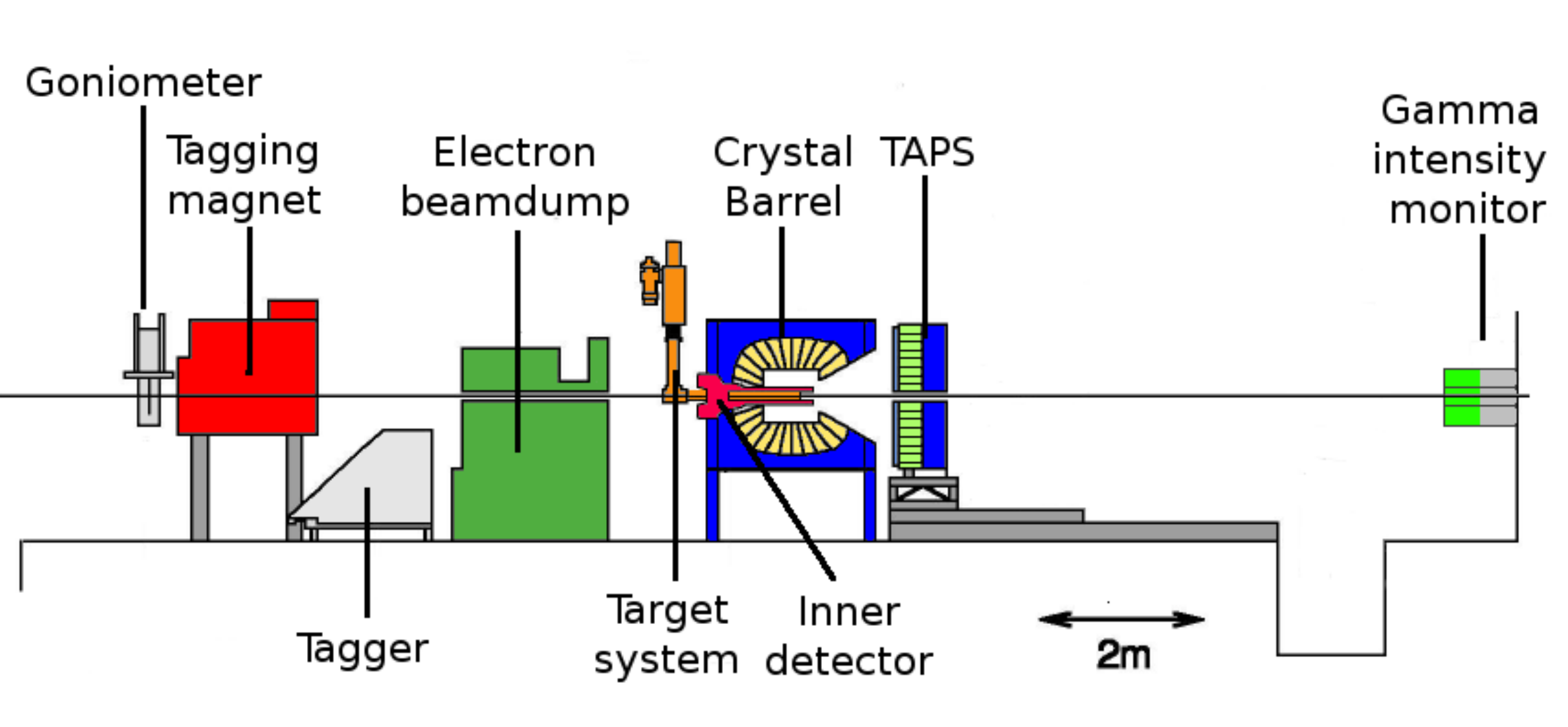}
\caption{Schematic picture of the CBELSA/TAPS experiment.}
\label{fig:beamline}
\end{figure*}

\section{\boldmath The reaction $\gamma p\to p\pi^0\eta$}

Multi-body final states can be expected to contribute to the search
for missing resonances and are interesting in their own right. In
this paper, data on the reaction
\begin{equation}
\gamma p \to p\;\pi^0\eta
\label{ppi0eta}
\end{equation}
are studied for photon energies from the $\pi^0\eta$ production
threshold up to $E_\gamma = 2.5$\,GeV. Compared to the data of the
CB-ELSA collaboration reported in \cite{Horn:2008qv,Horn:2007pp},
the statistics is increased by more than a factor of 12, and the solid
angle coverage is improved. 
In a subsequent paper \cite{Sokhoyan:2014}, we discuss the reaction $\gamma p \to
p\;\pi^0\pi^0$. Reaction (\ref{ppi0eta}) benefits from a strong
$\Delta(1232)\eta$ contribution. Here, the $\eta$ acts as an isospin
filter: resonances decaying into $\Delta(1232)\eta$ must have
isospin $I = 3/2$. The reaction (\ref{ppi0eta}) is hence easier to
analyze than $\gamma p \to p\;\pi^0\pi^0$.

The first report \cite{Weinheimer:2003ng} on $\pi^{0}\eta$
photoproduction demonstrated already the power of this reaction for
the study of nucleon and, in particular, of $\Delta$ resonances. The
data indicated that $\Delta$ resonances around 1.9\,GeV likely
contribute to this reaction. This conjecture was confirmed by a
partial wave analysis suggesting contributions from six $\Delta$
resonances, $\Delta(1600)3/2^+$, $\Delta(1920)3/2^+$,
$\Delta(1700)3/2^-$, \\ $\Delta(1940)3/2^-$, $\Delta(1905)5/2^+$,
$\Delta(2360)3/2^-$, and from two nucleon resonances, $N(1880)1/2^+$
and $N(2200)3/2^+$ \cite{Horn:2008qv}. Particularly interesting was
the observation of a parity doublet ($\Delta(1920)3/2^+$,
$\Delta(1940)3/2^-$) \cite{Horn:2007pp}, unexpected in quark models
\cite{Capstick:1986bm,Loring:2001kx} and in lattice gauge
calculations \cite{Edwards:2011jj}, but predicted
\cite{Forkel:2008un} by models based on AdS/QCD
\cite{deTeramond:2005su,Karch:2006pv,Forkel:2007cm,Brodsky:2010ev}
as well as by the conjecture that chiral symmetry might be restored
in excited baryons \cite{Glozman:1999tk}.

The GRAAL collaboration reported results on $\gamma p\to p\pi^0\eta$
and found that the low energy region is dominated by the formation of
$\Delta(1700)3/2^-$ and the sequential decay chain
$\Delta(1700)3/2^-\to\Delta(1232)\eta\to p\pi^0\eta$
\cite{Ajaka:2008zz}. This result was confirmed by precision data
taken with the Crystal Ball/TAPS detector and the tagged photon
facility at the MAMI C accelerator in Mainz in the $0.95 - 1.4$\,GeV
energy range \cite{Kashevarov:2009ww}. A ratio of the hadronic decay
widths $\Gamma_{\eta \Delta}/\Gamma_{\pi\rm\,N(1535)1/2^-}$ and the
ratio $A_{3/2}/A_{1/2}$ of the helicity amplitudes for this
resonance were reported.

Polarization observables are important in photoproduction to
disentangle the multitude of contributing resonances. In case of 
photoproduction of single mesons with linearly polarized photons
the cross section (equation 19) shows a cos $2\phi$ dependence where 
$\phi$ denotes the angle between the plane of linear polarization 
and the reaction plane. In three-body decays the angle can be defined 
using $p$, $\pi^0$, or $\eta$ as reference particle,
recoiling against the respective two-particle system. However,
additional observables, $I^{s}$ and $I^{c}$, can be extracted due to
the fact that an intermediate resonance can be aligned such that its
magnetic sub-states do not need to have a statistical population
\cite{Roberts:2004mn}. The alignment leads to an additional
dependence of the number of events on the angle between the reaction
plane and the decay plane of the three-body final state. The large
asymmetries reported in \cite{Gutz:2009zh} demonstrated the high
sensitivity of $I^s$ to the dynamics of the reaction. Further
confirmation of the dynamical nature of the $\Delta(1700){3/2^-}$
resonance and its S-wave decay into $\Delta(1232)\eta$ is derived
from studies of $I^{s}$ and $I^{c}$ in $\gamma p\to p\pi^0\eta$
\cite{Doring:2010fw}, within a chiral unitary approach for
$\eta$-$\Delta(1232)$ scattering \cite{Doring:2005bx}.

The paper is organized as follows: The experiment is described in
section~\ref{Section:ExperimentalSetup}, followed by sections
\ref{Calibration and reconstruction} on calibration and reconstruction
and \ref{Data and data selection} on data selection. In section
\ref{Extraction of observables}, we describe the technique how total
and differential cross sections and polarization observables 
are extracted from the data and present the results. An interpretation
of the data by means of a partial wave analysis is given in section
\ref{Partial wave analysis}. The paper ends with a short summary in
section \ref{Summary}.

\section{\label{Section:ExperimentalSetup}Experiment and Data}
\subsection{Overview}
The experiment was carried out at the electron accelerator facility
ELSA~\cite{Hillert:2006yb} at the University of Bonn using a
combination of the Crystal Barrel~\cite{Aker:1992} and
TAPS~\cite{Novotny:1991ht,Gabler:1994ay} detectors. The experimental
setup is shown in Fig.~\ref{fig:beamline}.

Electrons with an energy of $E_0=3.175$~GeV were extracted from ELSA via
slow resonant extraction, impinging on a diamond crystal producing
energy-tagged, linearly polarized photons. The photon beam hit a
liquid H$_2$ target. The direction of charged particles produced in
a photo-induced reaction could be measured by a three-layer
scintillating fiber detector inside the Crystal Barrel or plastic
scintillator tiles in front of the TAPS detector, respectively.
Charged and neutral particles were detected in the Crystal Barrel or
the TAPS detector.

\begin{figure}[pt]
\centering
{\includegraphics[width=0.4\textwidth,height=0.25\textwidth]{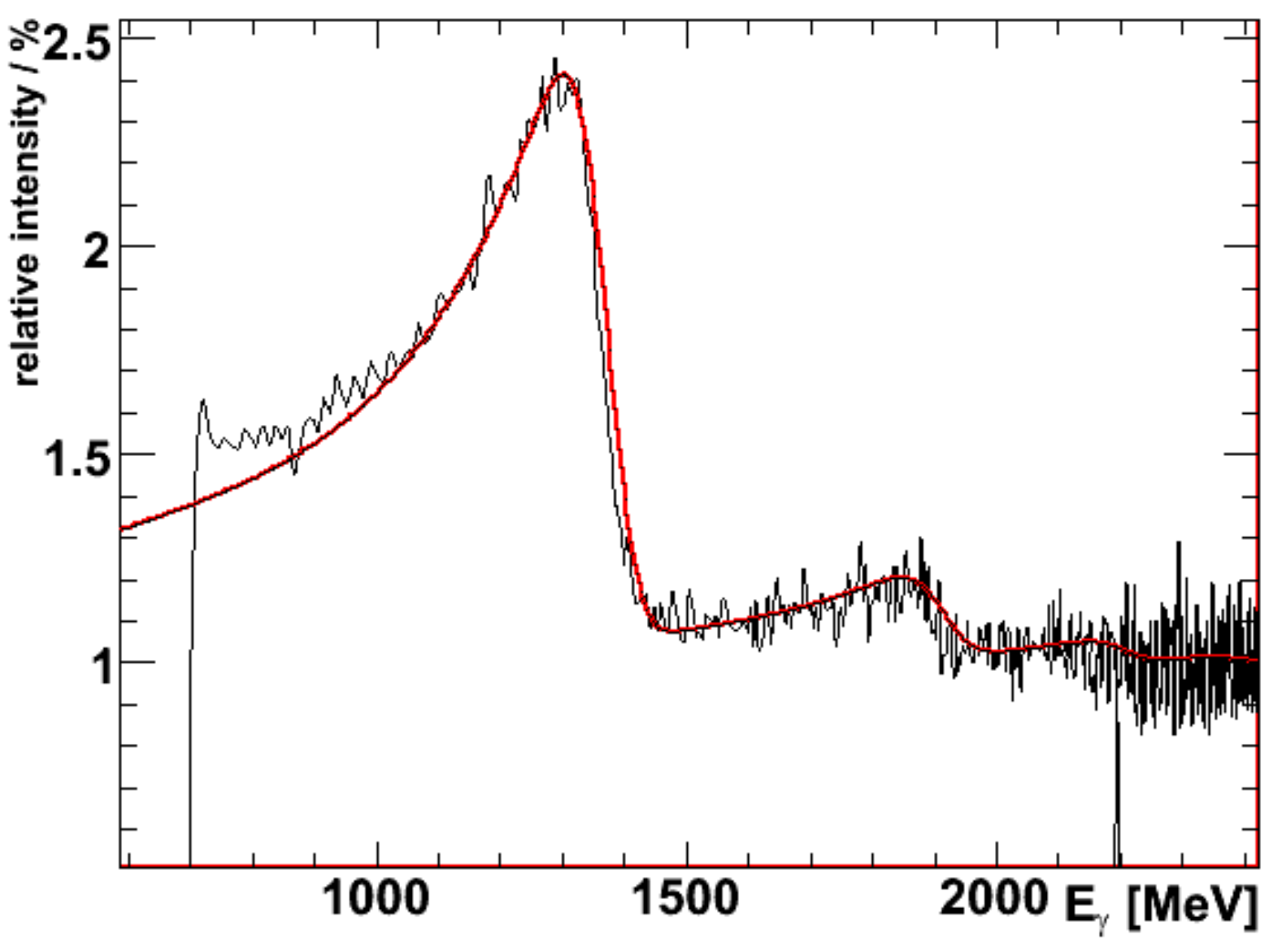}}
\caption{Relative intensity of coherent bremsstrahlung. The spectrum
was obtained with the tagging system and a diamond crystal radiator
and normalized to the corresponding incoherent spectrum
\cite{Elsner:2008sn}. The enhancements due to coherent processes are
clearly visible. Solid line: ANB calculation
\cite{Natter:2003,Phd-DE}.\vspace{-5mm}}
\label{fig:1overE}
\end{figure}
\begin{figure}[hbtp]
\centering
\vspace{5mm}
\includegraphics[width=0.48\textwidth]{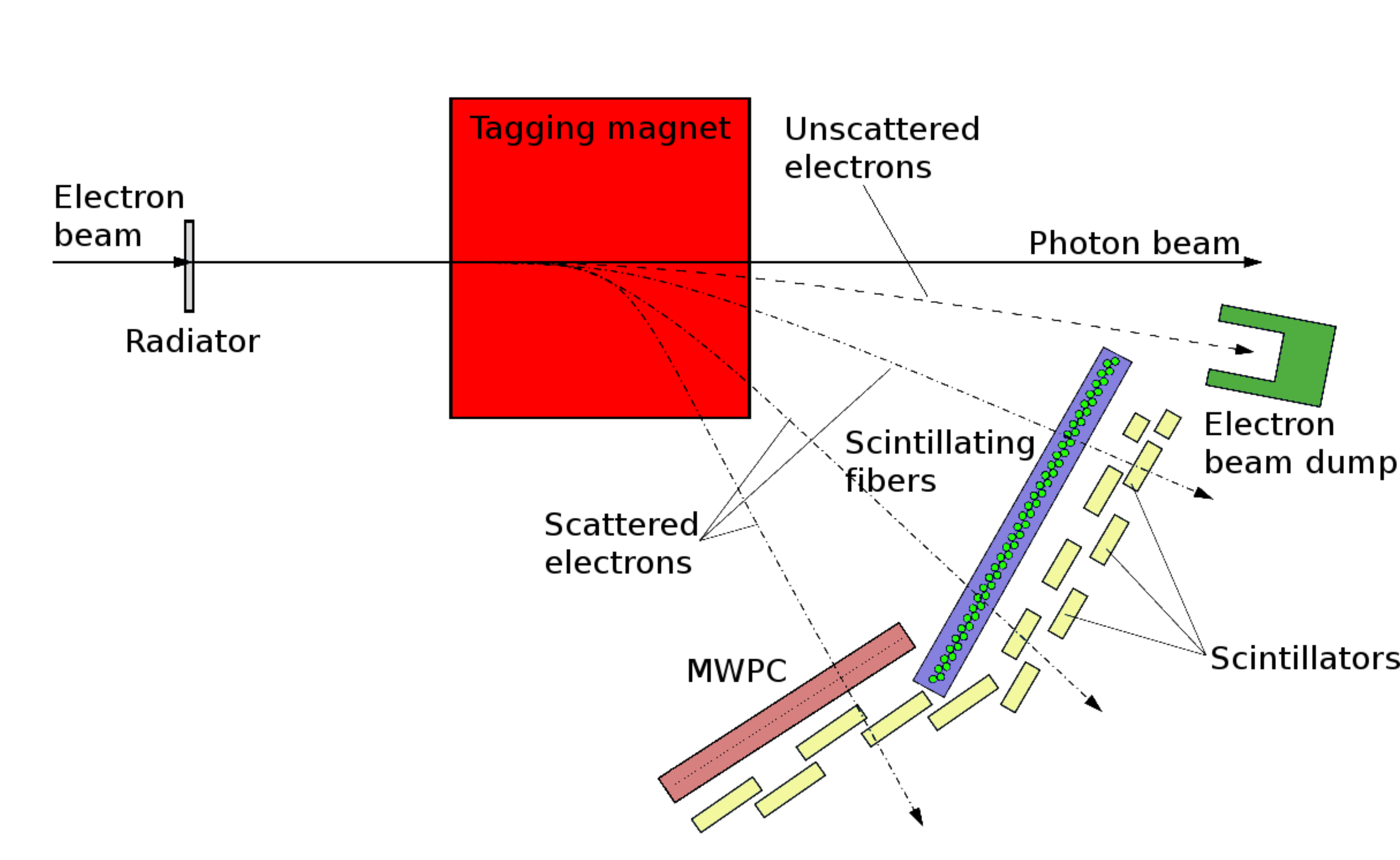}
\caption{Schematic view of the tagging system.\vspace{-3mm}}
\label{fig:tagger}
\end{figure}
\subsection{Photon beam and tagging system}
The photon-beam facility at ELSA delivered a
tagged photon beam in the energy range from 0.5 to 2.9~GeV.
The primary electron beam of energy $E_0$ passed either through a thin
copper radiator with a thickness of $(3.5/1000)\cdot X_R$ (radiation
length) or, alternatively, coherent
bremsstrahlung was produced from a 500\,$\mu$m thick diamond
crystal, corresponding to $(4/1000)\cdot X_R$. Fig. \ref{fig:1overE}
shows the intensity - normalized to the spectrum of an amorphous
copper radiator - as a function of photon energy for a typical
setting of the radiator. Technical details are given in
\cite{Elsner:2008sn}, the settings for the data presented here are
discussed in section \ref{sec: experim data}.

Electrons were deflected in the field of the tagger dipole magnet
according to their energy loss in the brems\-strahlung process.
Electrons not undergoing bremsstrahlung were deflected at small
angles and guided into a beam dump located behind the tagging
detectors. The beam dump consisted of layers of lead and iron,
combined with polyethylene and boron carbide. The beam dump was
built in a way to provide shielding of the detector system against
possible background produced by the electrons. Nevertheless, there
were background contributions in the calorimeters originating from
the beam dump. These contributions were suppressed in the offline
analysis, based on their unique topology.

The energy $E_{\rm e^-}$ of electrons which have undergone
bremsstrahlung was determined in a tagger detector consisting of
three components, as can be seen in Fig.~\ref{fig:tagger}. The
corresponding energy of a bremsstrahlung photon is $E_\gamma = E_0 -
E_{\rm e^-}$. A hodoscope of 480~scintillating fibers of 2\,mm
diameter covered the photon energy range from 18\% to 80\% of the
incoming electron energy. The energy resolution of the hodoscope
varied between 2 MeV (low electron energy) and 13\,MeV (high
electron energy). Highly deflected electrons were detected in a
multi-wire proportional chamber (208 wires), covering high-energy
photons with 80\% to 92\% of the incoming electron energy. In the
present analysis, these data were not used. Behind the scintillating
fibers and proportional chamber, 14~scintillation counters (tagger
bars) were mounted in a configuration with adjacent paddles
partially overlapping. The scintillator bars were read out by
Time-to-Digital Converters (TDC) and Charge-to-Digital Converters
(QDC).

\subsection{Target and scintillating-fiber detector}
The photons hit the liquid hydrogen target in the center of the
Crystal Barrel (CB) calorimeter. The cylindrical target cell
(52.75~mm in length, 30~mm in diameter) was built from kapton foil
of 125\,$\mu$m thickness; at both ends, a 80 $\mu$m kapton foil was
used. The target was positioned in an aluminum beam-pipe, which had
a thickness of 1\,mm. 
\begin{figure}[pb]
\centering
\includegraphics[width=0.4\textwidth]{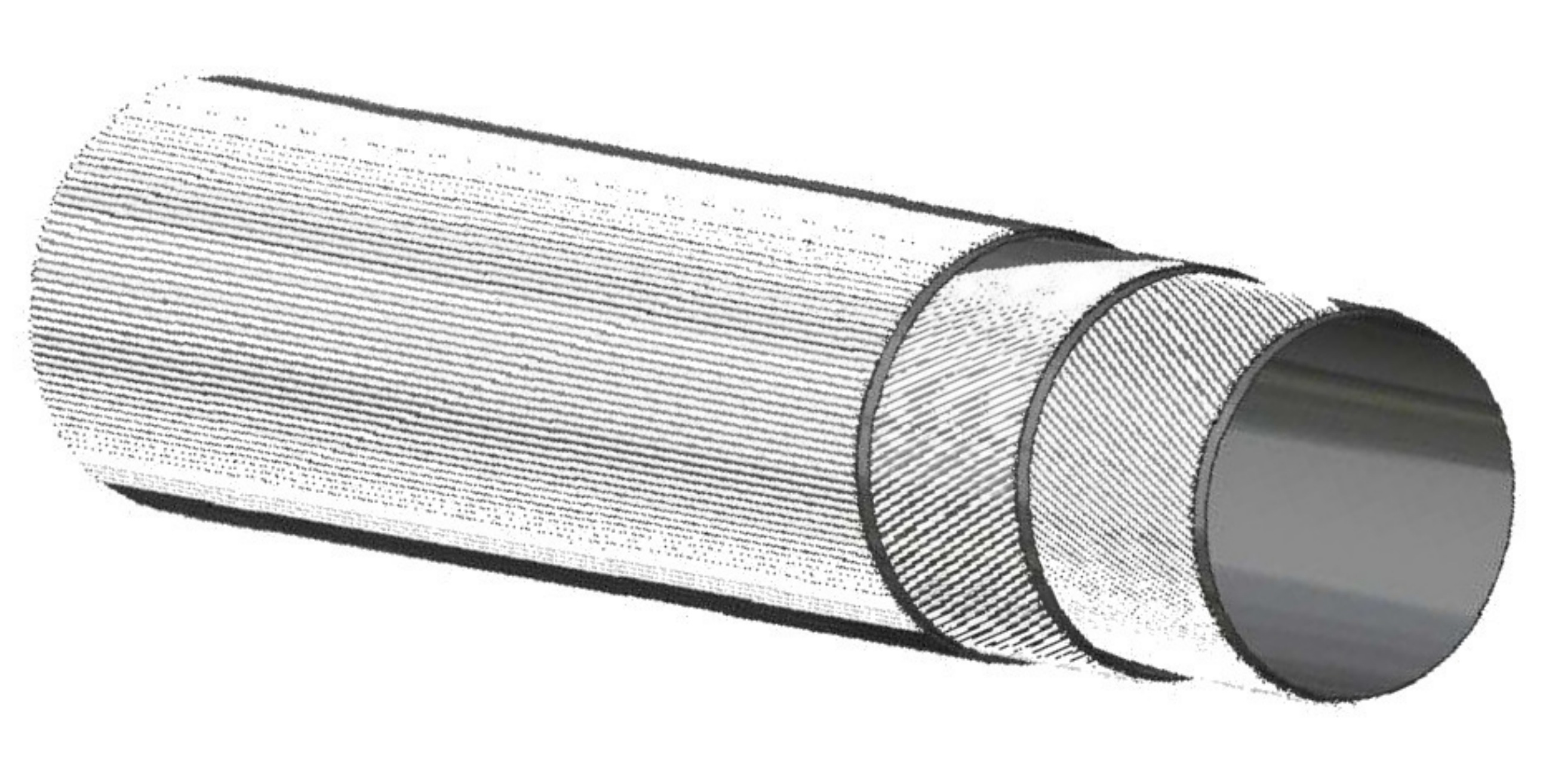}
\caption{The inner scintillating fiber detector with three layers
comprising a total of 513 fibers.}
\label{fig: Inner detector}
\end{figure}
The target was surrounded by a scintillating fiber
detector~\cite{Suft:2005cq,Phd-AF}, which provided an unambiguous
impact point for charged particles leaving the target. The detector
was 400\,mm long, had an outer diameter of 130\,mm and covered the
polar angle range of $28^{\circ} < \theta < 172^{\circ}$. It
consisted of 513 scintillating fibers with a diameter of 2\,mm,
which were arranged in three layers (see Fig. \ref{fig: Inner
detector}). For the detection of the charged particles, a
coincidence between two or three layers of the detector was used.
The outer layer (191 fibers) was positioned parallel to the beam
axis, the middle layer (165 fibers) was oriented at an angle of
$+25.7^\circ$, the innermost (157 fibers) at an angle of
$-24.5^\circ$ with regard to the beam axis. The angles resulted from
the requirement for the bent fibers to cover exactly halfway around
the detector. This arrangement allowed an unambiguous identification
of the position of the hits if only two of the fibers were fired.
The readout was organized via 16-channel photomultipliers connected
to the fibers via light guides. The efficiency of the detection in
case of the hits with two overlapping layers was 98.4\%, in the case
of three overlapping layers 77.6\%. 

\subsection{The Crystal Barrel detector}
\begin{figure}[pt]
\centering
\includegraphics[width=0.4\textwidth]{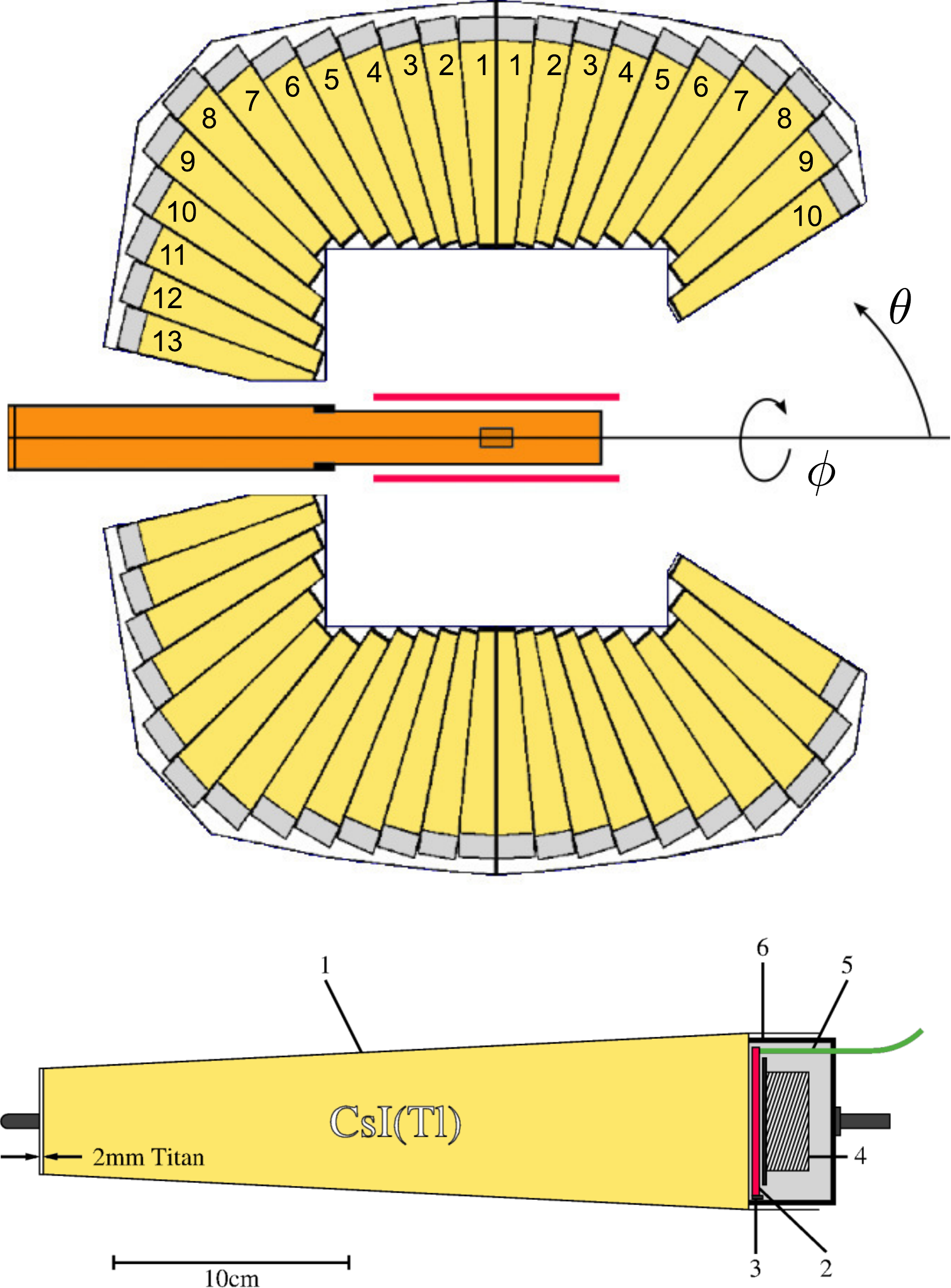}
\caption{Top: Schematic view of the Crystal Barrel calorimeter. The
photon beam enters from the left. Numbers denote $\phi$-symmetric
rings of identical module types. Also visible are the position of
the inner detector and the LH$_{2}$ target cell surrounded by the
beam pipe. Bottom: Schematic view of a CsI(Tl) module. 1: Titanium
casing, 2: Wavelength-shifter, 3: Photodiode, 4: Preamplifier, 5:
Optical fiber, 6: Electronics casing.}
\label{Figure:CB-Luzy-H2}
\end{figure}

The Crystal Barrel (see Fig. \ref{Figure:CB-Luzy-H2}) is an
electromagnetic calorimeter particularly well-suited for the
measurement of the energy and coordinates of photons. In its
CB\-ELSA/\-TAPS configuration of 2002/2003, it consisted of 1290
CsI(Tl) crystals arranged in 23 rings and covered the angular range
from $30^\circ$ to $168^\circ$ in the polar angle ($\theta$) and the
complete azimuthal range ($\phi$).
Each crystal had a length of 30\,cm corresponding to 16.1 radiation
lengths \cite{Aker:1992} and covered $6^\circ$ both in polar and
azimuthal angles. An exception were the crystals in the three
backward rings which covered $6^\circ$ in $\theta$ and $12^\circ$ in
$\phi$. The energy of the photons inducing showers in the Crystal
Barrel was determined via identification of clusters of hit crystals
and by summation of the energy deposits in a cluster. Details are
given in Section \ref{Reconstruction}. The energy resolution of the
Crystal Barrel depends on the energy of the photons as
\cite{Aker:1992}:
\begin{equation}
\frac{\sigma(E)}{E}  = \frac{2.5\%}{\sqrt[4]{E[\mathrm{GeV}]}}.
\end{equation}

The spatial resolution of the Crystal Barrel is related to the
energy of the photons, and for energies higher than 50\,MeV was
better than $1.5^{\circ}$ if a special weighting algorithm was used
for the determination of the photon position (see Section
\ref{Reconstruction}). The energy deposit in the  crystals of the
Crystal Barrel was measured via detection of the scintillation light
using PIN photodiode readout. Its wavelength was shifted using a
wavelength shifter to longer wavelengths matching the range of the
sensitivity of the photodiode. After detection by the photodiode the
signal was processed via preamplifier and dual-range ADC readout. To
control the functionality of the readout and stability of the
calibration, a light-pulser system was used, feeding light directly
into the wavelength shifter. For more details see \cite{Dipl-OB}.
\begin{figure}[pt]
\centering
\includegraphics[width=0.48\textwidth]{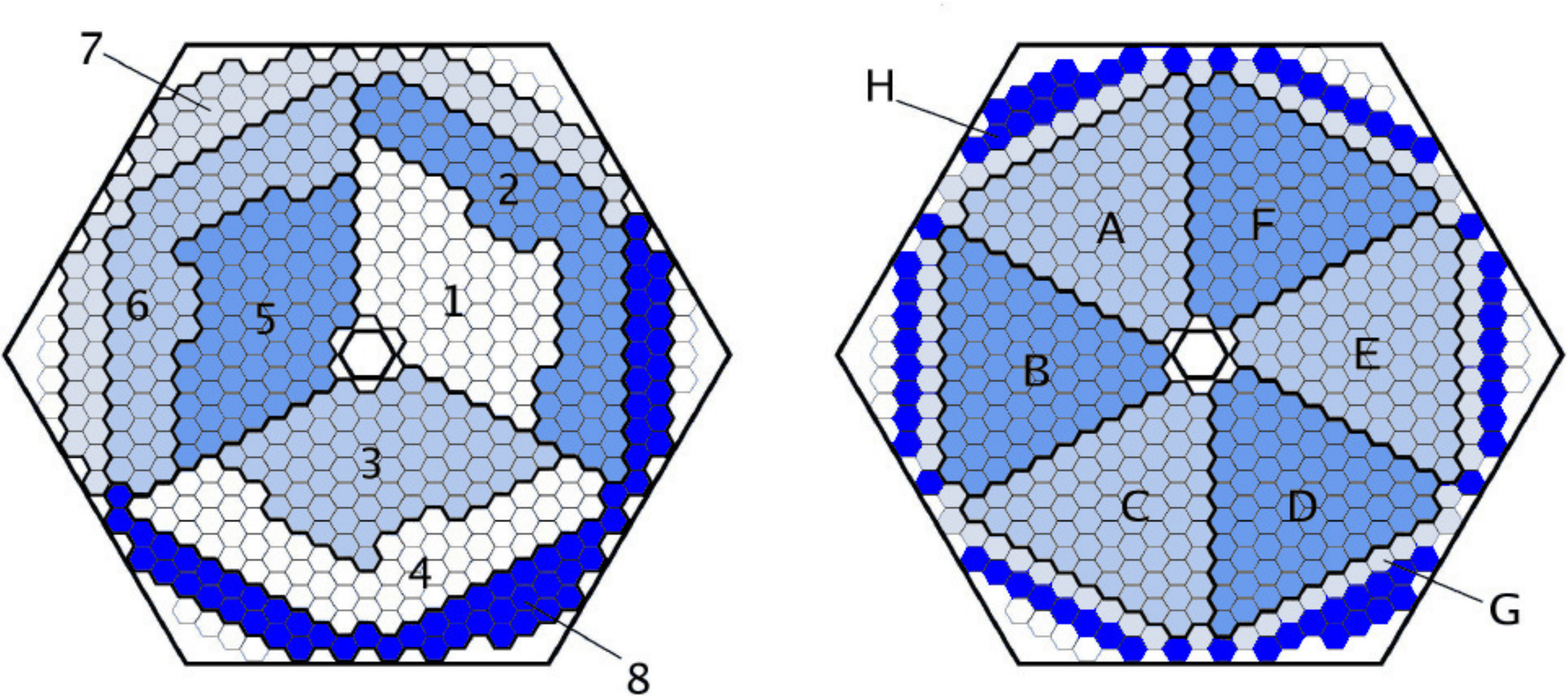}
\caption{TAPS trigger segmentation, on the left: segmentation used
for the LED low trigger,  on the right: the sectors used in the
formation of the LED high trigger.}
\label{fig: taps led}
\end{figure}
\subsection{The TAPS detector}
The TAPS calorimeter complemented the Crystal Barrel in the forward
direction, covering the polar angular range from $30^{\circ}$ down
to $5.8^{\circ}$ and the full azimuthal range. It consisted of 528
hexagonal BaF$_2$ modules in a forward wall setup, 1.18\,m from the
target center. A schematic picture of the TAPS crystal assembly and
its hexagonal shape is shown in Fig. \ref{fig: taps led}. The forward
opening of TAPS allowed for the passage of the photon beam.

The BaF$_2$ modules used in the TAPS calorimeter have a
width of 59\,mm and a length of 250\,mm, corresponding to 12
radiation lengths. Their cylindrical rear end with a diameter of 54\,mm
was attached to a photomultiplier. The front end of the crystal was
covered by a separate, 5\,mm thick plastic scintillator for the
identification of charged particles (mounted in a contiguous wall in
front of the calorimeter). These {\it charged particle vetos}
\cite{Dipl-SJ} were read out via optical fibers connected to
multi-anode photomultipliers. The readout of the BaF$_{2}$ modules
with photomultipliers allowed for signal processing without the use
of additional preamplifiers or shapers and could be used not only
for energy but also timing information. The signals of each module
were split and processed by constant-fraction discriminators (CFDs,
see section~\ref{TAPS calibration}) and leading-edge discriminators
(LEDs) for trigger purposes as described in the next paragraph.

The combination of the Crystal Barrel and TAPS calorimeters covered
99\,\% of the $4\pi$ solid angle and served as an excellent setup to
detect multi-photon final states. For photons up to an energy of
790\,MeV, the energy resolution of TAPS as a function of the photon
energy is given by \cite{Gabler:1994ay}
\begin{equation}
\frac{\sigma(E)}{E} = \frac{0.59\%}{\sqrt{E_{\gamma}}}+1.91\%\,.
\end{equation}

\subsection{The Photon Intensity Monitor}
The Photon Intensity Monitor consisted of a $3\times3$ matrix of
$\rm PbF_{2}$ crystals with photomultiplier readout, mounted at the
end of the beam line. It was used to measure the position of the
photon beam in the set-up period and to monitor the photon flux by
counting the coincidences between the Photon Intensity Monitor and
the tagging system.

\subsection{\label{sec:trig}The trigger}
The purpose of the trigger was to select events in which a tagged
photon of known energy induced a hadronic reaction in the target
with a minimum number of photons detected in the Crystal Barrel or
in TAPS, and to suppress $e^{+} e^{-}$ background from the
conversion of beam photons. The TAPS modules provided the
first-level trigger. A second-level trigger was based on a cellular
logic (FACE), which determined the number of clusters in the Crystal
Barrel. The trigger required either two hits above a low-energy
threshold in TAPS, or one hit above a higher-energy threshold in
TAPS in combination with at least one FACE cluster.

\paragraph{First-level trigger:}
The TAPS crystals were read out with two Leading Edge Discriminator
settings, one set at a high threshold (LED high) and one at a low
threshold (LED low). The shape of the logical segmentation for the
TAPS trigger is shown in Fig.~\ref{fig: taps led}. The thresholds
have been set ringwise, with values increasing with decreasing polar
angle. If two LED low signals from different segments are present,
the event was digitized and stored, and the decision of the second
level trigger was not required. If at least one TAPS crystal
provided a LED high signal, the event was digitized but the decision
about its final storage was deferred to the second level trigger.

\paragraph{Second-level trigger:}
The second level trigger was based on the relatively slow Crystal
Barrel signals. Two different trigger settings have been used in the
data presented here, demanding one or two clusters to be identified
by the cluster-finder algorithm FACE. The processing time varied
between $ 6 \, \mu \rm s$ and $10 \, \mu \rm s$, depending on the
size of the clusters. Accepted events were stored.

\section{\label{Calibration and reconstruction}Calibration and reconstruction}
\subsection{Crystal Barrel calibration}
\begin{figure}[pt]
\vspace{3mm} \centering
\includegraphics[width=0.42\textwidth]{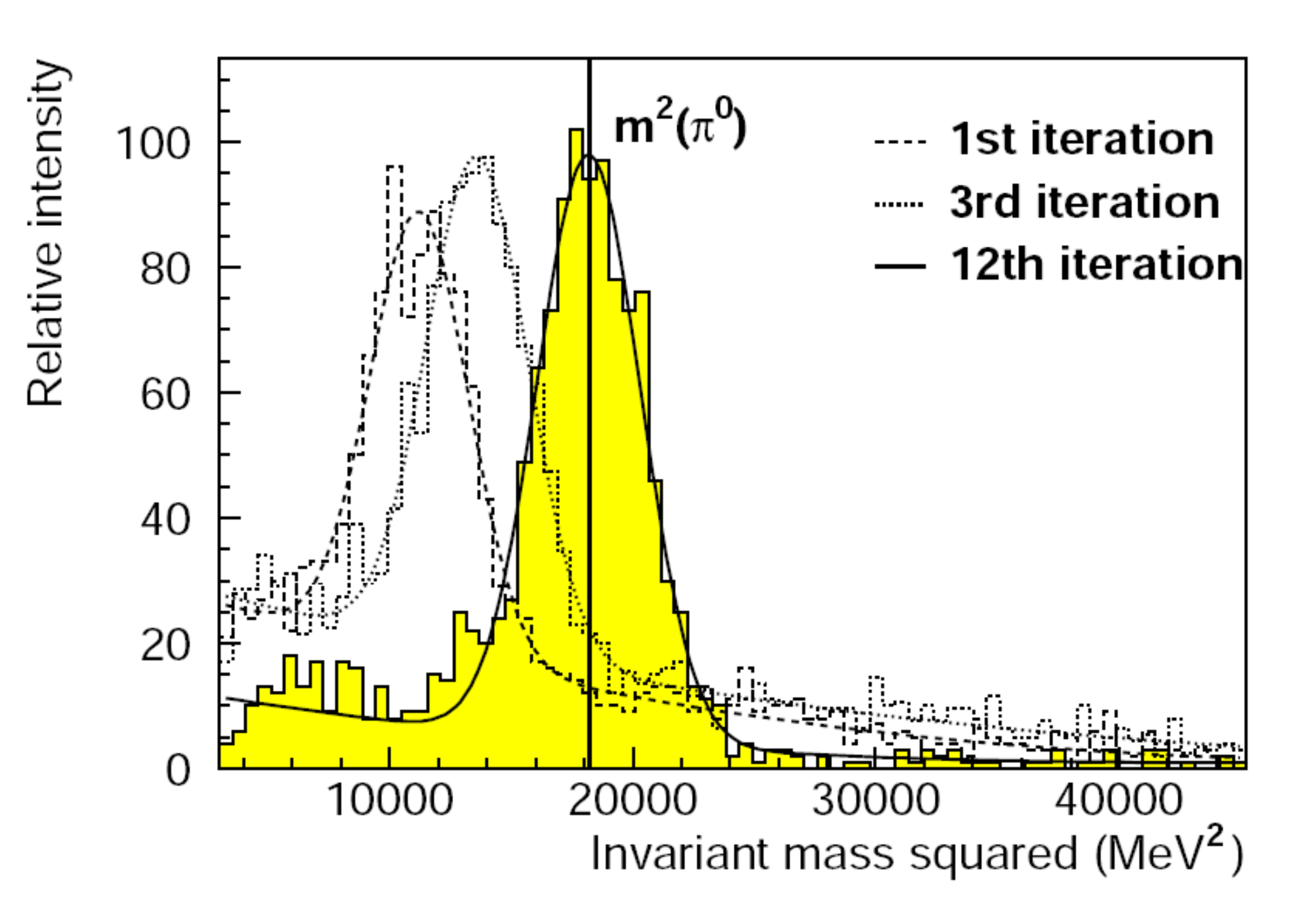}
\caption{$\pi^{0}$ peak used for the calibration of one CsI(Tl)
module in different iterations of the calibration procedure
\cite{Dipl-JJ}.}
 \label{fig:cbpion}
\end{figure}

The calibration of the Crystal Barrel used the position of the
$\pi^{0}$ peak in the $\gamma\gamma$ invariant mass spectrum (the
reconstruction of $\gamma$ energies from hits in the Barrel is
described below in section \ref{Reconstruction}). Fig.
\ref{fig:cbpion} shows the invariant mass spectra of photon pairs,
one of them hitting a selected crystal, for different iterations. As
long as the invariant mass was too low, the assigned photon energy
was increased. After all crystals were adjusted, the next iteration
was started until the results were stable. The calibration using the
$\pi^{0}$ mass was sufficient; no additional correction was required
to adjust the $\eta$ mass.
The CB  crystals were read out via dual range ADCs. The low ADC
range covered the photon energies up to about 130 MeV and was
calibrated using the $\pi^{0}$ mass as a reference. The high range
of the ADC covered the energies up to about 1100 MeV. The high range
was calibrated relative to the low range using  a light-pulser with
adjustable attenuations. The light was directly fed into the
wavelength shifter of the CsI(Tl) readout, and produced signals with
shapes simulating the response of the CB crystals. Using the known
gain of the ADC in the low range, the calibration was extended to
the high range.

\subsection{\label{TAPS calibration}TAPS calibration}
Since the fast photomultiplier readout of the TAPS
ca\-lo\-ri\-me\-ter allowed for timing signals to be branched off,
TAPS required a timing calibration as well as an energy calibration.

\begin{figure}[pt]
\centering
\includegraphics[width=0.40\textwidth]{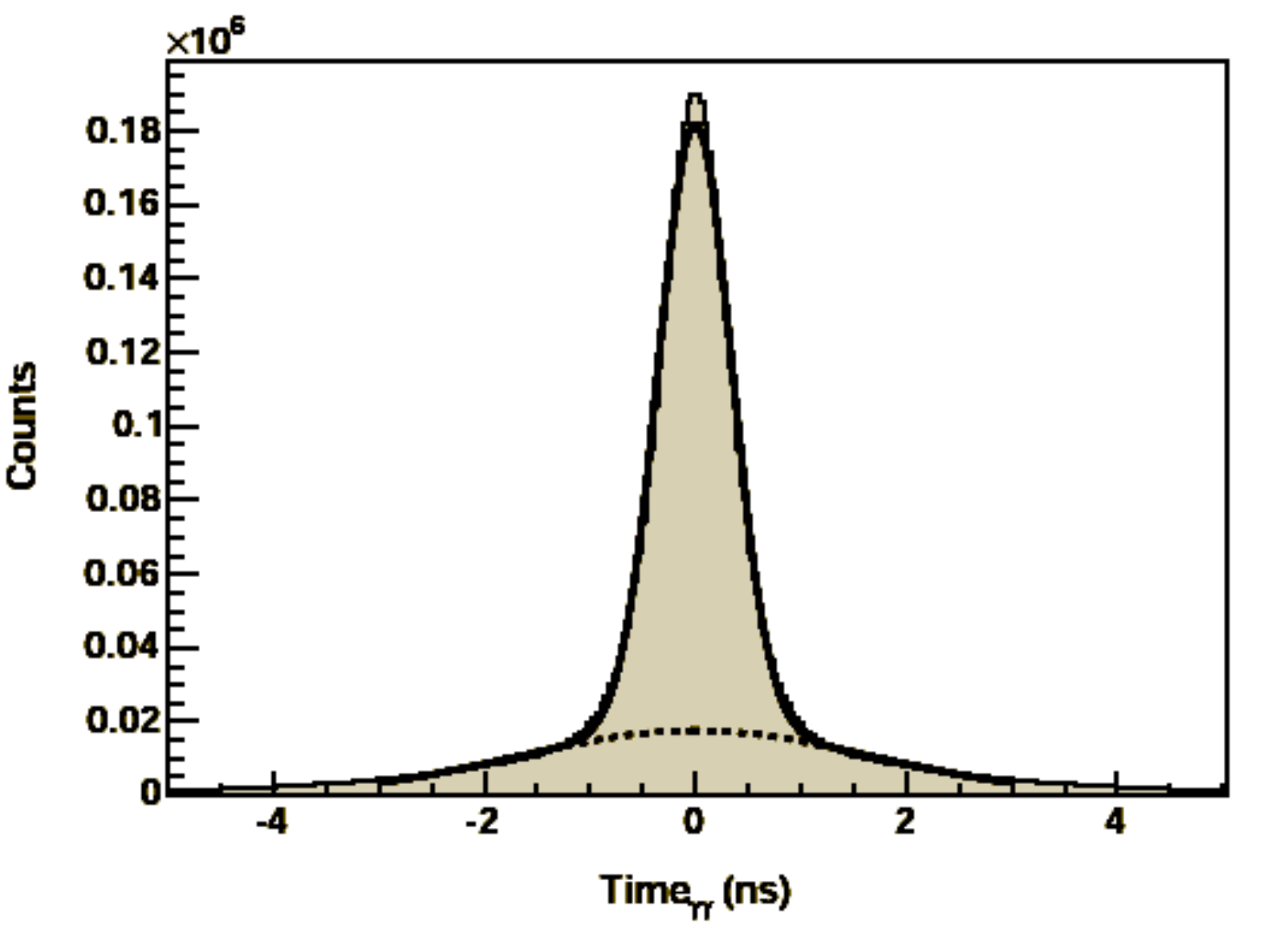}
\caption{Time difference between hits in the TAPS crystals. The
crystals for which the time difference was within the grey area were
considered as parts of the same~cluster,~taken~from~\cite{Phd-RC}.
\vspace{5mm}}
\label{fig: time rec taps}
\centering
\includegraphics[width=0.35\textwidth]{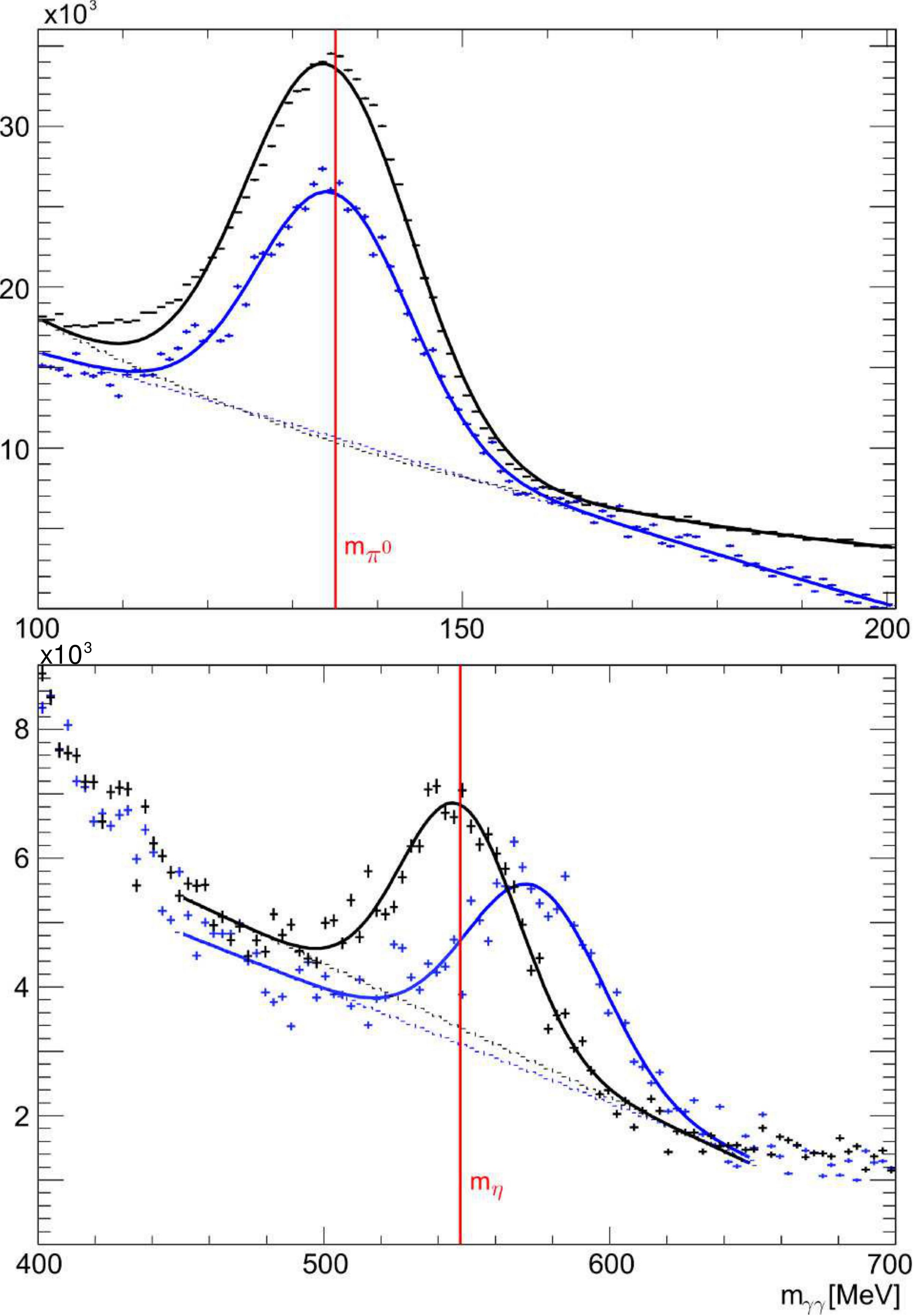}
\caption{Effect of the energy-dependent calibration (TAPS).
Reconstructed $\pi^{0}$- (top) and $\eta$-mass (bottom) before
(blue) and after (black) application of the correction function
(\ref{eq:endep}) (using different data samples, counts refer to
uncorrected spectra).
}
\label{fig:endep}
\end{figure}

\paragraph{Time calibration:}
Two steps were done for the time calibration of TAPS: calibration of
the gain and calibration of the time offsets of the TAPS TDC
modules. The gain of the TAPS TDC modules was calibrated using
pulses with known repetition frequencies, and by variation of the
pulse frequency. The TDC time offsets (due to different cable
lengths) were calibrated using the time differences between two
neutral hits detected in TAPS, belonging to the same $\pi^{0}$
decay. Fig. \ref{fig: time rec taps} shows the distribution of the
timing differences of different TAPS modules after calibration. The
time resolution was determined to $\sigma = 0.35$\,ns.

\paragraph{Energy calibration:}
The energy calibration of TAPS was performed in three steps: using
cosmic rays, the invariant mass of the photons from $\pi^{0}$
decays, and using the invariant mass of the photons from $\eta$
decays. Cosmic-ray muons deposit an average energy of 38.5\,MeV in
the 59\,mm wide TAPS crystals. The energy offset was determined
using electronic pulses of minimum pulse height. The positions of
the minimum ionizing peak and of the pedestal peak obtained with the
pulser determined a first approximate calibration of the
charge-to-digital converters using a linear function. In a second
step, the invariant mass of the two photons originating from
$\pi^{0}$ decays - with one photon hitting crystal $i$ as central
crystal of a cluster - was compared to the nominal $\pi^{0}$ mass,
and a gain correction was applied to the energy of crystal $i$. This
procedure was performed iteratively until the distribution of
invariant masses of photon pairs reproduced the $\pi^{0}$ mass. The
result of this calibration is shown in Fig. \ref{fig:endep}, top.
Good agreement between the calculated invariant mass and the nominal
$\pi^{0}$ mass was achieved.

Photons from $\eta$ decays were used in a third step. Of course,
here the statistics was comparatively low. Therefore an overall
correction gain factor was determined only. The correction function
was applied in the form:
\begin{equation}
 E_{new} = a\cdot E_{old} +b\cdot E_{old}^{2},
 \label{eq:endep}
\end{equation}
where the coefficients $\it a$ and $\it b$ were determined so that
both measured masses, the $\pi^{0}$ and $\eta$ mass, coincided with
the nominal values. Typical values for the coefficients were $a =
1.0165$, $b \rm = -5.6715 \cdot 10^{-5}$. Fig. \ref{fig:endep} shows
the effect of the correction of the invariant mass on two photons
using Eq.~\ref{eq:endep}.

\subsection{Tagger calibration}
\label{sec: tagger calib}
\begin{figure}[ph]
\centering
\includegraphics[width=0.40\textwidth]{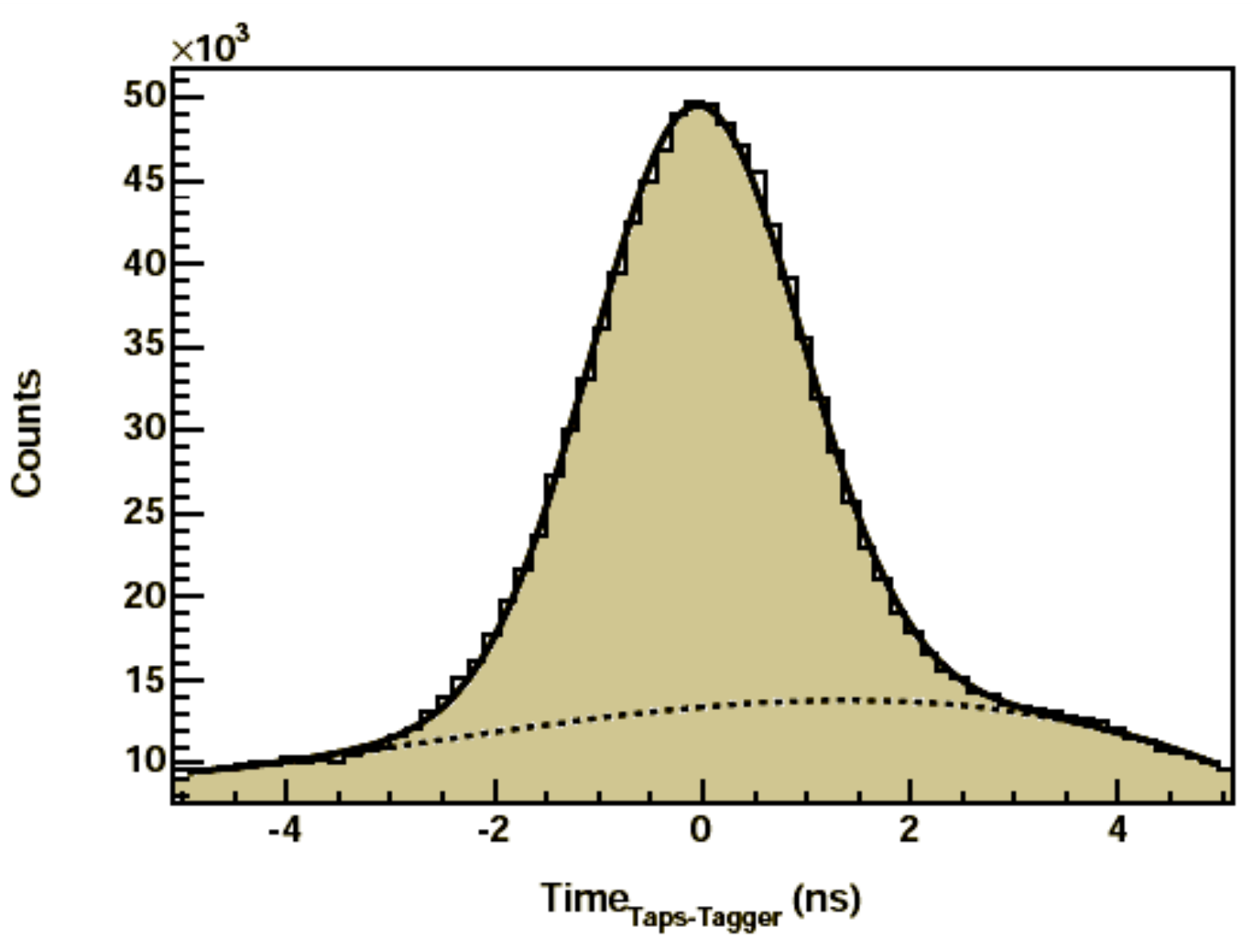}
\caption{Calibrated relative timing between the tagging hodoscope
and TAPS, taken from \cite{Phd-RC}.}
\label{fig:time}
\end{figure}

\paragraph{Time calibration:}
The time of the fibers of the tagger hodoscope relative to the TAPS
timing was measured by time-to-digital converters. Fig.
\ref{fig:time} shows the distribution of the time differences of
fiber signals and signals in TAPS after calibration. A tagger time
resolution of better than $\sigma = 1$\,ns was deduced.

\begin{figure}[ht!]
\centering
\includegraphics[width=0.40\textwidth]{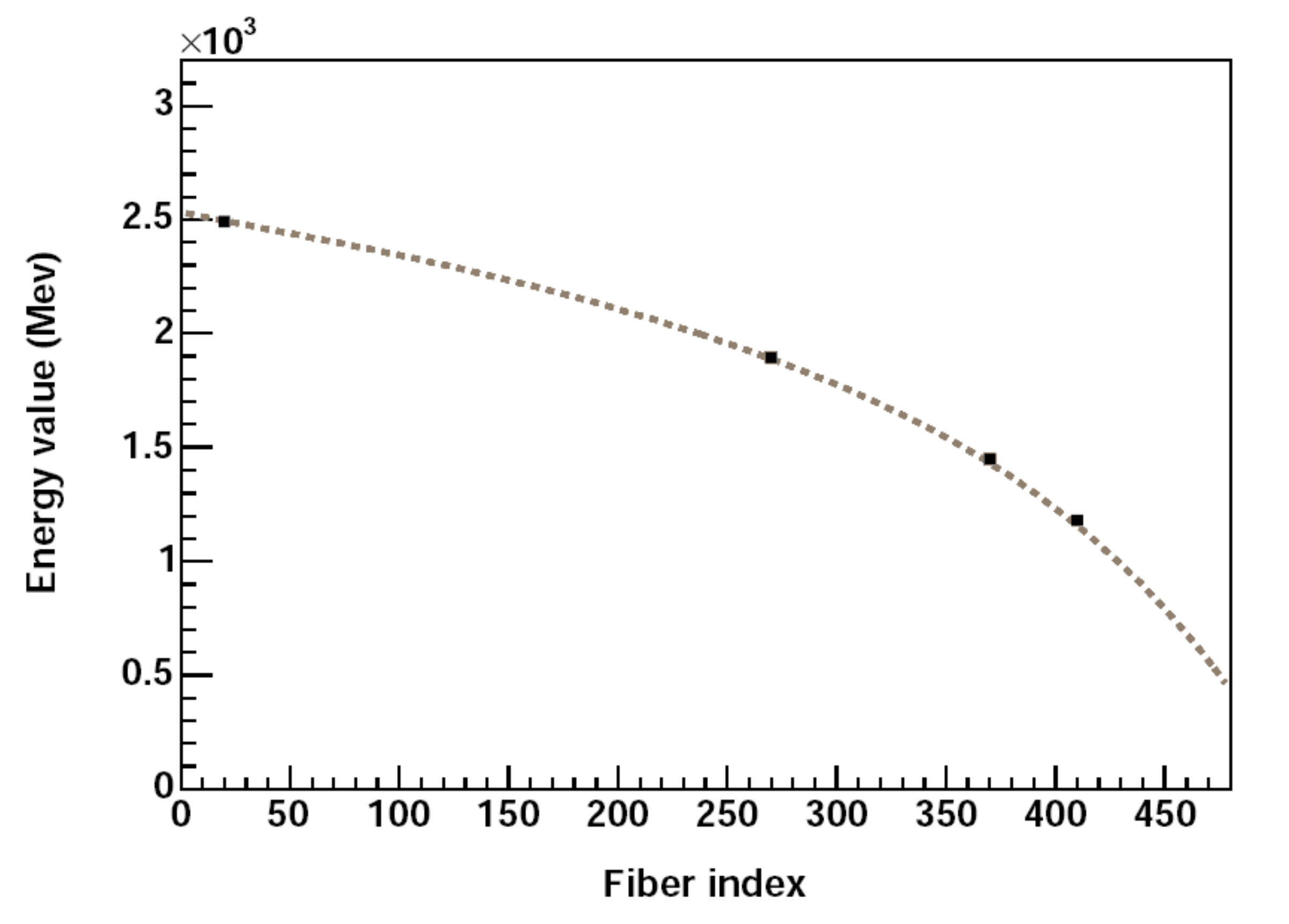}
\caption{Relation between photon energy and fiber number. Dashed
line: Tagger polynomial (\ref{tagpoly}), points: Data obtained by
direct injection.}
\label{fig:tagenergy}
\end{figure}

\paragraph{Energy calibration:}
Electrons hitting one of the fibers of the tagging hodoscope have
been deflected in the field of the tagger magnet. For electrons
which have passed the radiator, each fiber defined a small range of
energies. To determine the energy of electrons as a function of the
hit fiber, the relation between the energy of the electron and the
fiber number was calculated from the known geometry and the known
magnetic field map, resulting in a fifth degree polynomial. This
polynomial was then corrected using the direct injection of
electrons with four different energies (680, 1300, 1800, and
2500\,MeV) at a constant field of the tagging dipole of 1.413\,T.
The final polynomial used in this work is given by
\begin{eqnarray}
\label{tagpoly}
E & = & 2533.81 - 190.67\cdot10^{-2}x\\
& + &  28.86\cdot10^{-4}x^{2}   - 34.43\cdot10^{-6}x^{3}\nonumber\\
& + &  95.59\cdot10^{-9}x^{4} - 12.34\cdot10^{-11}x^{5}, \nonumber
\end{eqnarray}
where $E$ is the photon energy and $x$ the fiber index (see
Fig.~\ref{fig:tagenergy}).

\subsection{\label{Reconstruction}Reconstruction}
In this section we discuss how kinematic variables like energies and
momenta were calculated from the detector information using various
reconstruction routines.

\paragraph{Crystal Barrel reconstruction:}
The reconstruction of photons in the Crystal Barrel detector was
performed in the same way as described in \cite{Pee-pi0}, forming
clusters of contiguous CsI(Tl) crystals with individual energy
deposits above 1\,MeV and an energy sum of at least 20\,MeV to
reduce the contributions from split-offs, separate clusters of
energy, which might have originated from fluctuations of the
electromagnetic shower induced by photons. If there was only one
local maximum in the cluster, the energy measured by the
participating crystals $i$ was added directly to get the {\it
Particle Energy Deposit} (PED)
\begin{equation}
 E_{\mathrm{PED}}= E_{\mathrm{cl}} = \sum\limits_{i} E_{i},
\end{equation}
where $E_{\mathrm{cl}}$ is the cluster energy. There may be,
however, two or more local maxima (central crystals $k$) in one
cluster. Then, the sum of energy contents of each of the central
crystals and its eight neighbors, the {\em nine-energy} $E_{9}$, was
formed
\begin{equation}
\label{eq:nine}
 E_{9}^{k}=E_{\mathrm{cen}}^{k}+\sum\limits_{j=1}^{8}E_{j}^{k},
\end{equation}
and the total cluster energy was shared accordingly,
\begin{equation}
 E_{\mathrm{PED}}^{i}=\frac{E^{i}_{9}}{\sum\limits_{k}^{}E^{k}_{9}}\cdot E_{\mathrm{cl}}.
\end{equation}
If a crystal $i$ was adjacent to several ($j$) local maxima, its
energy contribution to the nine-energies $E_{9}^{i}$ was shared
proportional to the energy deposits in the central crystals. For
central crystal $k$, the fraction reads
\begin{equation}
\label{eq:9frac}
 E^{i}_{9k,\mathrm{frac}} = \frac{E_{\mathrm{cen}}^{k}}{\sum\limits_{j}^{}E^{j}_{\mathrm{cen}}}\cdot E^{i}_{k},
\end{equation}
where the summation extends over the $j$ adjacent local maxima.\\
Some loss of the shower energy was unavoidable due to edge effects
and some inactive material like the aluminum holding structure of
the Crystal Barrel. 

A $\theta$-dependent correction function to the
PED energy improved the energy resolution:
\begin{equation}
E_{\mathrm{PED}}^{\mathrm{corr}} = \left(a(\theta)+b(\theta)\cdot
e^{-c(\theta)\cdot E_{\mathrm{PED}}}\right)\cdot E_{\mathrm{PED}}.
 \label{eq: energy corr}
\end{equation}
The parameters were determined using Monte Carlo simulations, with typical values of 
$a \approx 1.05$, $b \approx 0.05$, $c \approx 0.007$.

The position of PEDs in the Crystal Barrel calorimeter was
determined by a weighted average of the polar and azimuthal angles of the crystal
centers, $\theta_{i}$ and $\phi_{i}$,
\begin{equation}
\theta_{\mathrm{PED}} =
\frac{\sum_{i}w_{i}\theta_{i}}{\sum_{i}w_{i}},\;\;\;
\phi_{\mathrm{PED}} = \frac{\sum_{i}w_{i}\phi_{i}}{\sum_{i}w_{i}},
\end{equation}
where the factors $w_{i}$ are defined as
\begin{equation}
 w_{i} = {\mathrm{max}}\left\{0;W_{0}+\ln\frac{E_{i}}{\sum_{i}E_{i}}\right\},
 \label{formula: weight}
\end{equation}
with a constant $W_{0} = 4.25$.  In case of multi-PED clusters, only
the central crystal and its direct neighbors were used for the
position reconstruction, so the sum over all crystal energies,
$\sum_{i}E_{i}$, in (\ref{formula: weight}) was replaced by the
nine-energy (\ref{eq:nine}).

A spatial resolution for photons of $1^{\circ} - 1.5^{\circ}$ in
$\theta_{\mathrm{lab}}$ and $\phi_{\mathrm{lab}}$, depending on the
energy and polar angle of the incident photon, was achieved. A
shower depth correction (see below) was not necessary since all
crystals point to the center of the target.

\paragraph{TAPS reconstruction:}
\label{sec: taps recon}
The energies and the coordinates of the photons in TAPS were
determined in a way similar to the Crystal Barrel described above.
Here, a cluster was defined as any contiguous group of BaF$_{2}$
crystals registering an energy deposit above their CFD threshold.
The hardware thresholds were set to 10\,MeV.
Again, the crystal with the largest energy deposit within the
cluster was taken as the central crystal. The times of the central
crystal and other crystals in the cluster were compared to reject
contributions from other clusters, not correlated in time; a time
window of 5\,ns was chosen (see Fig. \ref{fig: time rec taps}). The
total energy of the cluster had to exceed 25\,MeV. This cut reduced
the probability of fake photons due to split-offs. However, for the
polarized data a software value of 30\,MeV was used for both, the
CFD threshold and the maximum cluster energy, to avoid $\phi$
dependent systematic effects due to drifts.
Due to the geometry of the TAPS calorimeter, the position of the
photons was calculated using the cartesian coordinates of the
crystals which formed the cluster, again with energy-dependent
weighting factors \cite{Phd-RC}:
\begin{equation}
\label{eq:sumweight}
X = \frac{\sum_{i}w_{i}x_{i}}{\sum_{i}w_{i}},\;\;\; Y =
\frac{\sum_{i}w_{i}y_{i}}{\sum_{i}w_{i}},
\end{equation}
where the factors $w_{i}$ are defined as in eq.~(\ref{formula:
weight}), with a constant $W_{0} = 4.0$. Since the crystals were
mounted in a plane and photons originating from the target developed
a shower within the crystals, the center of the shower had a
displacement (in the outer direction) which increased with
penetration depth and hence with the photon energy. This effect was
accounted for by a shift of the reconstructed impact point, $s$, (on
the crystal surface) of
\begin{equation}
\frac{\Delta X}{X}=\frac{\Delta Y}{Y}=\left(\frac sZ
+1\right)^{-1};\, Z =
X_{0}\left(\ln\frac{E}{E_{c}}+C_{\gamma}\right),
\label{formula: taps rec}
\end{equation}
with $C_{\gamma}=2.0$ as determined via Monte Carlo simulations.
$X_0$ is the radiation length of BaF$_2$, $E$ the photon energy and
$E_c = 12.78$\,MeV is the critical energy for BaF$_2$. The
resolution in the polar angle was determined to be better than
$1.3^{\circ}$.\vspace{-2mm}
\begin{figure}[pb]
\centering
\includegraphics[width=0.45\textwidth]{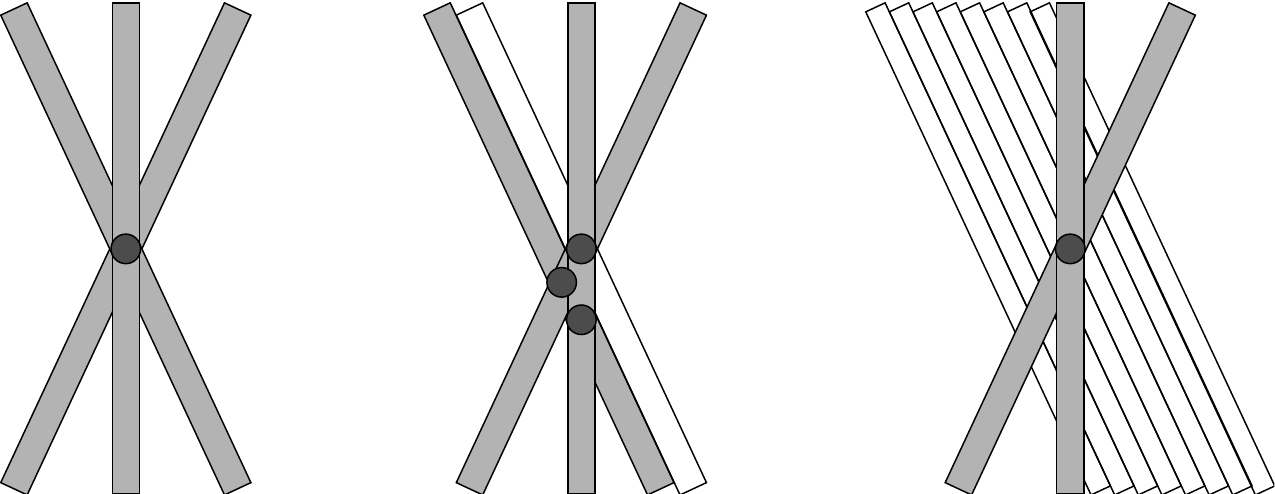}
\caption{Charged particles crossing the inner fiber detector produce
different hit patterns. A hit in three layers may result in one
unique intersection point (left) or in three intersection points
(middle). One inefficiency still yields a defined intersection point
(right).}
\label{kreuzung}
\end{figure}

\paragraph{Inner detector reconstruction:}

A charged particle traversing the fiber detector could fire one or
two adjacent fibers in each layer forming a {\it British-flag}-like
pattern which defined the impact point (see Fig. \ref{kreuzung},
left). A single charged particle may also fake three intersection
points (see Fig. \ref{kreuzung}, middle). In order not to lose these
events, up to three intersection points were accepted in the
reconstruction. One inefficiency (one broken or missing fiber)
leading to a pattern shown in Fig. \ref{kreuzung}, right, was
accepted by the reconstruction routine. The accuracy of the
reconstruction of the impact point was studied using simulations and
was determined, for a pointlike target, to $\pm0.5$ mm in the $X$-
and $Y$-coordinates (representing the resolution in $\phi$). The
uncertainty in the $Z$ coordinate was determined to be 1.6\,mm. 
The angular resolution was $\Delta\phi = 0.4^\circ$ and up
to $\Delta\theta = 0.1^\circ$.\vspace{-2mm}

\begin{figure}[bt]
\centering
\includegraphics[width=0.40\textwidth,height=0.25\textwidth]{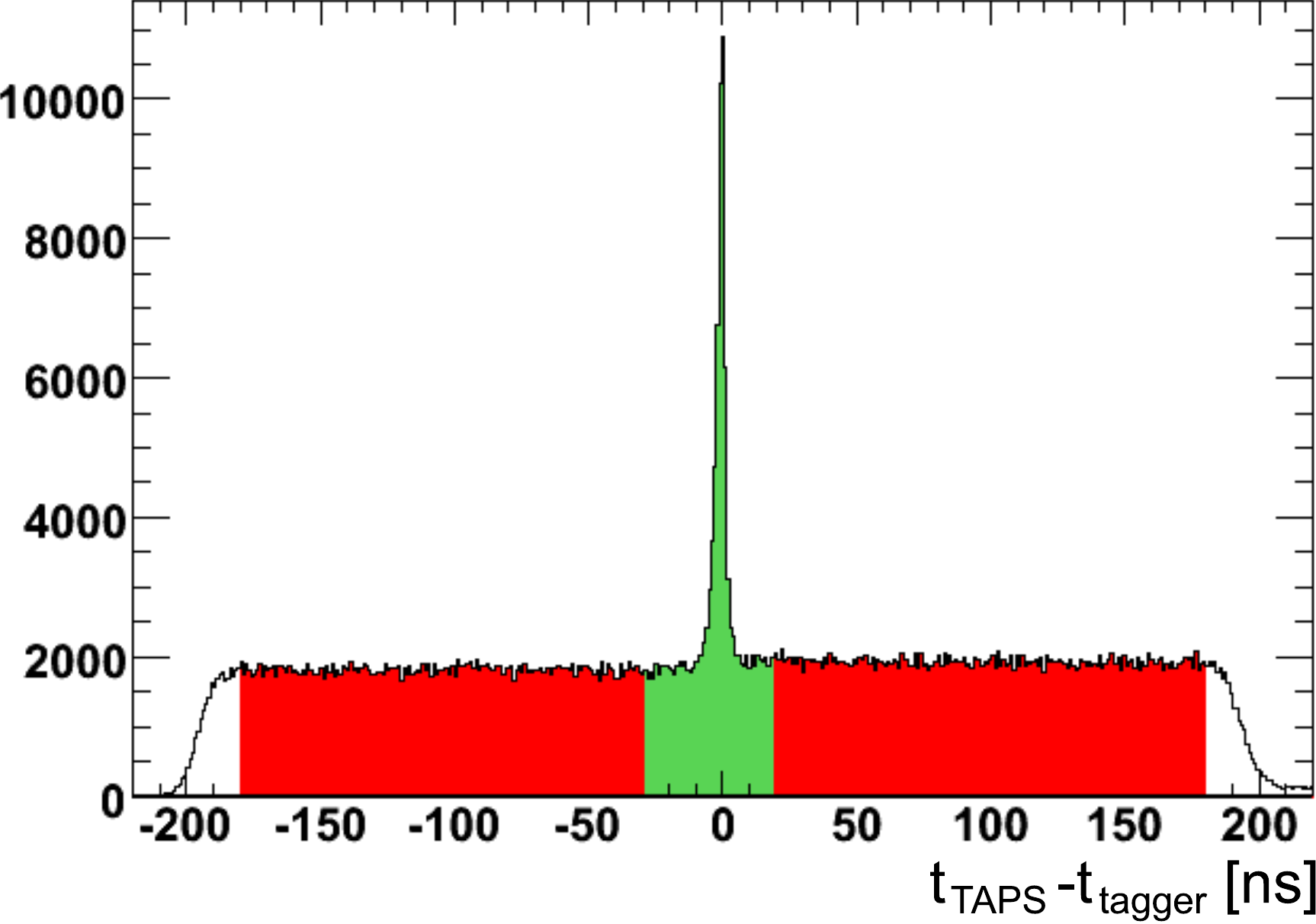}
\caption{TAPS-tagger time spectrum. Background events in the signal
region (green) were accounted for by sideband subtraction using
events from the uniform part in the random background region (red).
The signal region was chosen asymmetrically to account for timing
signals produced by protons in the TAPS calorimeter.}
\label{fig: time}
\end{figure}

\paragraph{Tagger reconstruction}
Electrons hitting the tagger could fire one or more fibers. In case
of more than one fiber hit, contiguous groups of tagger fibers
forming clusters were identified. Afterwards the fiber numbers were
averaged, and the mean value was used to calculate the electron
energy, see Eq.~(\ref{tagpoly}). The time of the cluster was taken
to be the mean time of the fibers participating in the cluster. For
fibers combined to one cluster, a 2\,ns time window was imposed. To
account for inefficiencies, a group of fibers was still considered
as a cluster when one fiber in between two fired fibers had no
signal.

Coincidences between tagger and TAPS were imposed in the offline
analysis to maintain control of random coincidences. The
distribution of time differences of tagger and TAPS hits,
$t_{\mathrm{TAPS}}-t_{\mathrm{tagger}}$, is shown in Fig. \ref{fig:
time}. The peak around 0\,ns corresponds to the coincident hits, the
evenly distributed events to the uncorrelated accidental background.
The coincidence region was selected with a (-30, 20) ns wide cut.
The cut is asymmetrical to accept events with slow protons which may
have triggered TAPS. The remaining random background under the
coincidence peak was subtracted using events in the sidebands,
scaled accordingly.\vspace{-2mm}

\subsection{\label{Section:MonteCarloSimulations}Monte Carlo simulations}
The performance of the detector was simulated in GEANT-3-based
Monte Carlo studies. The program package used for CBELSA/TAPS is
built upon a program developed for the CB-ELSA experiment. The Monte
Carlo program reproduces accurately the response of the TAPS and
Crystal Barrel crystals when hit by a photon. For charged particles,
the detector response is reasonably well understood. The
bremsstrahlung process, the tagger, and the emerging photon beam are
not simulated; the experimental data are taken as input. The
hadronic reactions under study are produced by an event generator
according to the available phase space. Dynamical effects, like the
formation of resonant states, are not simulated. The created
particles are tracked through the insensitive material and the
detector components in small steps and their interactions calculated
on a statistical basis. Possible processes include ionization,
Coulomb-scattering, shower formation and decays. The results of the
simulation are then digitized based on the properties of the
sensitive detector components and stored in the same format as real
experimental data, allowing for the use of the same analysis
framework for data and simulation. Additionally, the Monte Carlo
data has also been subjected to a trigger simulation (see
section~\ref{sec:trig}). For the reaction discussed here, $\gamma
p\to p\,\pi^{0}\eta$, in total 3 million Monte Carlo events were
used. These events served to understand possible background
contributions as well as the acceptance and reconstruction
efficiency.
\section{\label{Data and data selection}Data and data selection}
\subsection{Data}
\label{sec: experim data}
\begin{figure}[pt]
\centering
\includegraphics[width=0.45\textwidth]{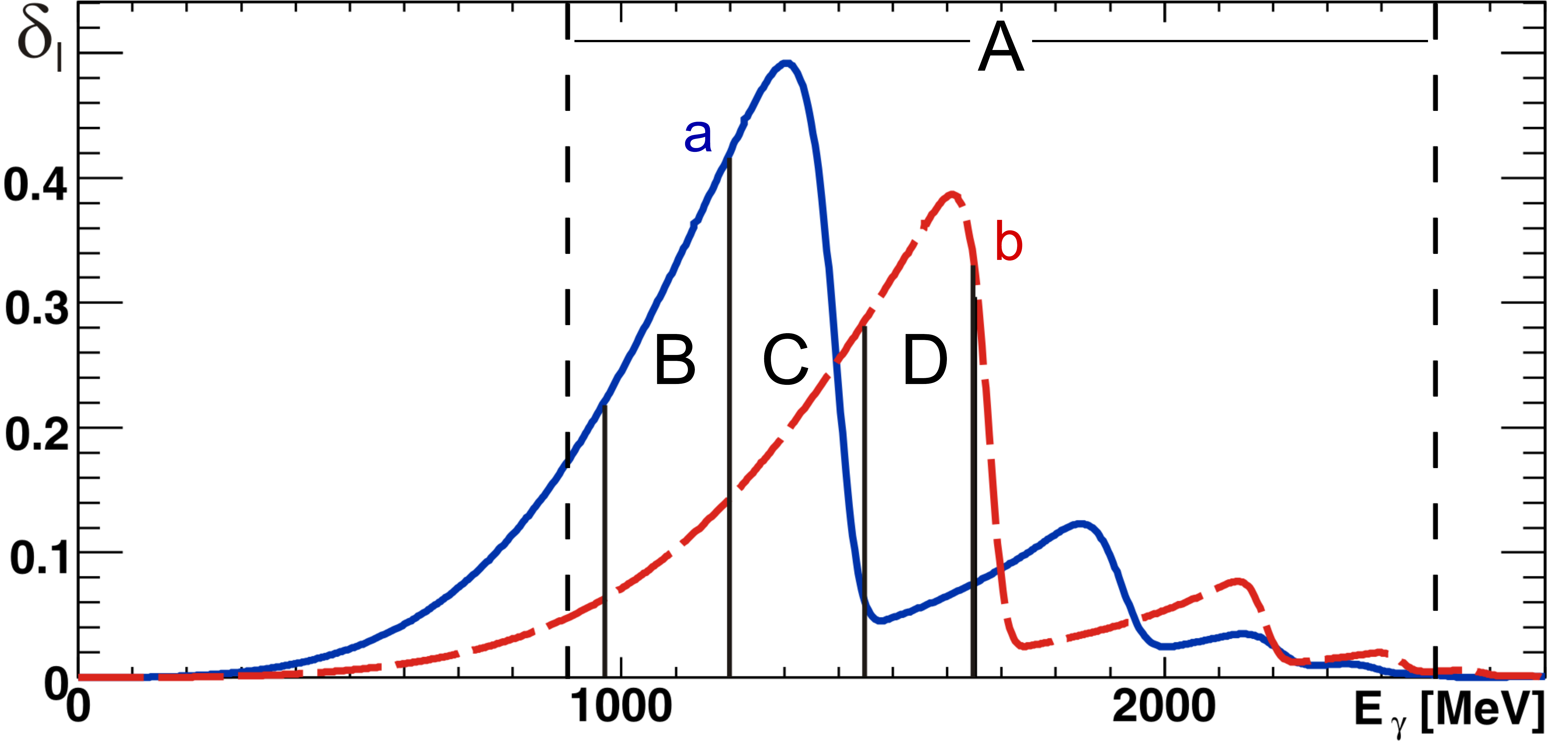}
\caption{The degree of linear polarization for the two beam times,
(a) and (b). The calculation relies on an Analytic Bremsstrahlung
Calculation and on measured photon intensity distributions like the
one shown in Fig. \ref{fig:1overE}. The highest polarizations were
49.2\% at $E_\gamma = 1305$\,MeV (a, March) and 38.7\% at 1610\,MeV
(b, May), respectively (see \cite{Elsner:2008sn} for details).
Vertical lines indicate the energy ranges chosen for the extraction
of cross sections (A, CB/TAPS1, May only) and polarization
observables (B, C, D), respectively.}
\label{fig:pol}
\end{figure}

Polarization data have been acquired in two different run periods in
2003, referred to as (a) and (b). Both datasets were taken with a
diamond crystal to produce linearly polarized bremsstrahlung. The
coherent edges for (a) and (b) were set to achieve maximal
polarization of 49.2\% at $E_{\gamma}=1300$\,MeV (a) or 38.7\% at
1600\,MeV (b), respectively. These data were divided into three
groups (B), (C), and (D), as indicated by the vertical lines in
Fig.~\ref{fig:pol}. Set (B) used the setting (a), restricted to the
photon beam energy range $E_{\gamma}=970-1200 \rm \, MeV$, set (D)
used setting (b) for $E_{\gamma}=1450-1650 \rm \, MeV$, and set (C)
used data from (a) and (b) in the overlap region,
$E_{\gamma}=1200-1450 \rm \, MeV$. These data are presented in
section \ref{sec:polobs}.

For the extraction of total and differential cross sections as well
as Dalitz plots, shown in sections \ref{sec:cross} to
\ref{sec:comp}, an incoming photon energy range from threshold up to
2.5\,GeV has been selected (A in Fig.~\ref{fig:pol}). However, due
to normalization issues, only parts of the May dataset could be used
for the extraction of cross sections. These data will be referred to
as data set CB/TAPS1. Additionally, data taken in a run period in
October and November 2002, using unpolarized photons, is presented
and referred to as data set CB/TAPS2.

\subsection{\label{Selection}Selection}
In the following, the selection process for the dataset CB/ TAPS1 is
explained in detail. In case of the CB/TAPS2 dataset, a slightly
different method was applied and is addressed in a concluding
paragraph. The data selection aims to identify events due to
reactions $\gamma p\to p\;\pi^0\pi^0$ and $\gamma p\to
p\,\pi^0\eta$, both with four photons and one proton in the final
state. The initial state of the reaction was completely known
kinematically (incoming photon momentum, proton mass and momentum),
so it was generally possible to reconstruct the final state even if
one of the particles escaped detection. This option was not used
however, since the contribution of such 4-PED events was found to be
negligible. A cut on not more than one charged cluster was applied
on the data. Events with all five calorimeter hits marked as neutral
were, however, admitted to avoid systematic effects caused by
non-uniform charge identification efficiencies of the relevant
subdetectors.  If five clusters of energy deposits were detected,
each of them was then tentatively treated as a proton candidate. The
four photons were grouped pairwise to find $\pi^0$ and $\eta$
candidates. The assignment of the observed clusters of energy
deposits to protons and $\pi^0$ and $\eta$ mesons was the aim of the
combinatorial analysis.

\paragraph{Kinematic cuts:}
\label{sec:cuts}
After the preselection of the data detailed above, a combinatorial
analysis with respect to the relevant kinematical constraints of the
reaction under consideration was performed in order to ascertain the
final state particles. Since no proton identification on detector
level has been used, each event entered the analysis-chain five
times, each time with a different cluster tested against the
hypothesis of being the final-state proton.

The coplanarity of a three-body final state poses constraints on the
angles between the proton candidate and the remaining four-photon
system. In case of the azimuthal angle, the difference
$\Delta\phi=\phi_{4\gamma}-\phi_{p}$ had to agree with $180^{\circ}$
within $\pm 10^{\circ}$. For the polar angle, the different
resolutions of the CB and TAPS calorimeters were taken into account
in such a way that the difference between the angle of the missing
proton calculated from the $4\gamma$-system and the measured fifth
cluster, $\Delta\theta = \theta_{p}^{miss}-\theta_{p}^{meas}$, had
not to exceed $\pm 15^{\circ}$ for the fifth cluster being detected
in the Crystal Barrel, and $\pm 5^{\circ}$ in TAPS, respectively.

As for the additional constraints posed by the known masses of the
final state particles, the corresponding distributions have been
fitted assuming a Gaussian signal on a polynomial background. The
half-width of the respective cuts has been set to 3.89 times the
$\sigma$ of the Gaussian, translating to a net loss of signal of not
more than $10^{-4}$ or a confidence interval of 99.99\%. The
corresponding numbers are given in Table~\ref{tab:cuts}.
\begin{table}[pt]
\renewcommand{\arraystretch}{1.5}
\begin{center}
\begin{tabular}{|c|c|c|c|c|}\hline
 Variable & Fit & Mean & $\sigma$ & Cut\\
  & function & [MeV] & [MeV] & width\\\hline
 Miss. mass & G+pol(3) & 936.4 & 31.8 & $\pm123.6$\,MeV\\
 $m(\gamma\gamma)$, $\pi^{0}$ & G+pol(3) & 135.3 & 8.6 & $\pm33.6$\,MeV\\
 $m(\gamma\gamma)$, $\eta$ & G+pol(1) & 548.2 & 20.9 & $\pm81.4$\,MeV\\\hline\hline
 $\Delta\phi$ & $-$ & $\pm180^{\circ}$ & $-$ & $\pm10^{\circ}$\\
 $\Delta\theta$ & $-$ & $0^{\circ}$  & $-$ & $\pm15^{\circ}$ (CB)\\
  & & & & $\pm5^{\circ}$ (TAPS) \\\hline
 \end{tabular}
 \caption{Width of the kinematic cuts.}
\label{tab:cuts}
\end{center}\vspace{-5mm}
\end{table}

It is unlikely, yet possible, for an event to pass these cuts in
more than one combination of final state particles. To eliminate any
residual combinatorial background due to such ambiguities in the
proton determination, the preselected data were subjected to a
preliminary kinematic fit to the $p\pi^{0}\eta$ final state,
described below. Here the only condition to be met is the
convergence of the fit. In case of an event passing through this
stage multiple times, only the one with the highest confidence level
(CL) was retained.

\begin{figure}[pb]
\centering
\includegraphics[width=0.4\textwidth,height=0.32\textwidth]{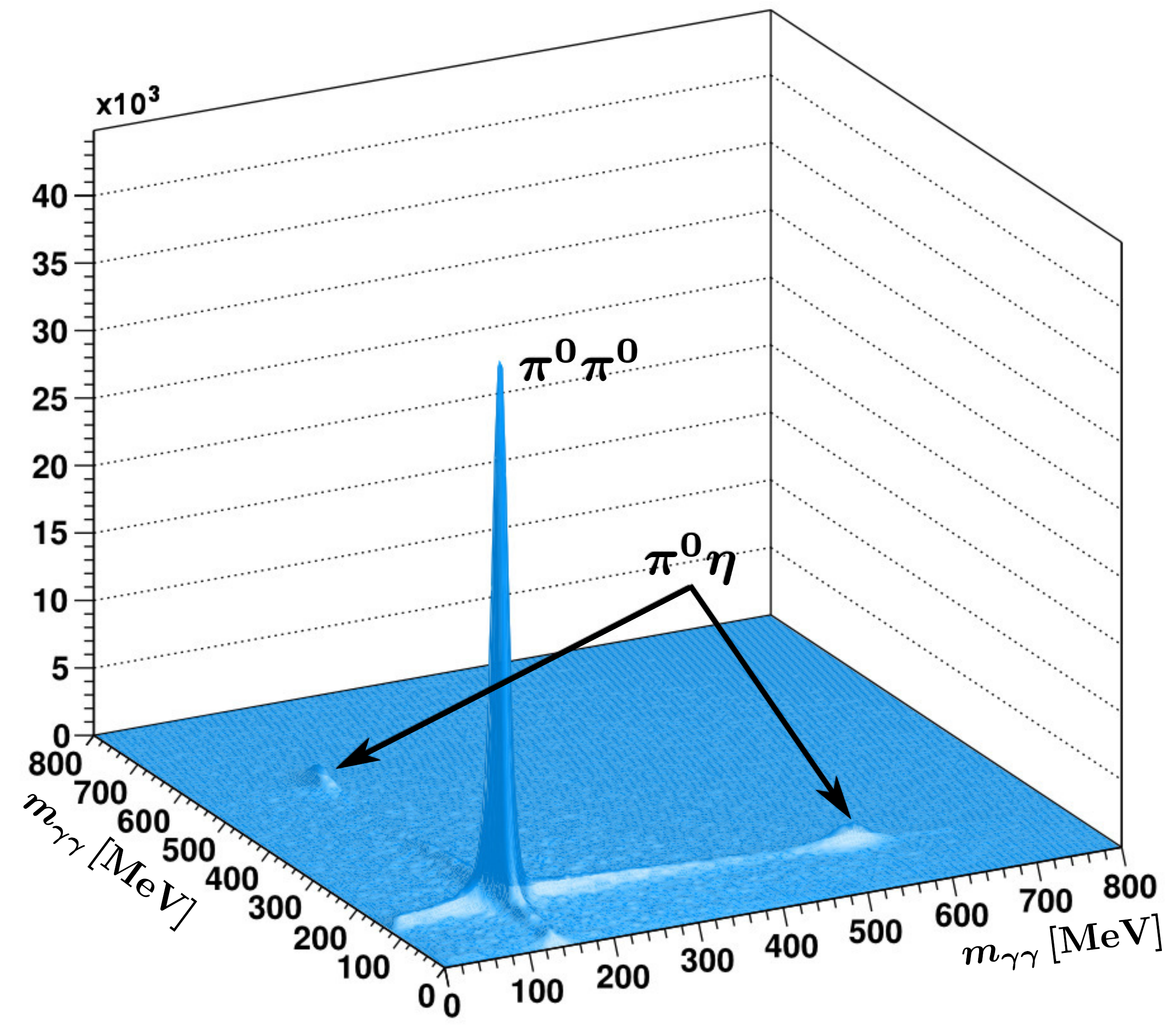}
\caption{$\gamma\gamma$ invariant masses after selection of the $p
4\gamma$ final state, before kinematic fitting. Cuts on the meson
masses have not yet been applied.\vspace{-7mm}}
\label{fig:ggvsgg}
\end{figure}
\begin{figure}[pt]
\centering
\includegraphics[width=0.45\textwidth]{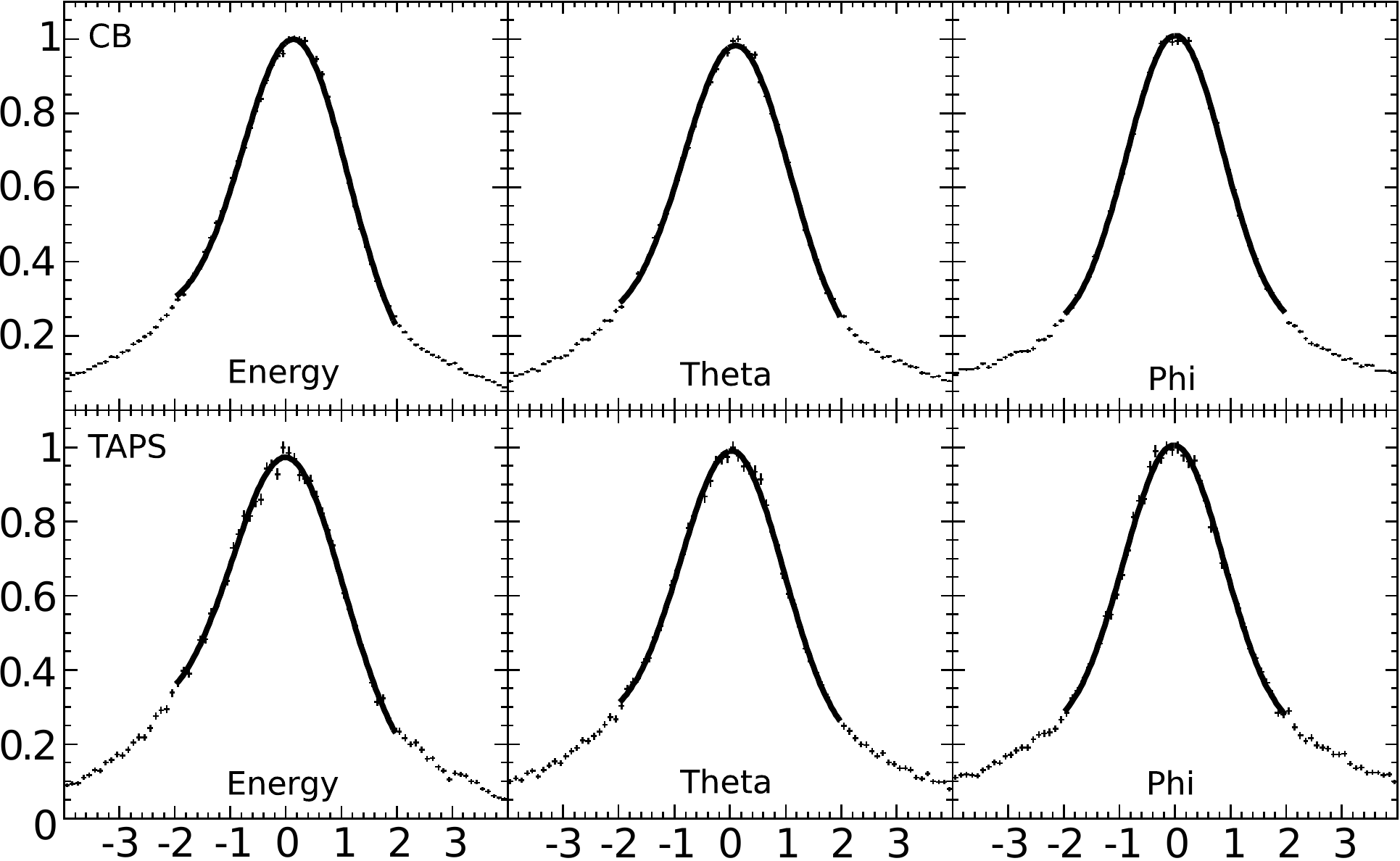}
\caption{Pull distributions from the $\gamma p \rightarrow
p_{\mathrm{miss}} \pi^{0}\eta$ fit to the data over the full energy
range analyzed. Top row: Particles in the Crystal Barrel, bottom
row: Particles in TAPS. Left to right: Pulls in energy, $\theta$,
$\phi$. The data are well compatible with the expected Gaussian
shape with mean $\mu=0$ and $\sigma=1$.}
\label{fig:pulls}
\end{figure}
Fig.~\ref{fig:ggvsgg} shows the invariant mass of one
$\gamma\gamma$-pair versus the invariant mass of the other, after
the application of the selection procedure described above. Clear
signals for both the dominant production of $2\pi^{0}$ as well as
for $\pi^{0}\eta$ production are observed.

\paragraph{Kinematic fits:}
\label{sec:kinfit}

After the preselection described above, the remaining data sample
was subjected to a kinematic fit, imposing energy and momentum
conservation as well as the masses of final-state particles as
constraints. Analogous to the selection procedure prior to the fit,
the proton was treated as a missing particle.

Only a brief description of the fit is given here, for a detailed
description of the procedure, see \cite{Pee-pi0}. The kinematic fit
used the measured parameters of the reaction and varied them within
given error-margins to satisfy the given constraints. The method,
apart from improving accuracy by returning corrected quantities,
provided means to control the data with respect to systematic
effects. The deviations between the measured values used in the fit,
e.g. energies and angles of the particles, and the results of the
kinematic fit, normalized to the respective measurement errors,
should form a Gaussian centered around zero and with unit width
($\sigma$). Such so-called {\em pull}-distributions are shown in
Fig.~\ref{fig:pulls}, separately for particles detected in one of
the two calorimeters for the quantities energy, $\theta$ and $\phi$,
and for the fit hypothesis $\gamma p \rightarrow p_{\mathrm{miss}}
\pi^{0}\eta$ (see below). The presence of systematic effects in the
data, not accounted for in the error margins given to the fit,
should cause a shift of the distribution. A wrong estimation of the
error margins themselves would lead to the width deviating from
unity. The distributions were obtained from a fit to the data over the
full energy range under consideration and nicely agree with the
expected values.

\begin{figure}[pt]
\centering
\includegraphics[width=0.45\textwidth]{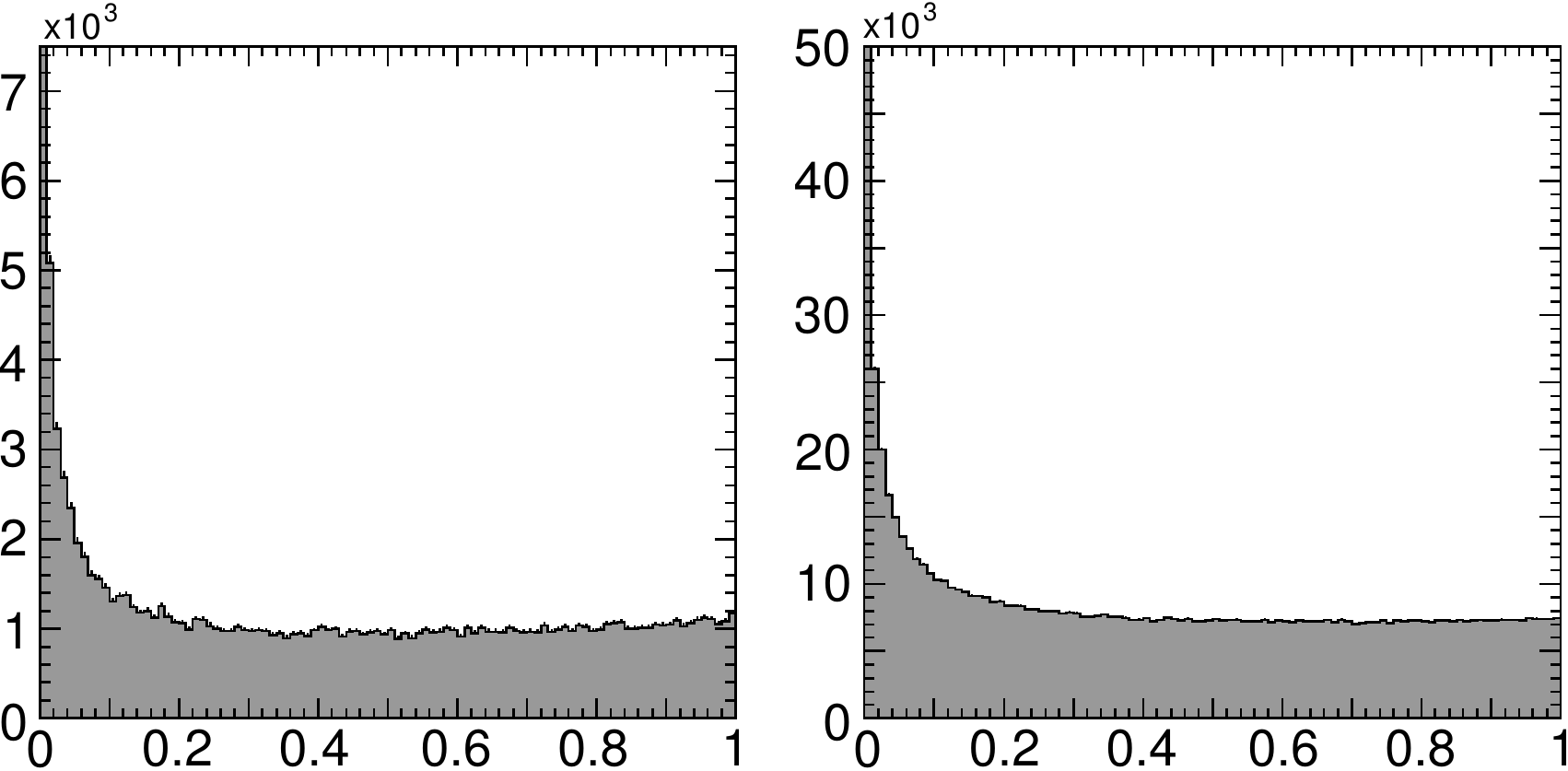}
\caption{Confidence level distributions for the hypothesis $\gamma p
\rightarrow p_{\mathrm{miss}} \pi^{0}\eta$, imposing energy and
momentum conservation and additionally using the meson masses as
constraints. The final state proton is treated as a missing
particle. Left: Data, right: Monte Carlo.\vspace{3mm}}
\label{fig:cls}
\end{figure}

A convenient value to quantify the quality of the fit is the so-called {\em
confidence level} (CL), the integral over the $\chi^{2}$
probability, varying between 0 and 1. For correctly determined
errors, this distribution should be flat. However, background
events, leading to bad fits and therefore high $\chi^{2}$, should
result in a peaking of the CL distribution towards 0. This can be
seen in Fig.~\ref{fig:cls}, where the CL distributions for the
$\gamma p \rightarrow p_{\mathrm{miss}} \pi^{0}\eta$ fit hypothesis
are shown for data and Monte Carlo simulations. The generally flat
distributions show a steep rise towards 0, indicating background
contributions in the data or efficiency effects in the
reconstruction for some subset of the data and Monte Carlo (the MC
has been generated background-free). To avoid contamination of the
final data sample by such effects, a cut on the CL can be applied.
\begin{figure}[pb]
\vspace{-3mm} \centering
\includegraphics[width=0.35\textwidth,height=0.27\textwidth]{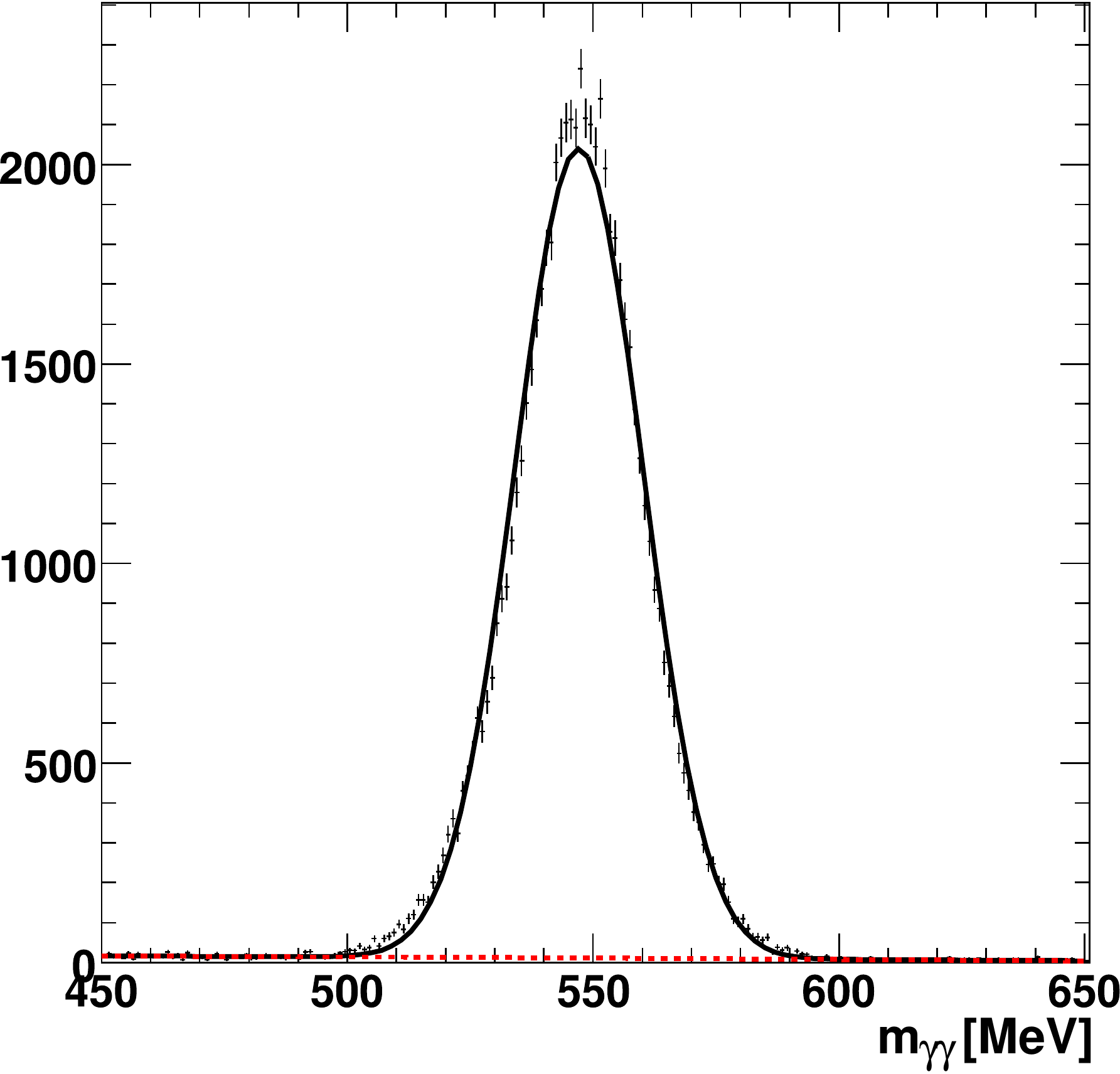}\vspace{3mm}\\
\caption{Invariant mass of the free $\gamma\gamma$-pair after the
$\gamma p \rightarrow p_{\mathrm{miss}}\pi^{0}\gamma\gamma$ fit,
from threshold up to $E_{\gamma}=2500$\,MeV. An anticut on the
$\pi^{0}\pi^{0}$-hypothesis has been performed. The cut on the
$\pi^{0}\eta$-CL rejects most of the background (dashed line).}
\label{fig:eta}
\end{figure}

\begin{figure*}[pt] \centering
\includegraphics[width=0.7\textwidth,height=0.45\textwidth]{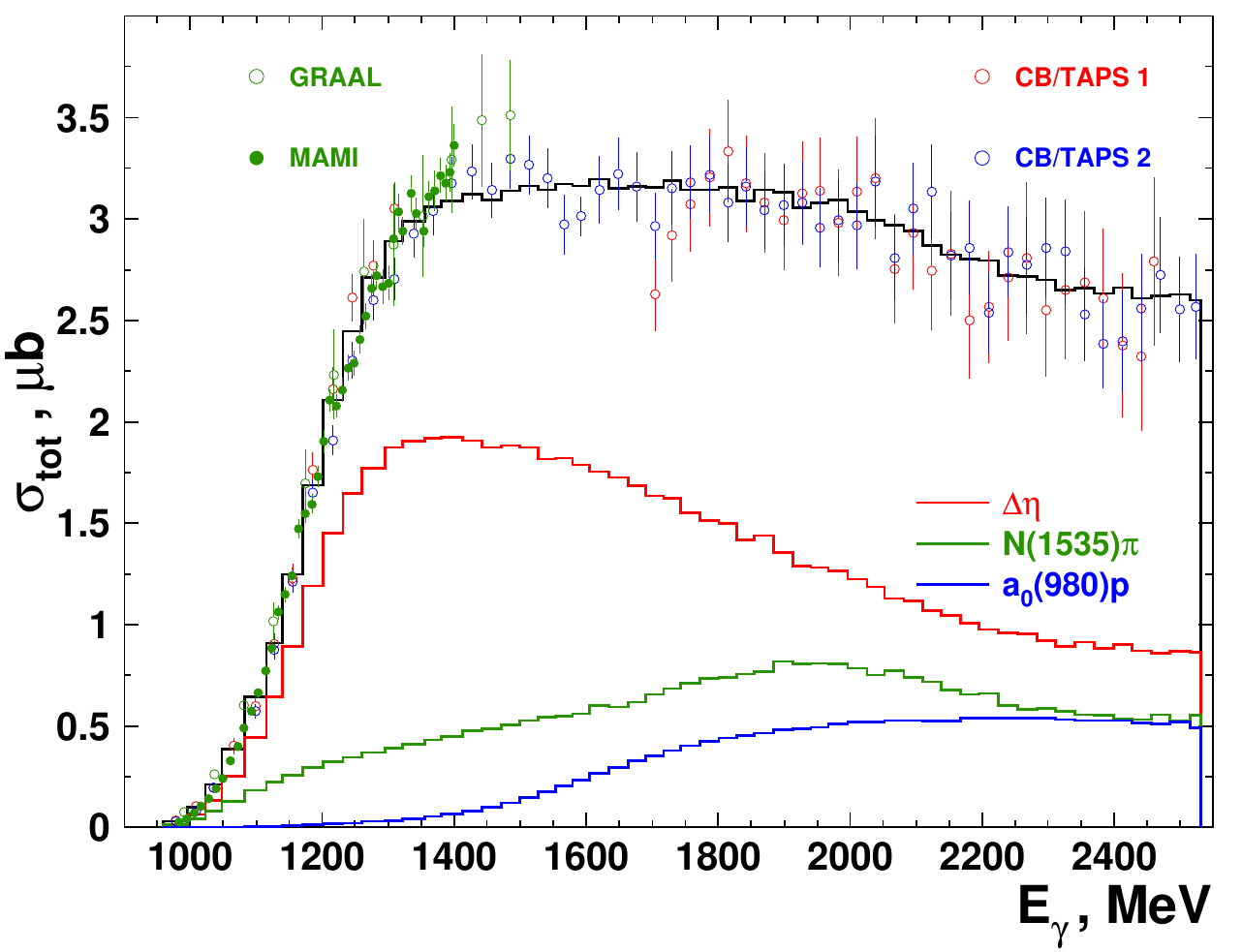}
\caption{Total cross sections for the reaction $\gamma p \rightarrow
p \pi^{0}\eta$ (normalized to the MAMI data, see text) and excitation 
functions for the most important
isobars according to the BnGa PWA fits. This work (open circles,
blue/red), compared to measurements from GRAAL \cite{Ajaka:2008zz}
(open circles, green) and MAMI \cite{Kashevarov:2009ww} (full
circles, green).   
 \vspace{3mm}}
\label{fig:tot_wq}
\begin{tabular}{cccc}
\includegraphics[width=0.24\textwidth]{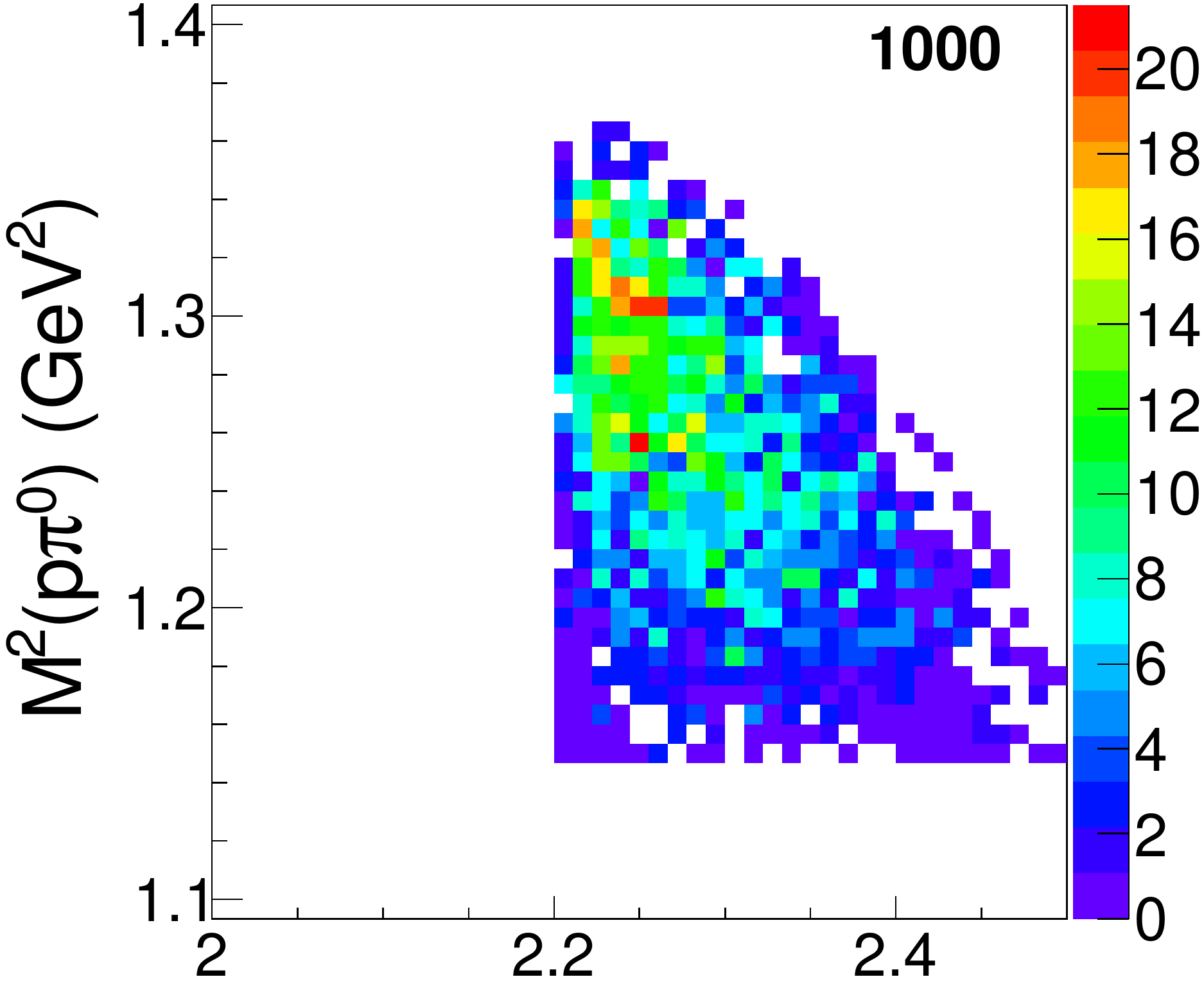}&
\hspace{-2mm}\includegraphics[width=0.24\textwidth]{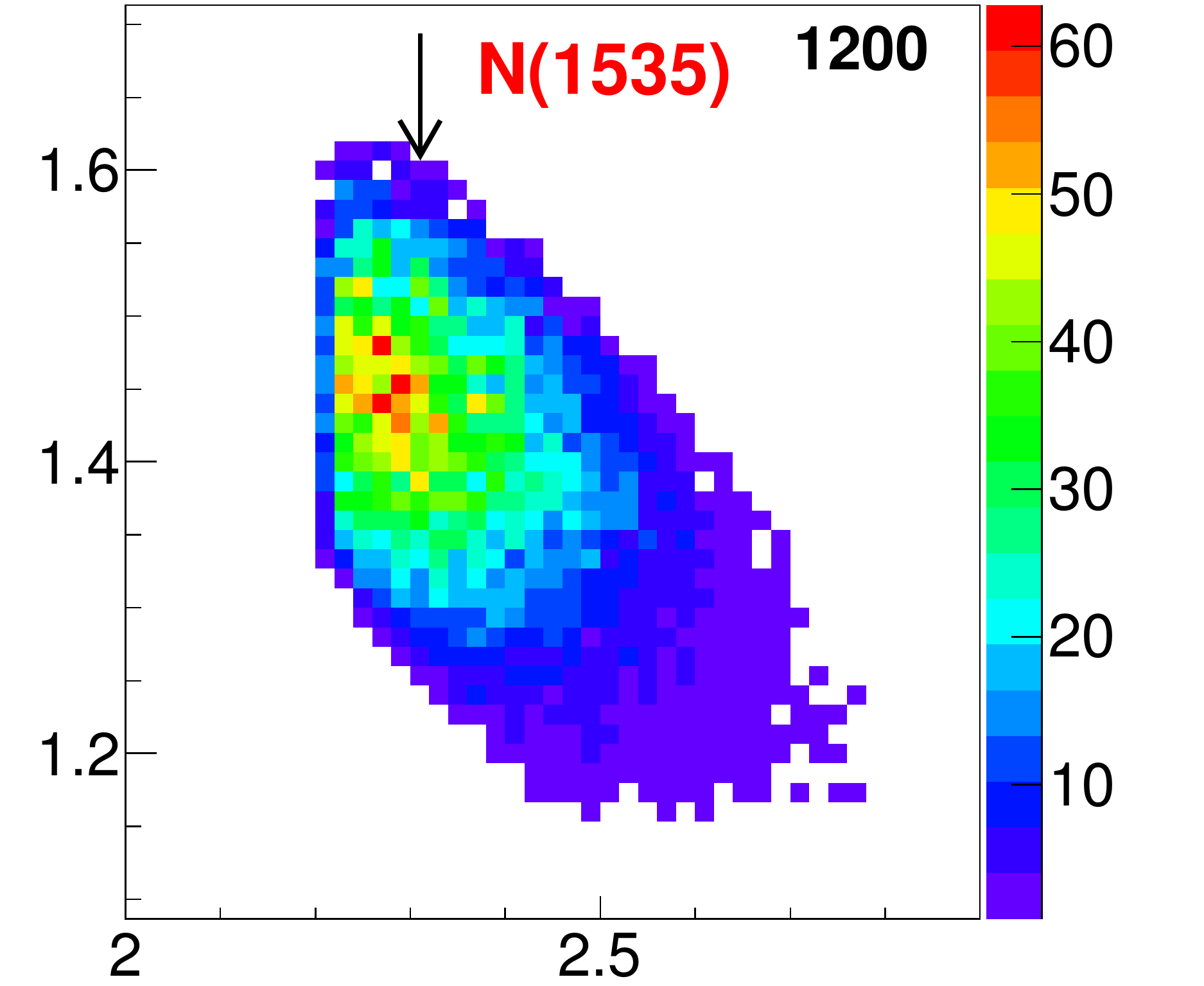}&
\hspace{-6mm}\includegraphics[width=0.24\textwidth]{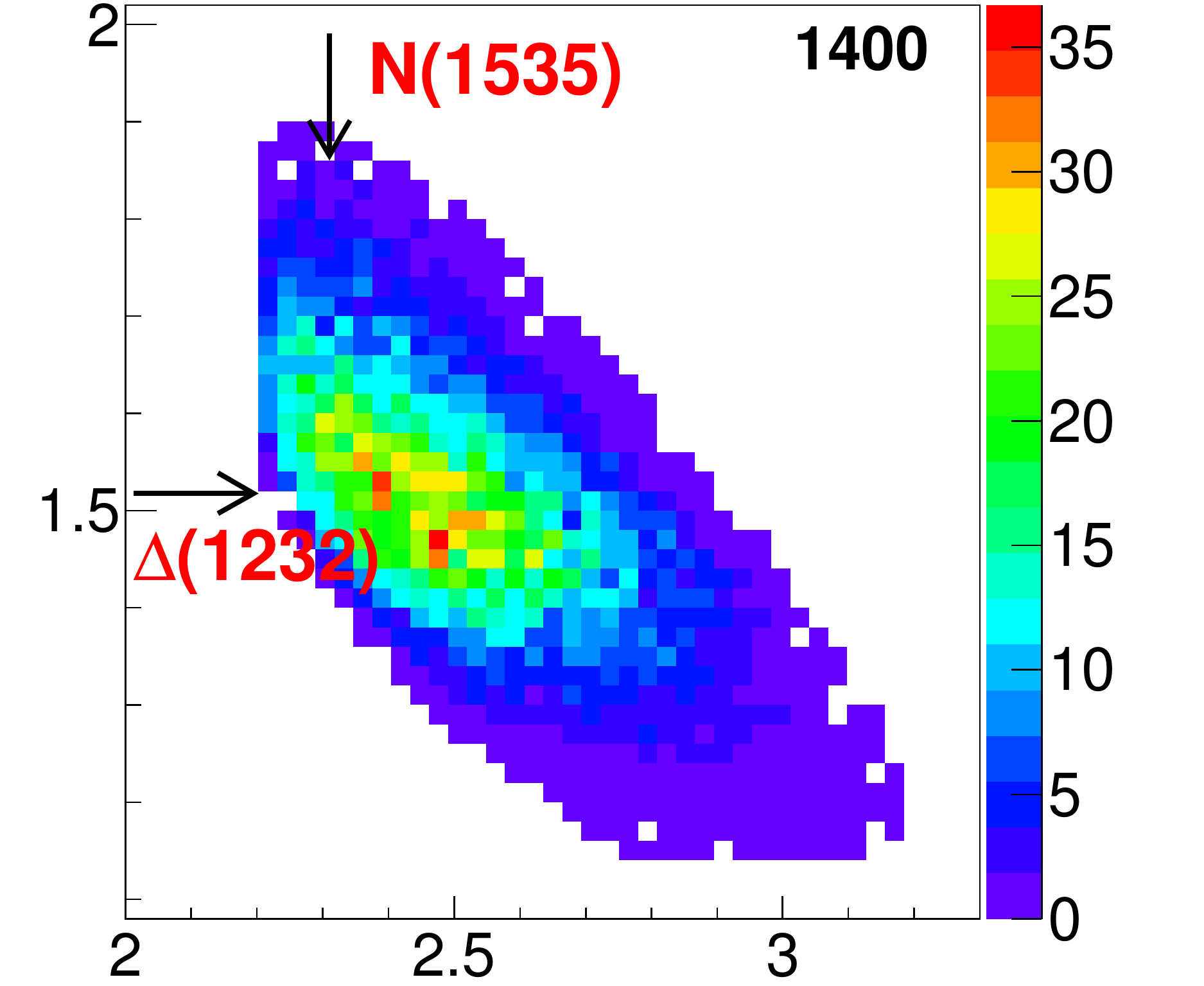}&
\hspace{-6mm}\includegraphics[width=0.24\textwidth]{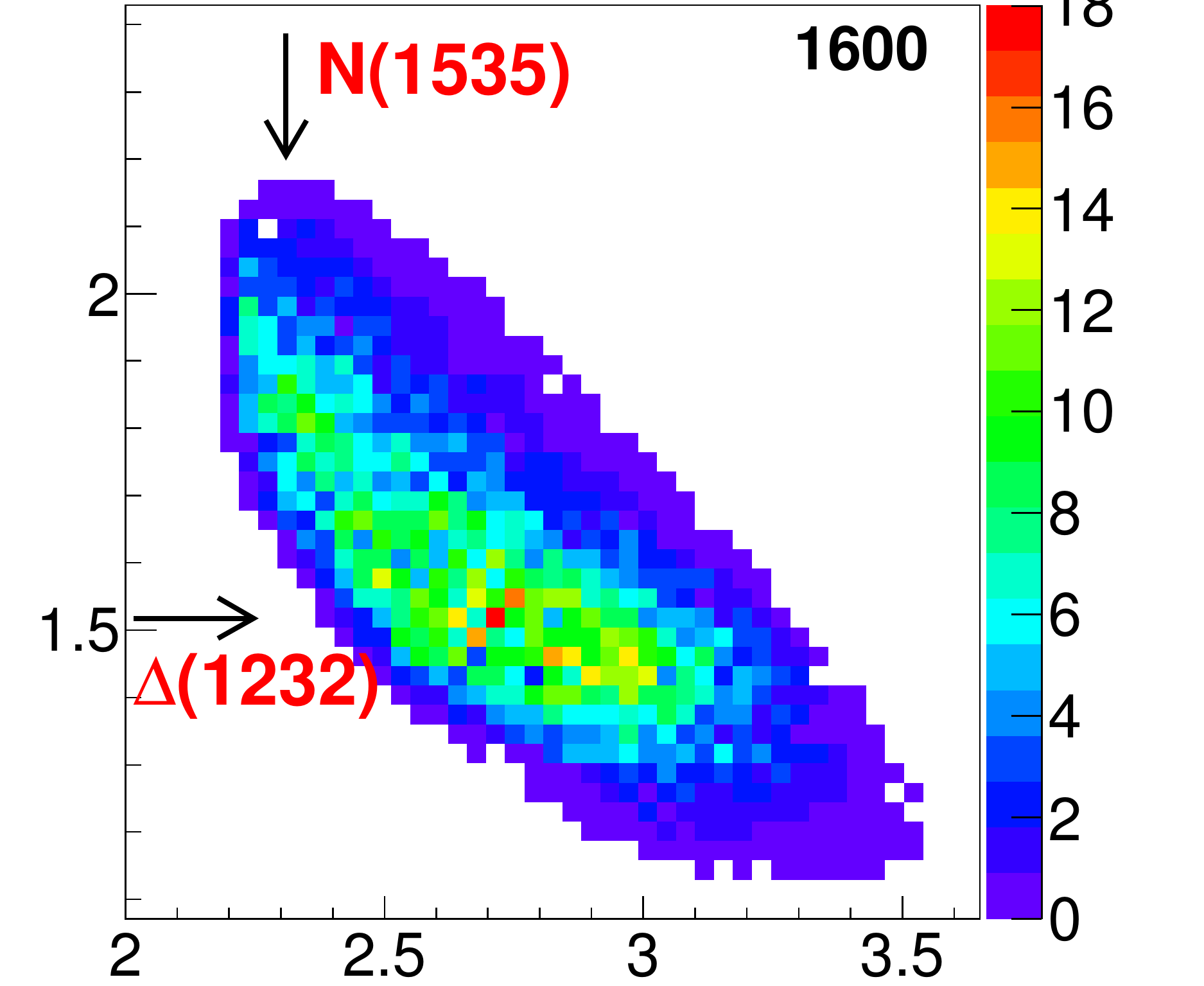}\\
\hspace{-1mm}\includegraphics[width=0.24\textwidth]{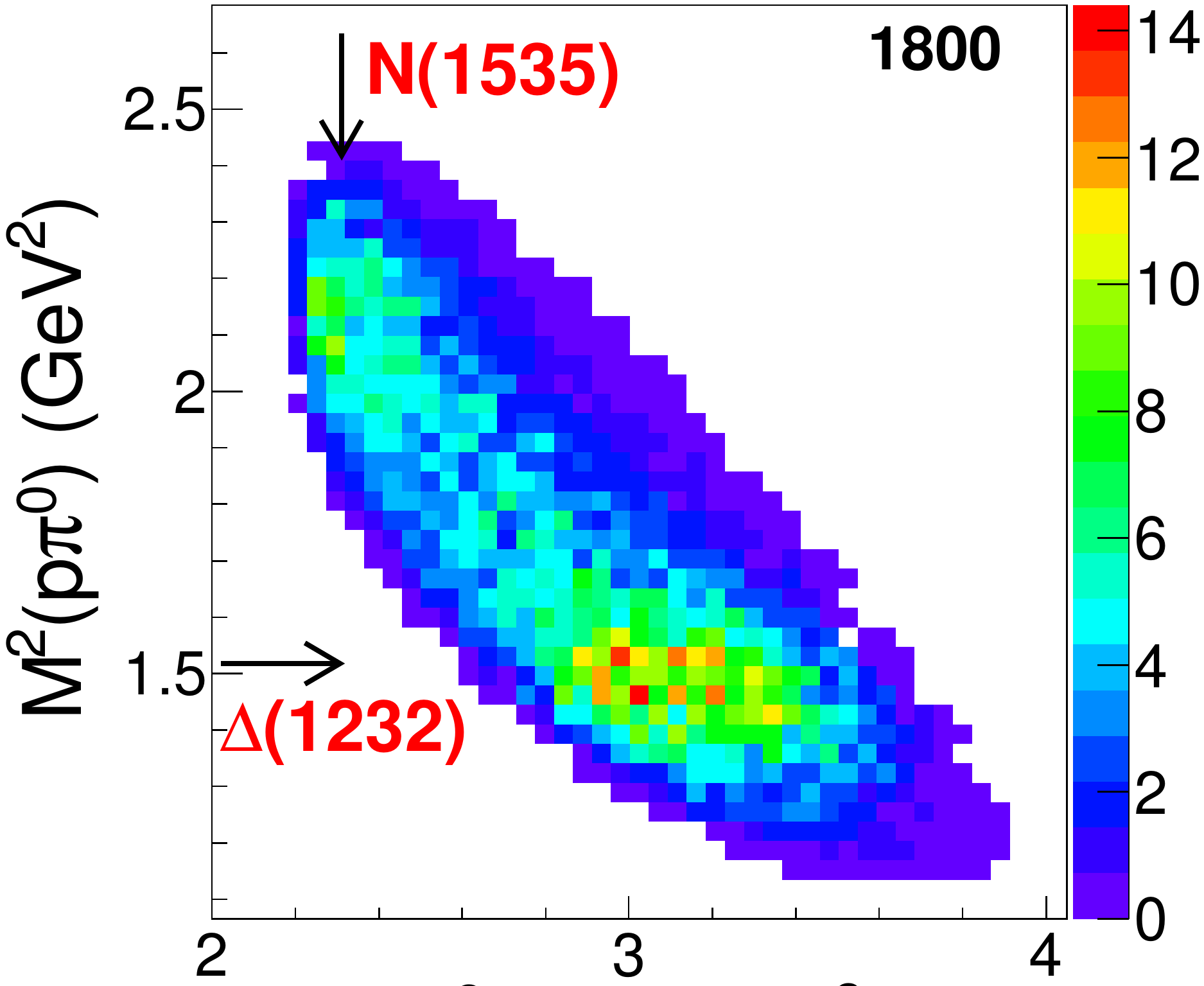}&
\hspace{-2mm}\includegraphics[width=0.24\textwidth]{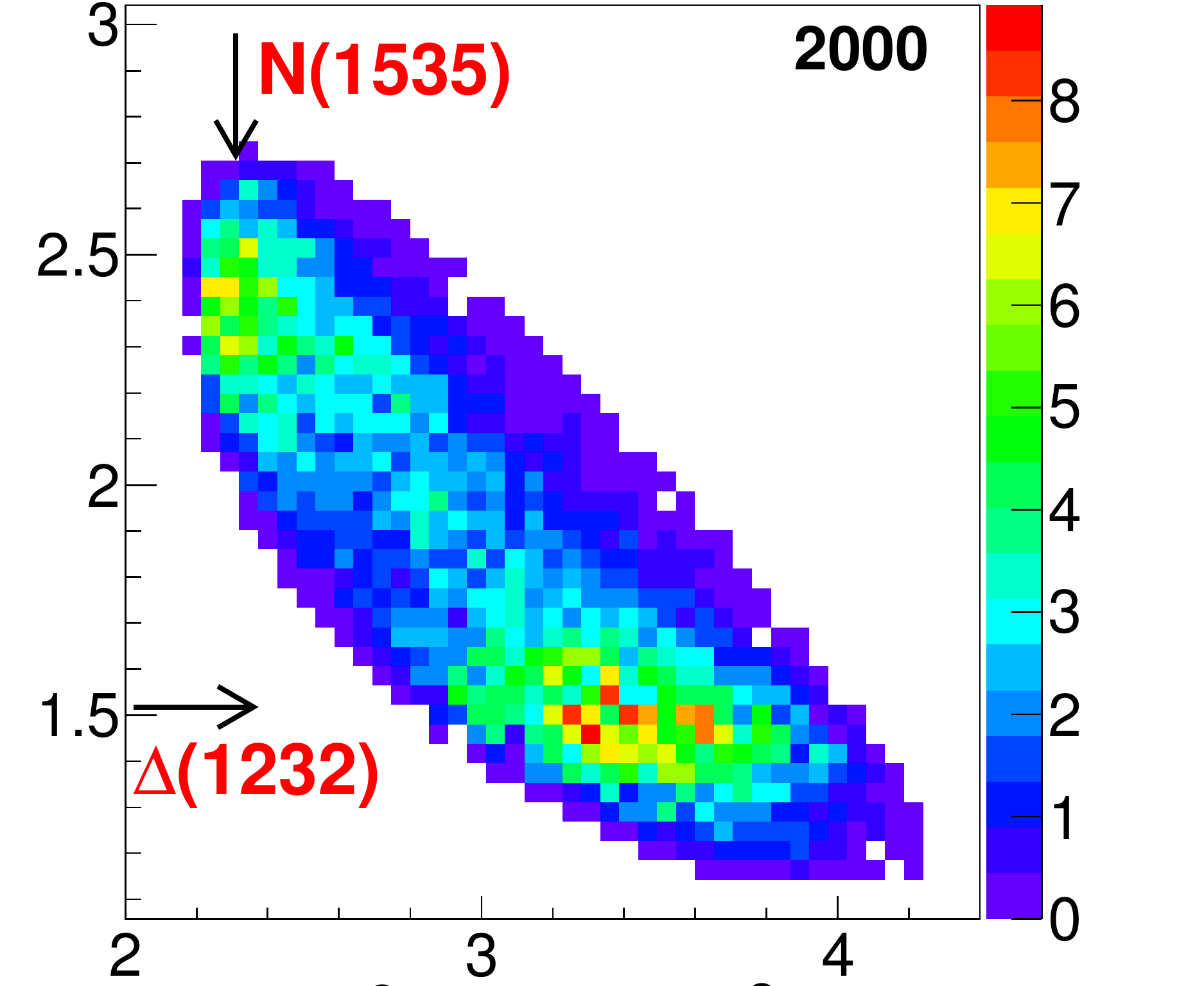}&
\hspace{-6mm}\includegraphics[width=0.24\textwidth]{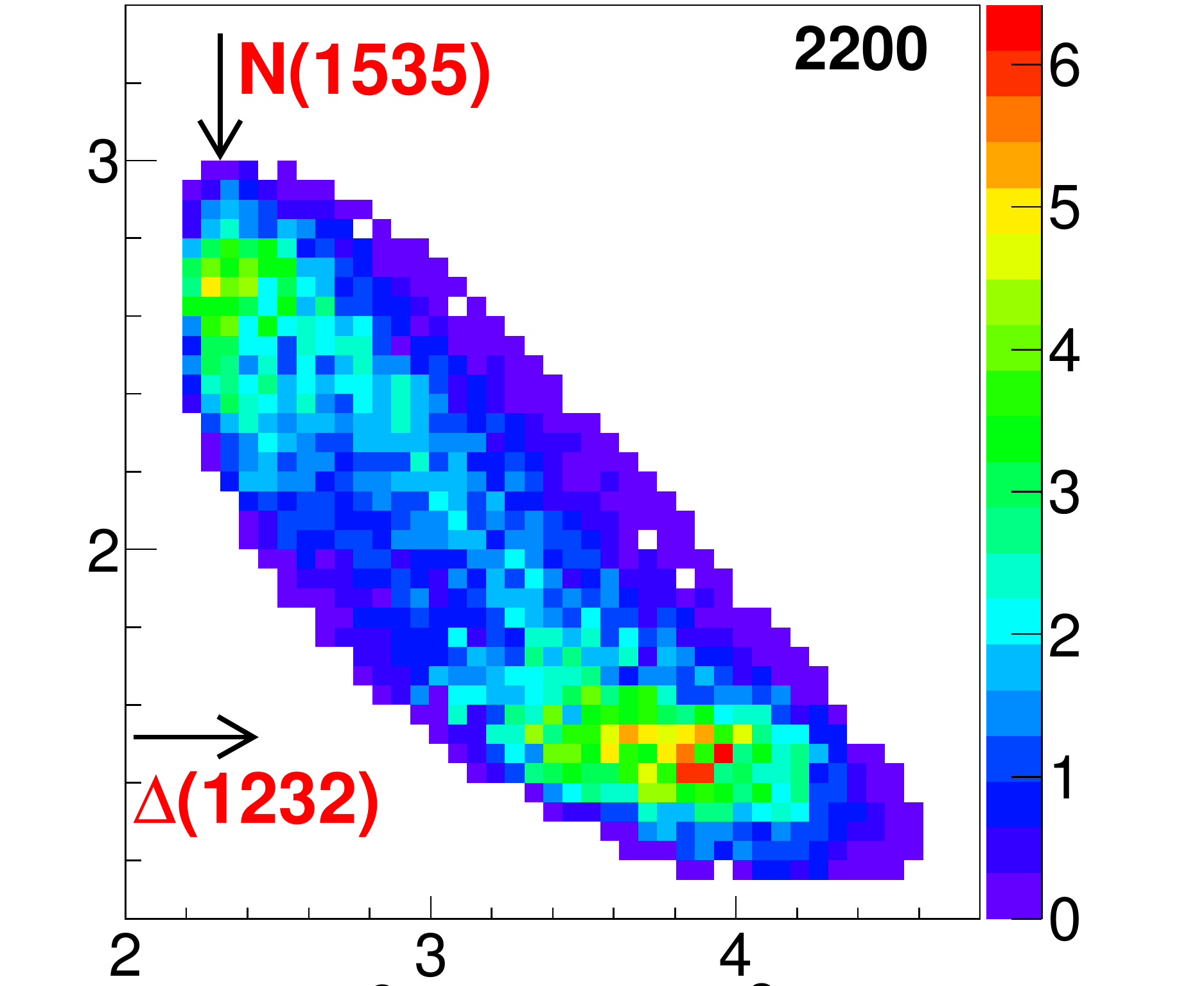}&
\hspace{-6mm}\includegraphics[width=0.24\textwidth]{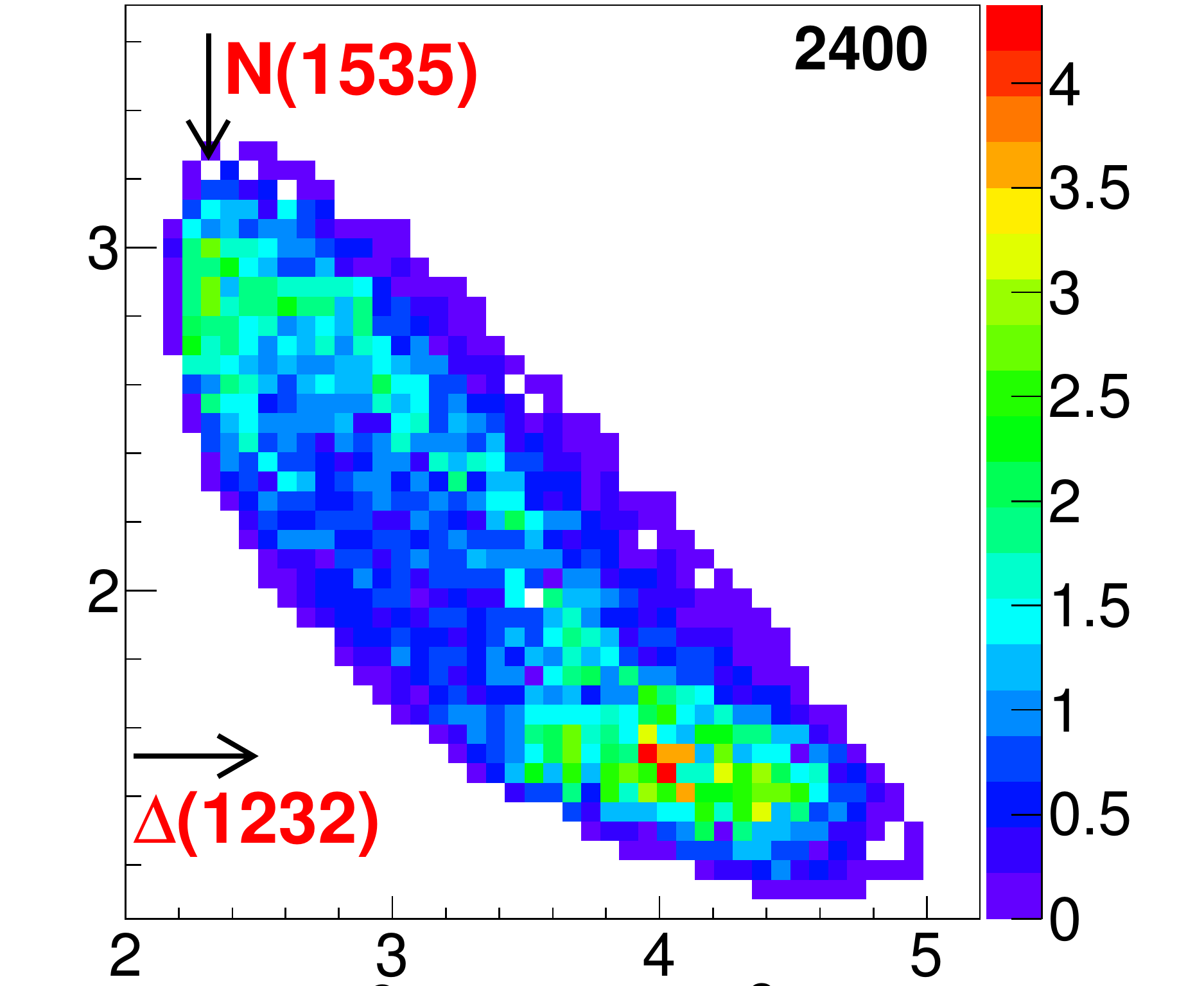}\\[2ex]
\end{tabular}
\caption{Dalitz plots $M^2(p\pi^0)$ versus $M^2(p\eta)$ for the
incoming photon energy ranges $1000\pm100$\,MeV to
$2400\pm100$\,MeV.}
\label{fig:dalitz-c}
\end{figure*}

In the given analysis, the preselected data have been subjected to
four differently constrained fits:\vspace{-2mm}
\begin{eqnarray}
\gamma p & \rightarrow & p_{\mathrm{miss}}\,4\gamma\\
\gamma p & \rightarrow & p_{\mathrm{miss}}\,\pi^{0}2\gamma\\
\gamma p & \rightarrow & p_{\mathrm{miss}}\,\pi^{0}\pi^{0}\\
\gamma p & \rightarrow &
p_{\mathrm{miss}}\,\pi^{0}\eta.\vspace{-2mm}
\end{eqnarray}
Here $p_{\mathrm{miss}}$ denotes that the proton, identified in the
preceding selection, was treated as a missing particle. The first
two hypotheses, introducing one and two constraints, respectively,
were used for control purposes and background studies only. The latter 
two hypotheses include energy and momentum
conservation and two mass constraints. Convergent fits
could be observed for the $\pi^{0}\pi^{0}$ hypothesis, some with
high confidence levels, even after the selection of the
$p\pi^{0}\eta$ final state described above. Few events 
fulfilled both the $\pi^{0}\pi^{0}$ as well as the $\pi^{0}\eta$
hypothesis with CLs up to 1. Therefore not only a cut on the CL of
the desired reaction, $\mathrm{CL}_{\pi^{0}\eta}>0.06$, but also an
anticut on the CL of the competing reaction,
$\mathrm{CL}_{\pi^{0}\pi^{0}}<0.01$ has been applied to the data.

Finally, a cross-check between the results of the kinematic fit and
the initial selection procedure was performed by comparing the
azimuthal and, again independently for the two calorimeters, polar
angles of the proton identified at the two stages. Cuts on
$\Delta\phi = \pm 4^{\circ}$, $\Delta\theta_{\mathrm{CB}} = \pm
5^{\circ}$ and $\Delta\theta_{\mathrm{TAPS}} = \pm 2^{\circ}$ have
been applied to reduce the effects of events, in which the fit had
to overly adjust the proton direction, since this might affect the
results with respect to the extraction of beam asymmetries.

The final event sample comprised a total of approx. 65,500 events
for the extraction of polarization observables and, in two data
taking periods, approximately 40,000 (CB/TAPS1 using polarized
photons) and 145,000 (CB/ TAPS2 using unpolarized photons) for the
cross sections and Dalitz plots. In both cases, the final background
contamination, derived from the $\eta$-signal using different cuts
on the kinematic fit and confirmed by comparison to the Monte Carlo
simulations (see Fig.~\ref{fig:eta}), amounted to $\approx
1\%$.\vspace{-3mm}

\paragraph{Data with unpolarized photons} For the CB/TAPS2 data set,
the selection procedure was nearly identical to the one described
above. Here, however, the proton was allowed to be missing. Again,
the contribution of 4-PED events was found to be negligible. Charged
and neutral particles were identified using the information from the
inner detector and the TAPS vetos, since cross sections are less
sensitive to small variations in the charge identification. Each
preselected event was subjected to the kinematic fit once. As for
the tagger reconstruction, each fiber hit in the tagging detector
was treated as an individual photon, clustering of fibers was not
applied. The selection of the unpolarized data follows the procedure
discussed in \cite{Crede:2011dc}.

\section{Extraction of observables}
\label{Extraction of observables}
\subsection{The total cross section}
\label{sec:cross}

\begin{figure*}[pt]
\begin{tabular}{cccc}
\includegraphics[width=0.24\textwidth]{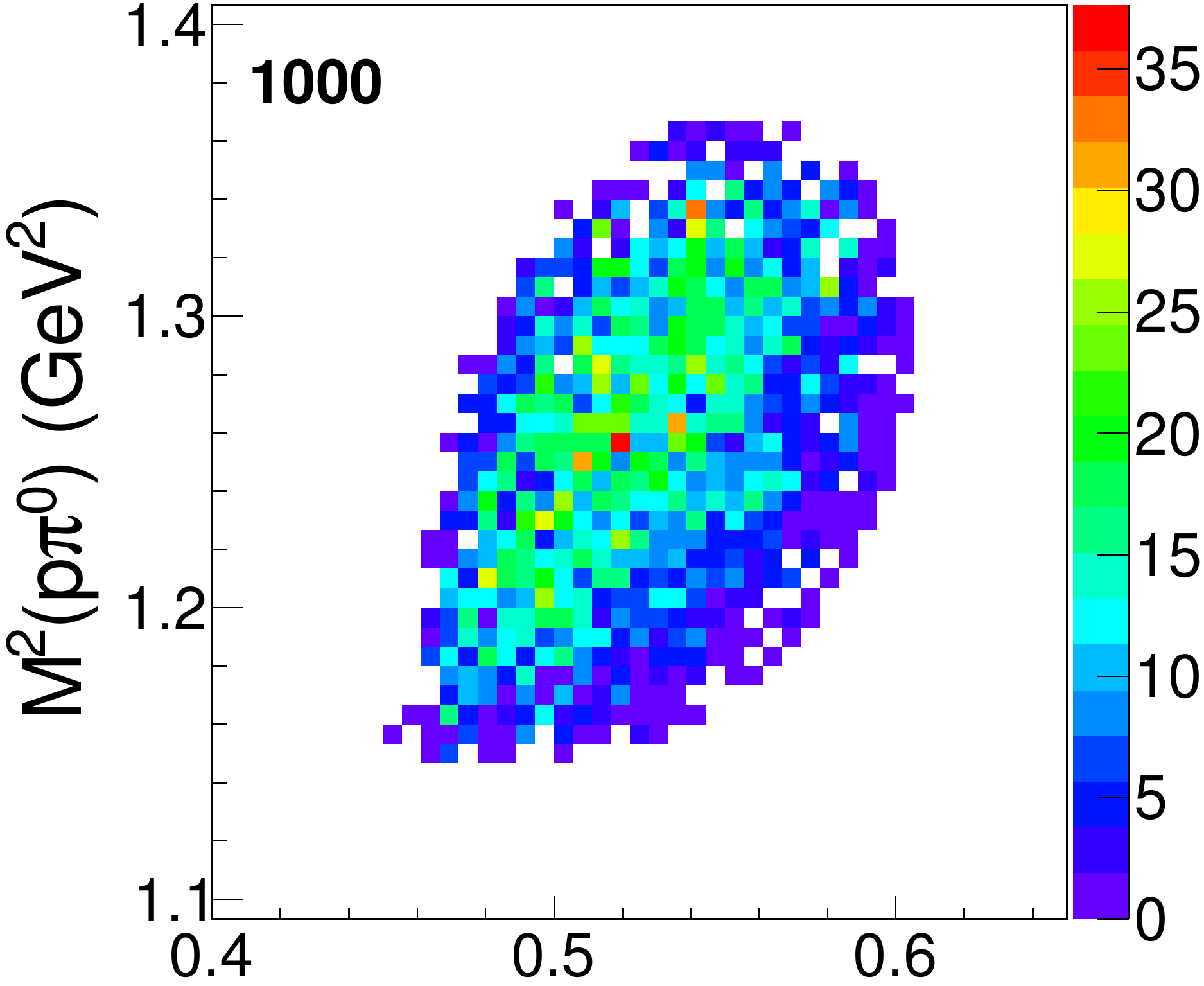}&
\hspace{-2mm}\includegraphics[width=0.24\textwidth]{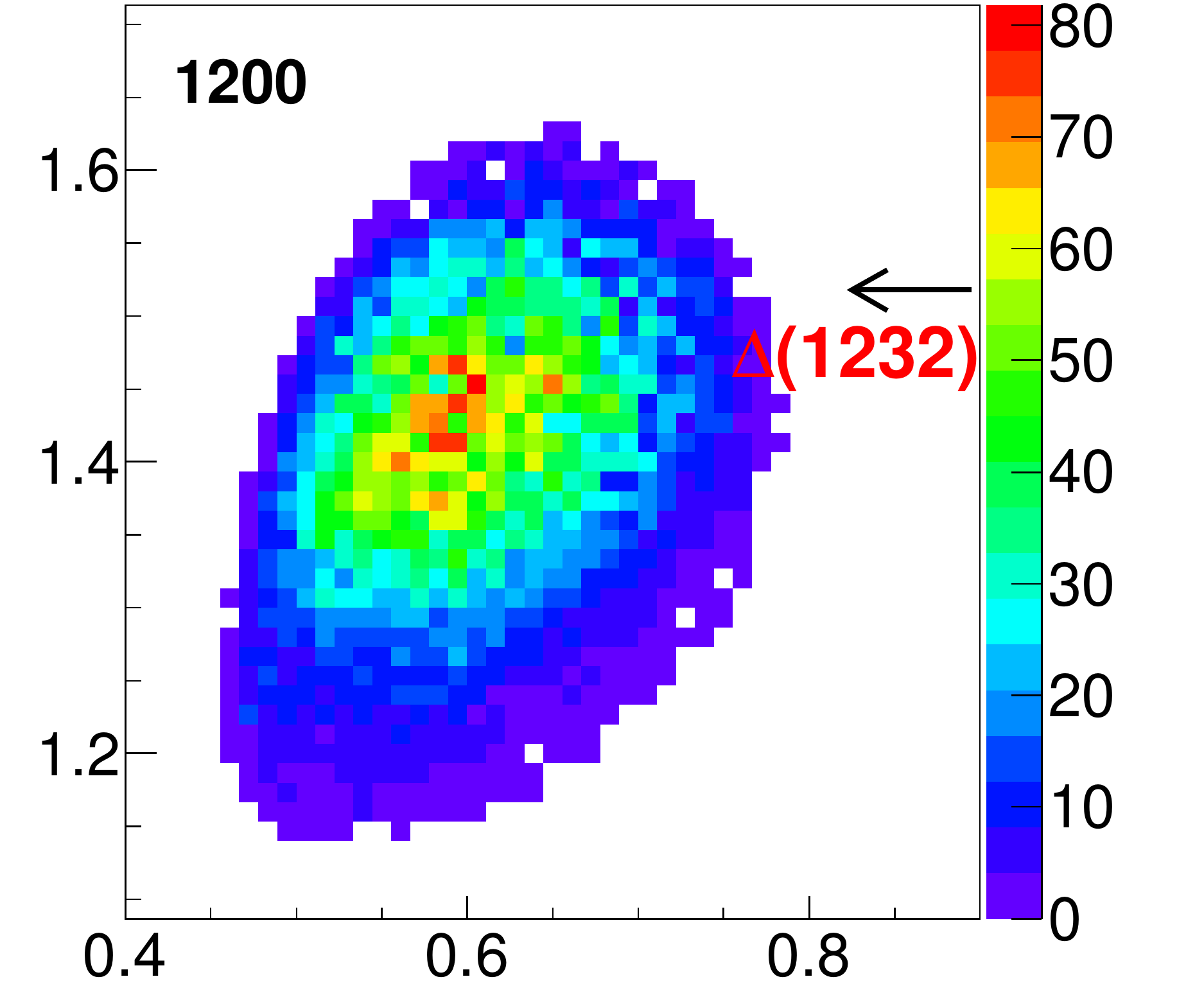}&
\hspace{-6mm}\includegraphics[width=0.24\textwidth]{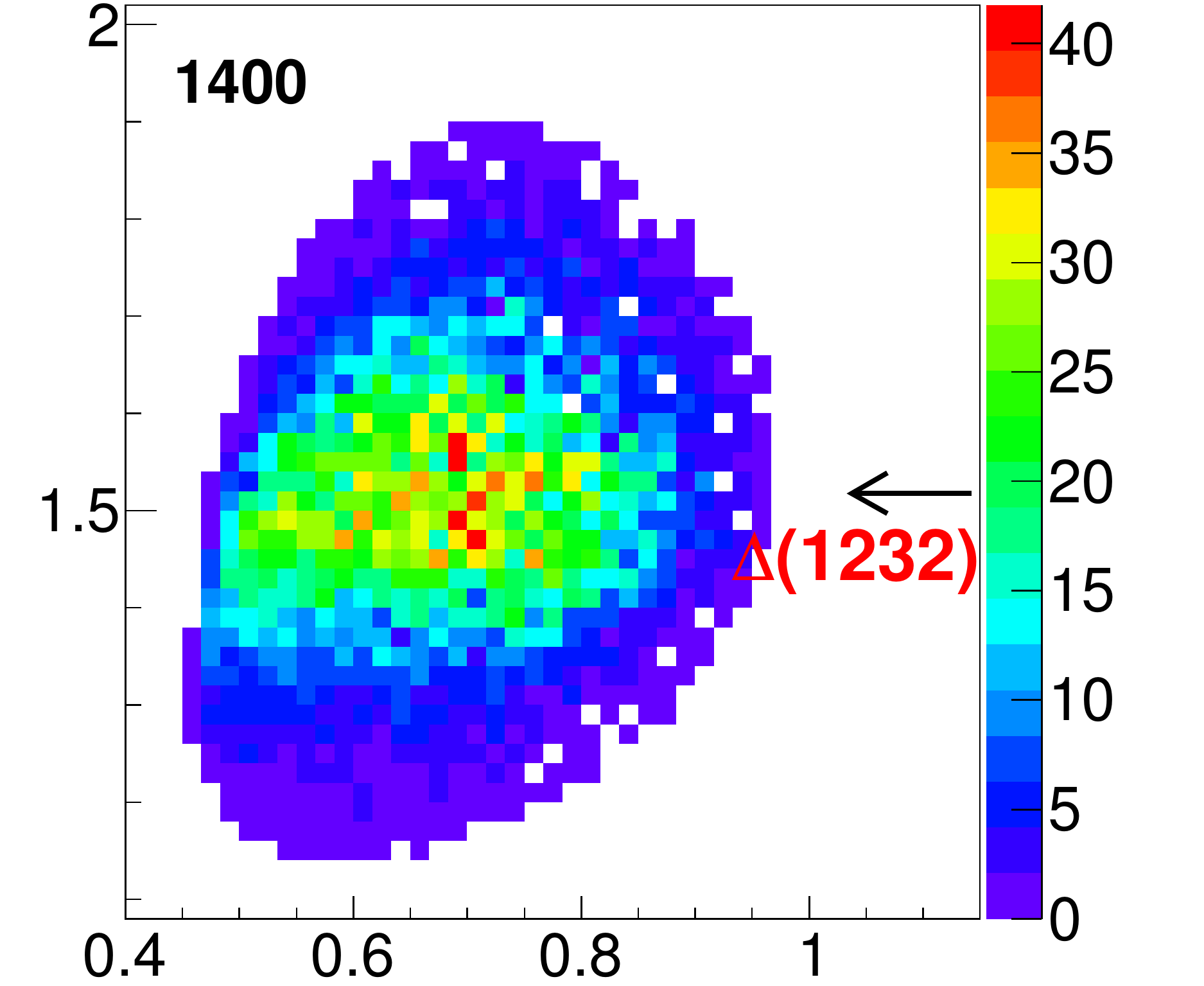}&
\hspace{-6mm}\includegraphics[width=0.24\textwidth]{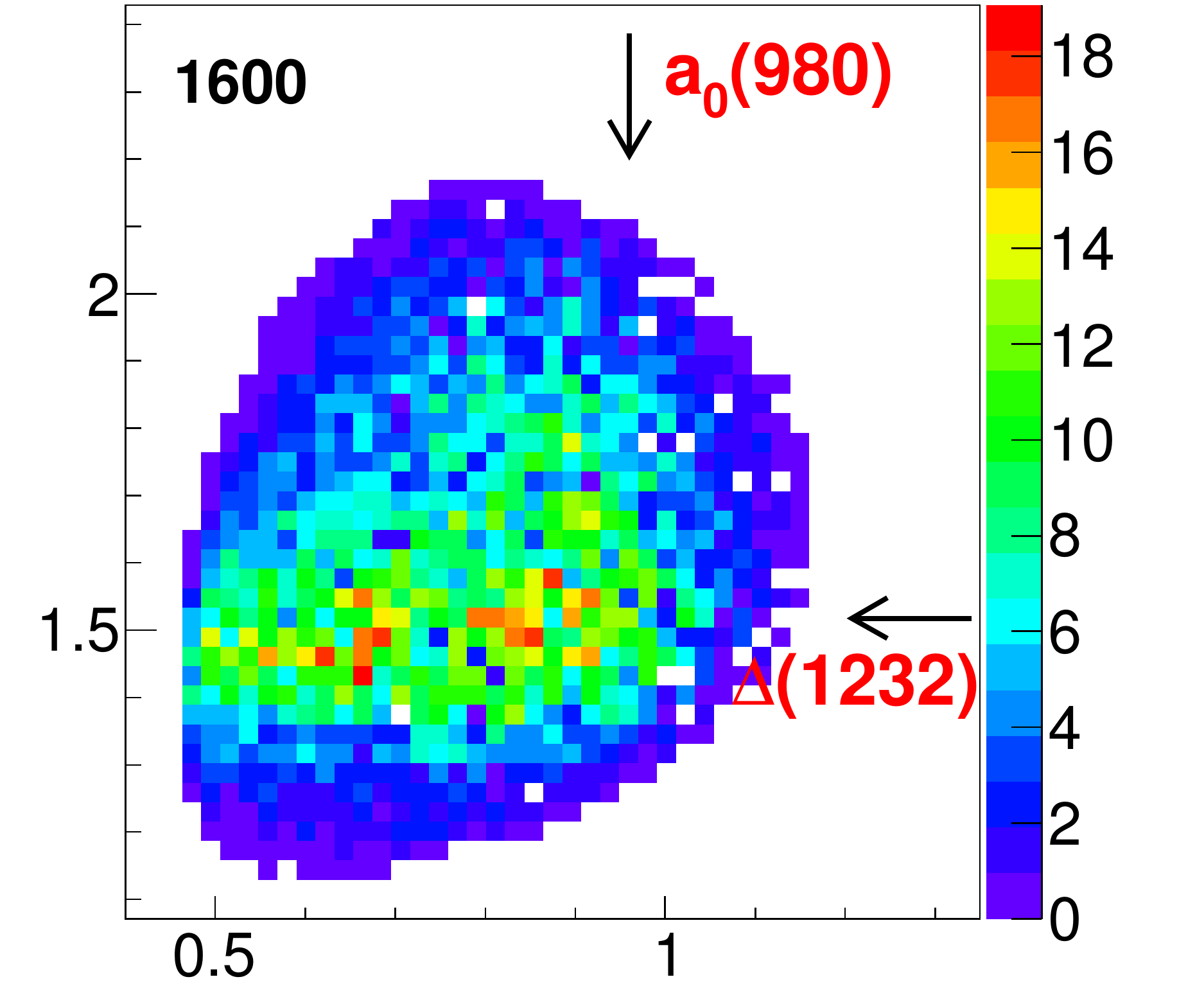}\\
\hspace{-1mm}\includegraphics[width=0.24\textwidth]{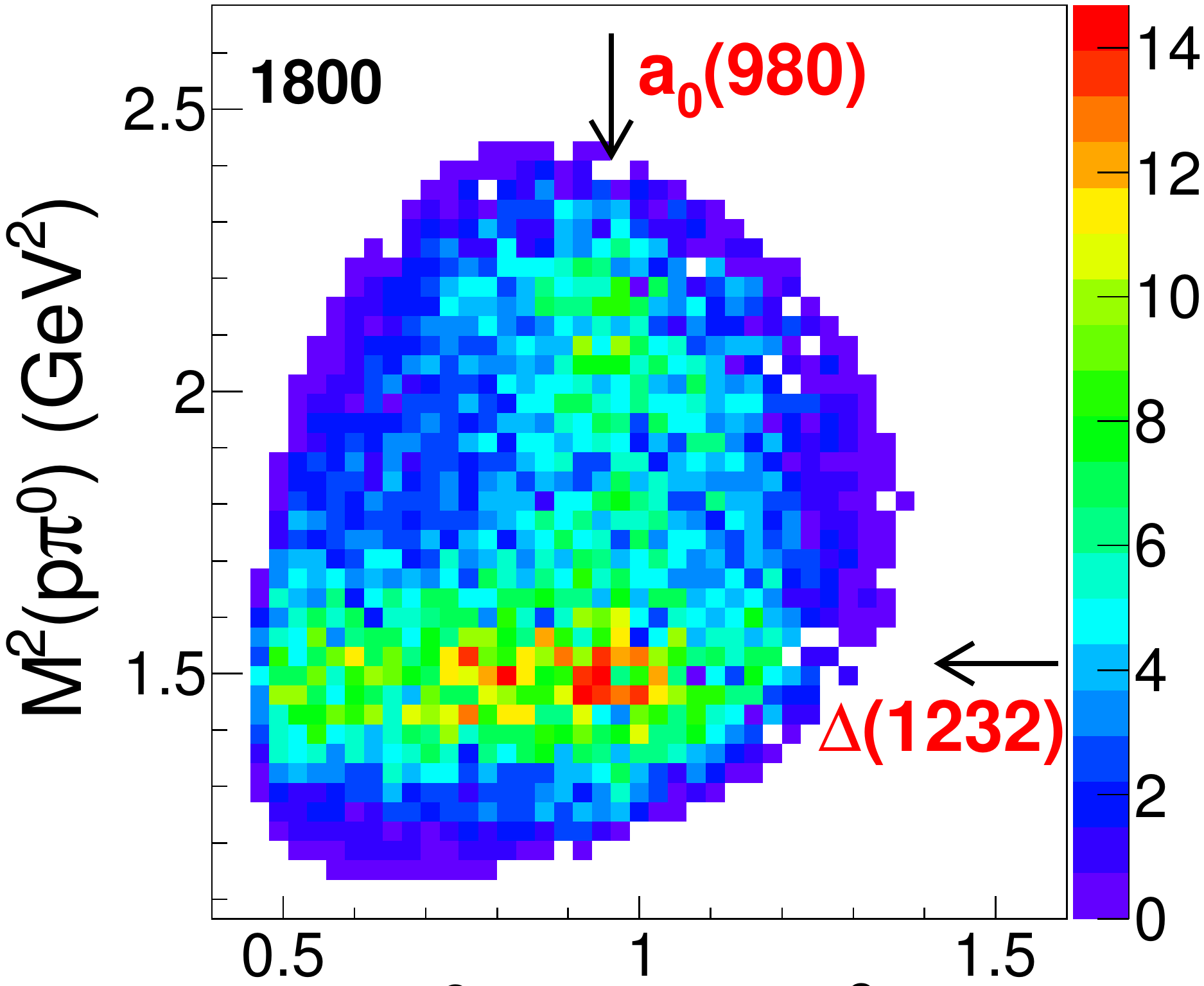}&
\hspace{-2mm}\includegraphics[width=0.24\textwidth]{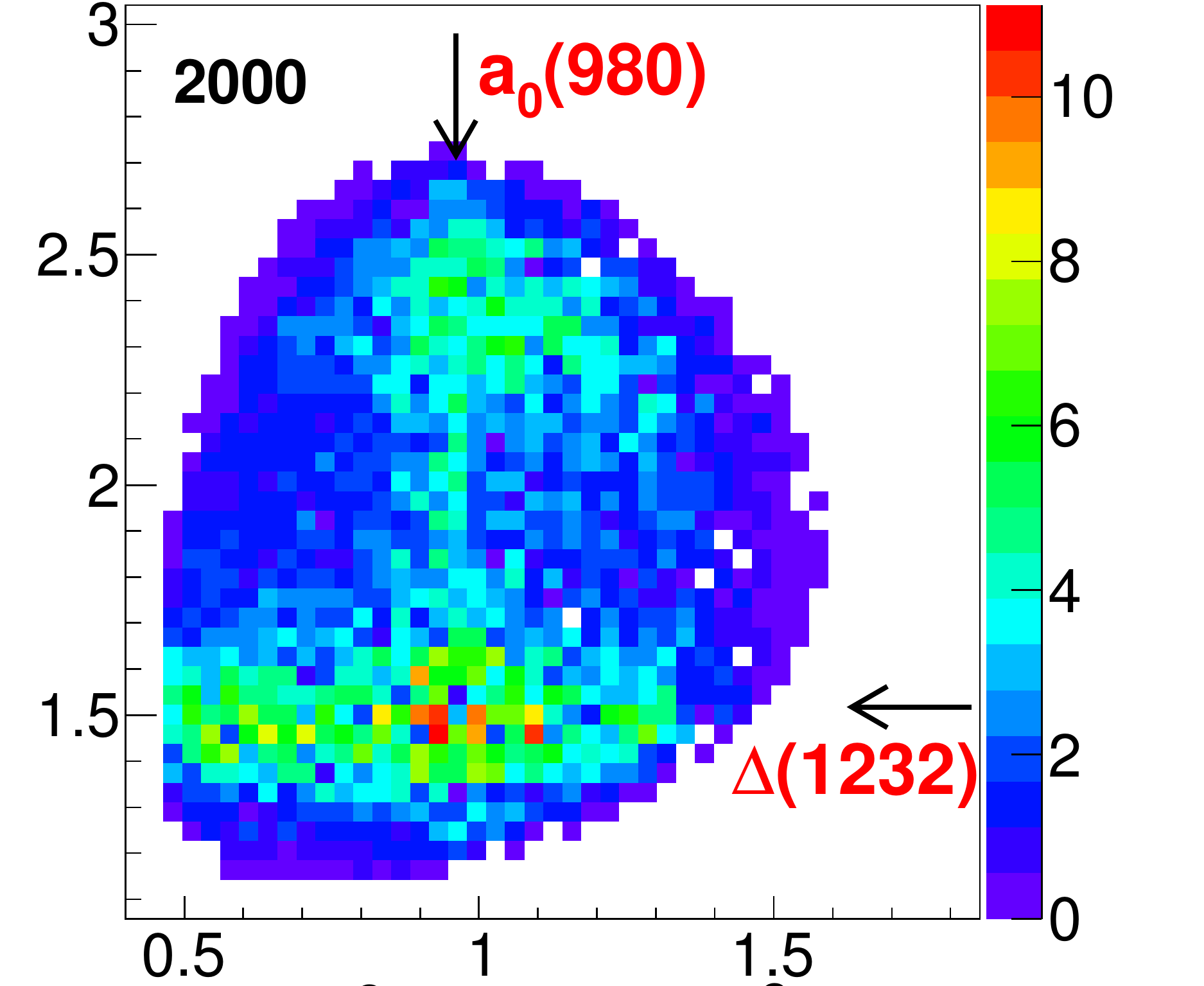}&
\hspace{-6mm}\includegraphics[width=0.24\textwidth]{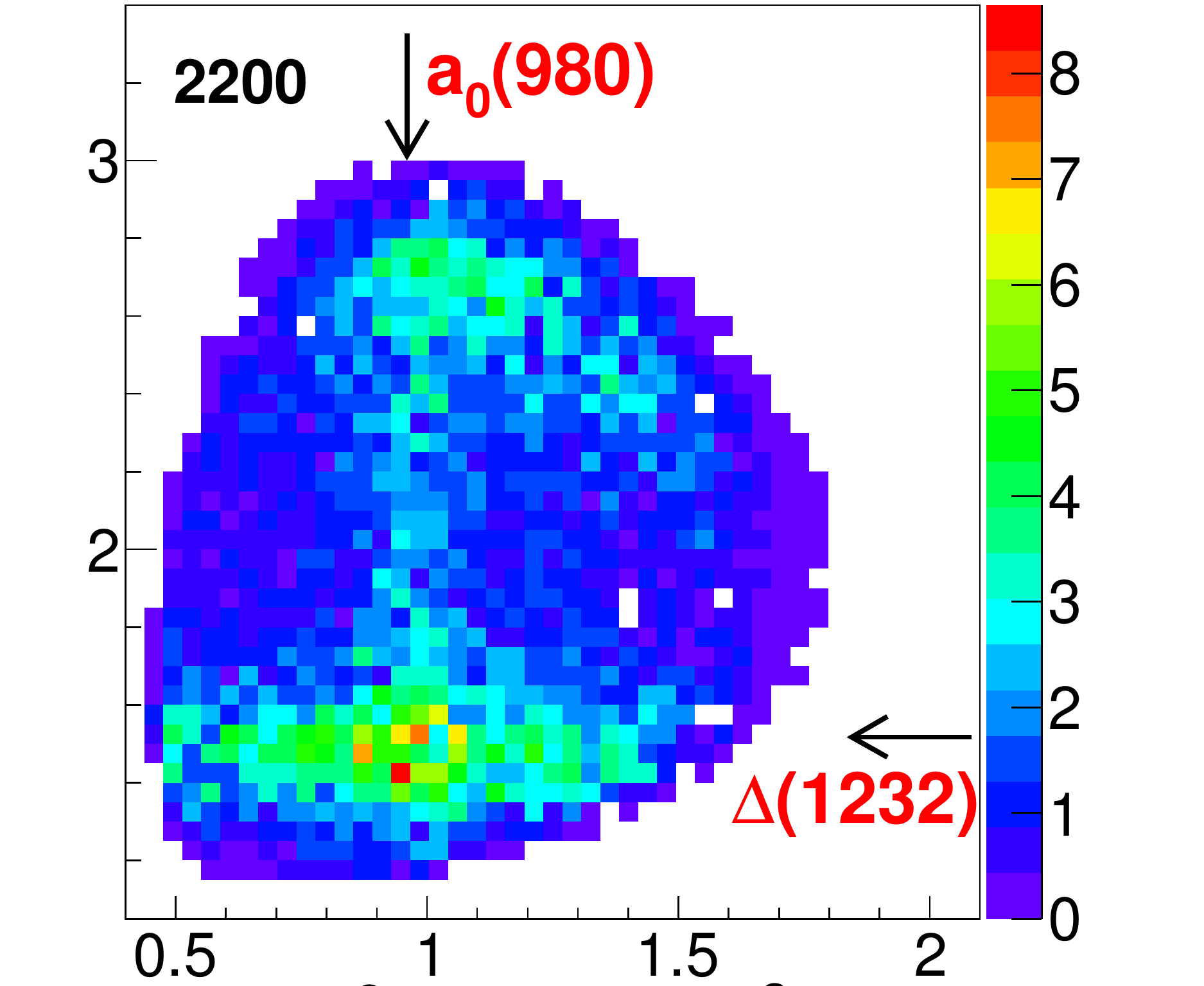}&
\hspace{-6mm}\includegraphics[width=0.24\textwidth]{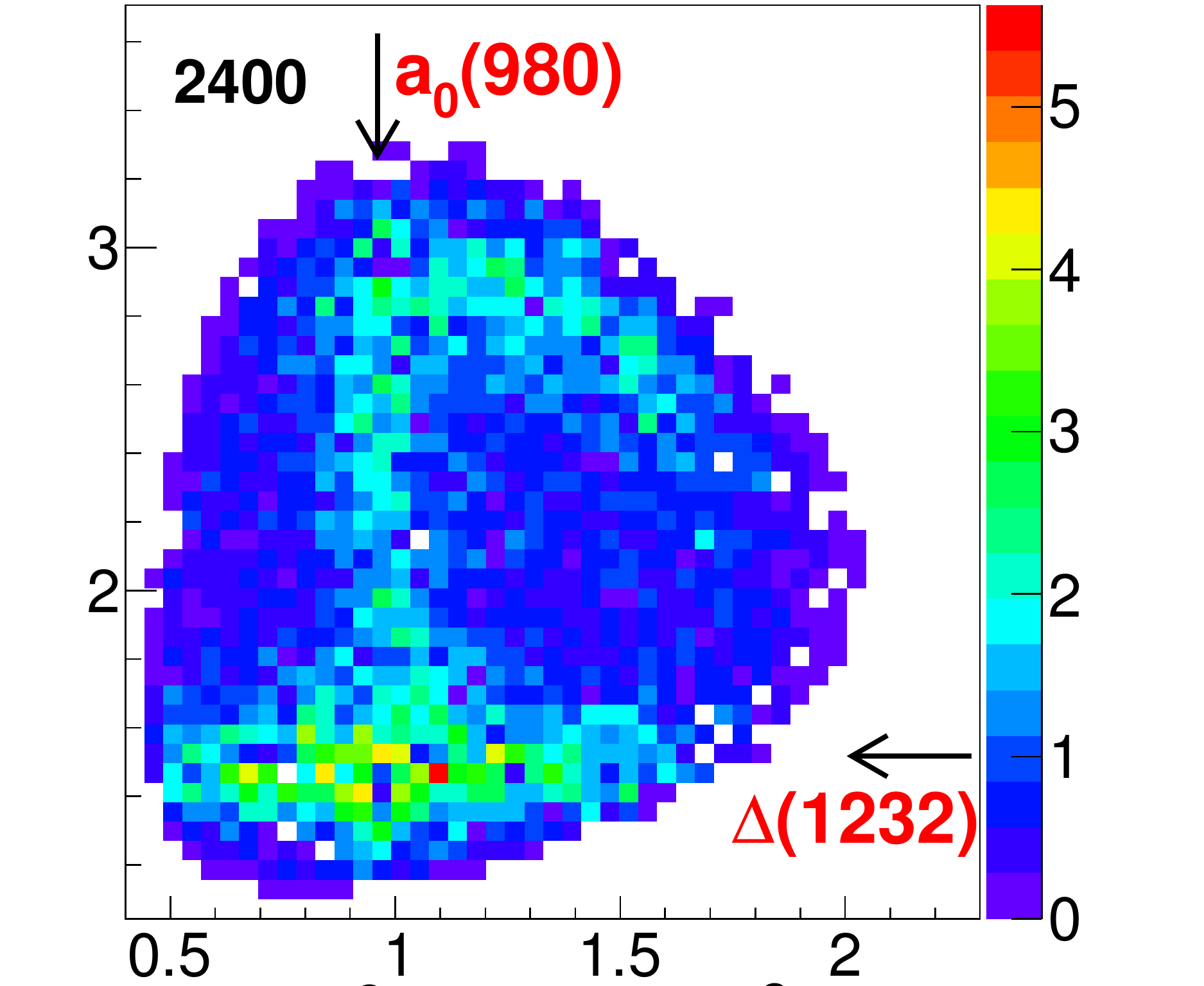}\\[2ex]
\end{tabular}
\caption{Dalitz plots $M^2(p\pi^0)$ versus $M^2(\pi^0\eta)$ for the
incoming photon energy ranges $1000\pm100$\,MeV to
$2400\pm100$\,MeV.  \vspace{2mm}}
\label{fig:dalitz-a}
\begin{tabular}{cccc}
\includegraphics[width=0.24\textwidth]{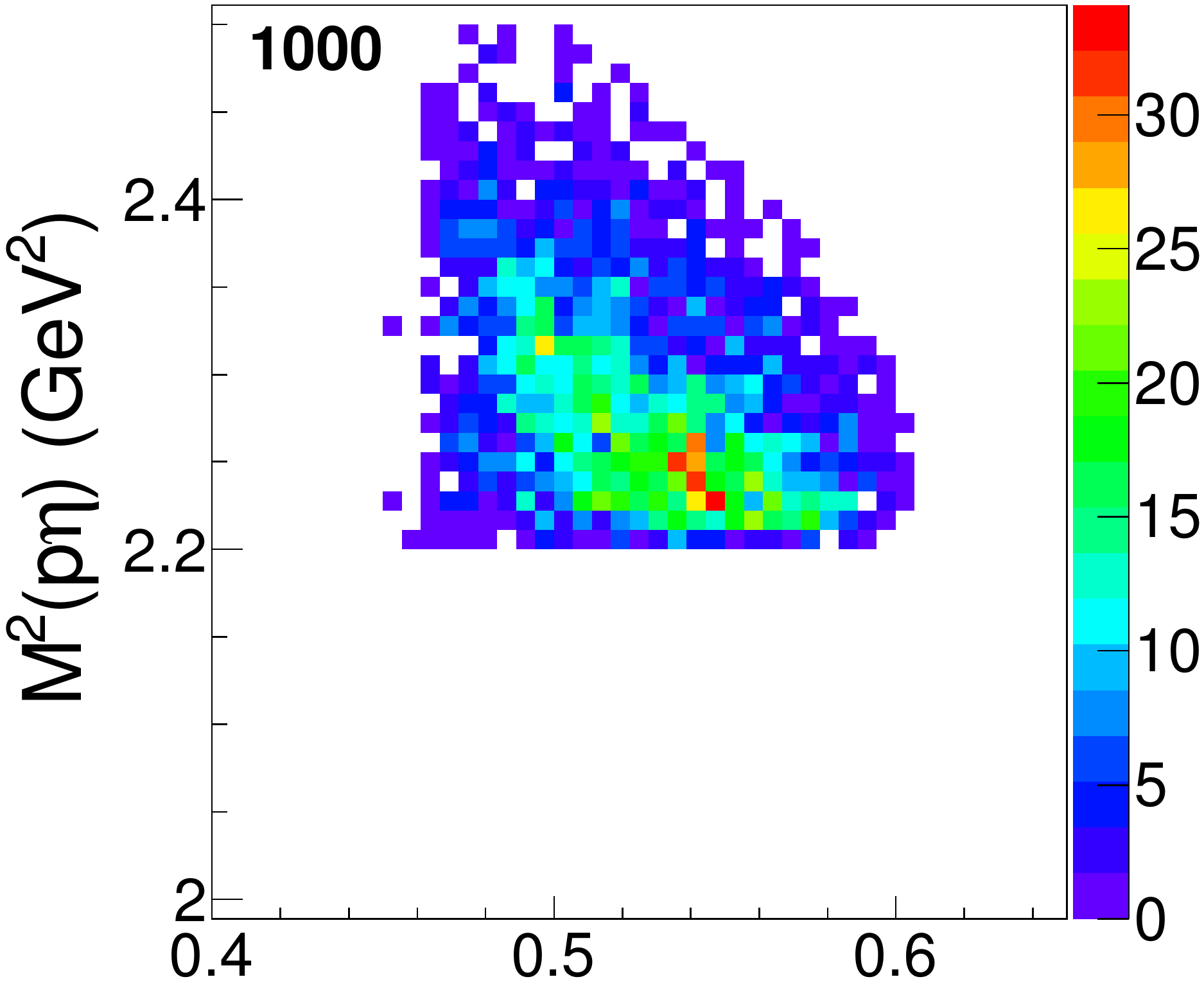}&
\hspace{-2mm}\includegraphics[width=0.24\textwidth]{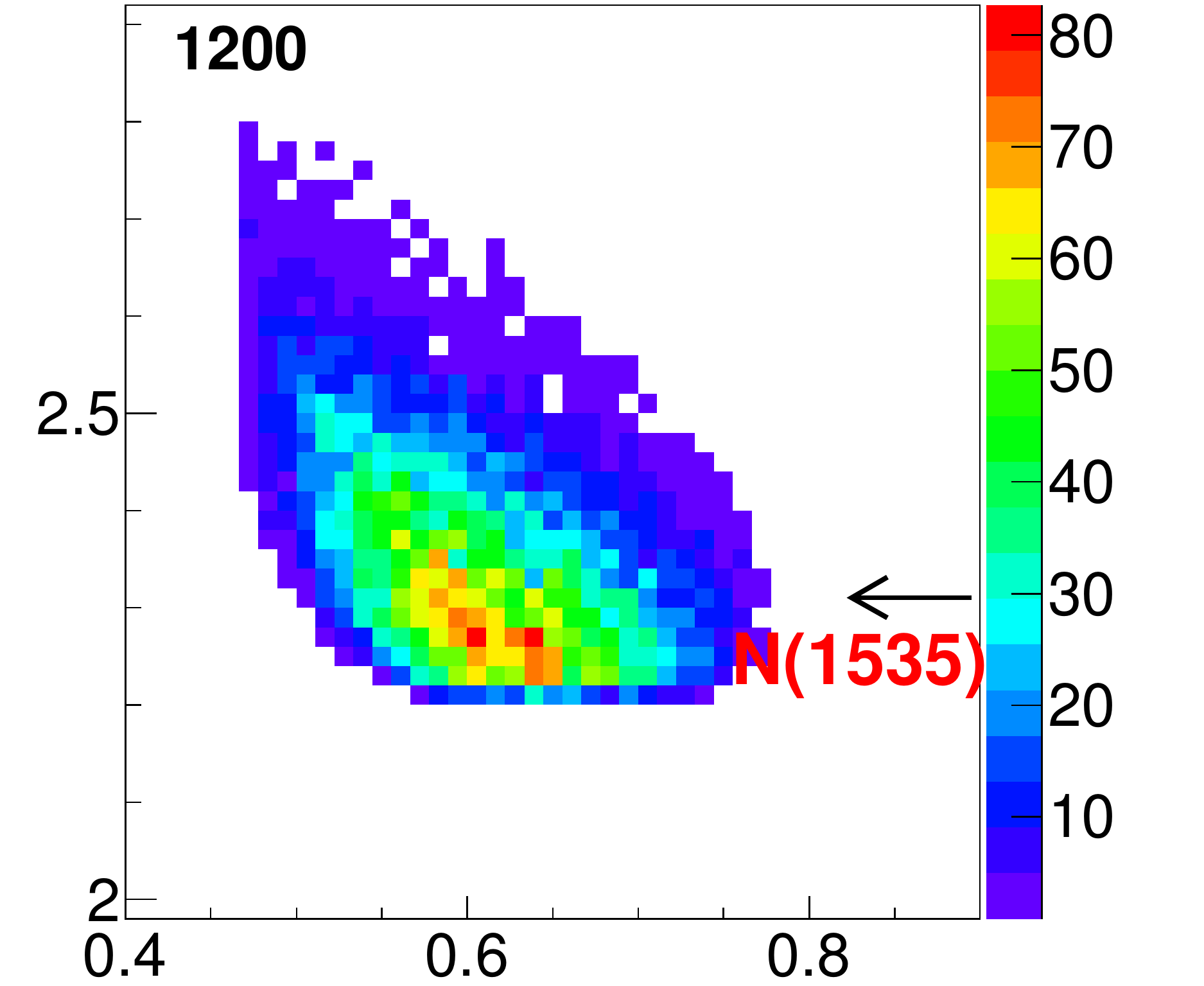}&
\hspace{-6mm}\includegraphics[width=0.24\textwidth]{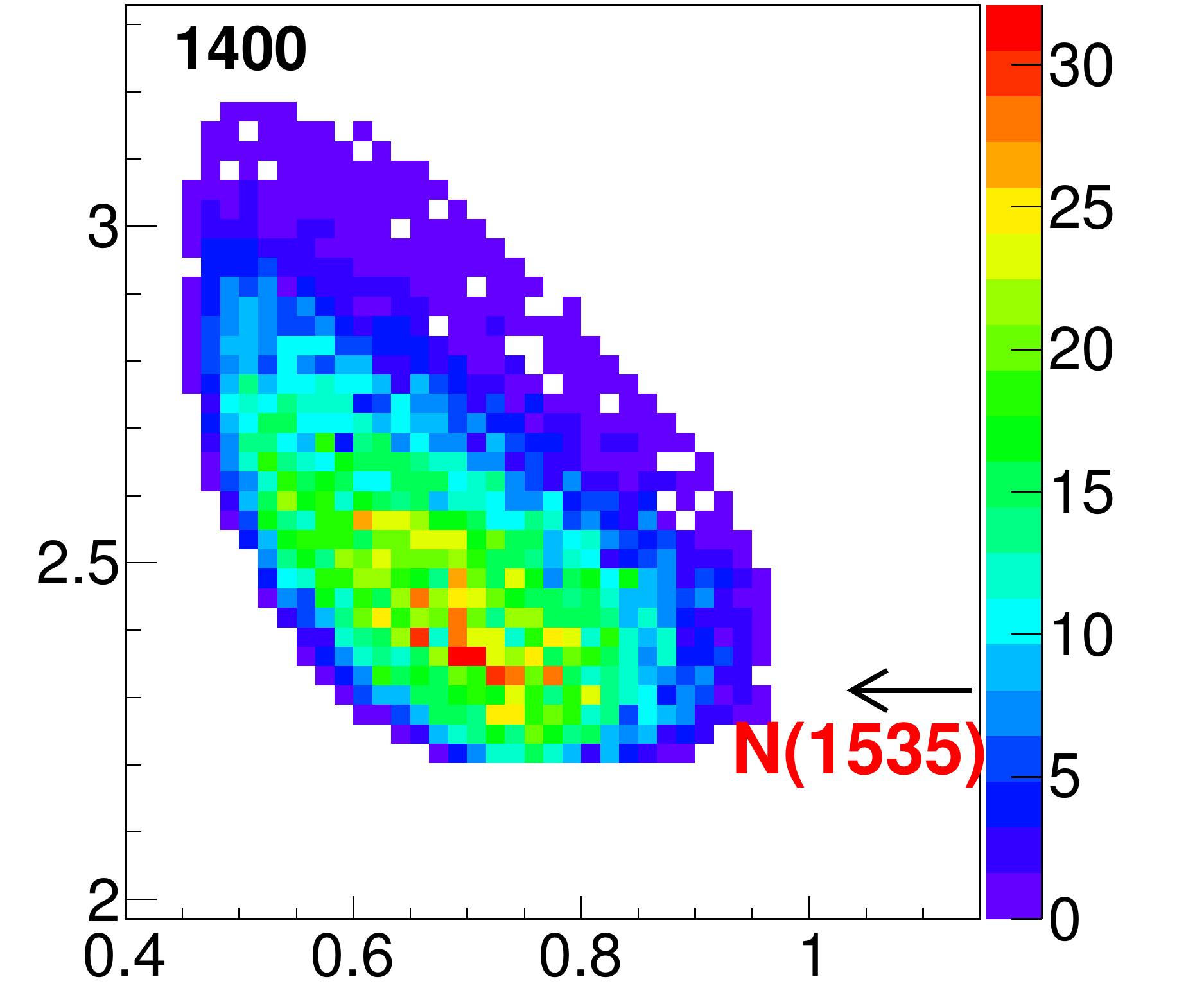}&
\hspace{-6mm}\includegraphics[width=0.24\textwidth]{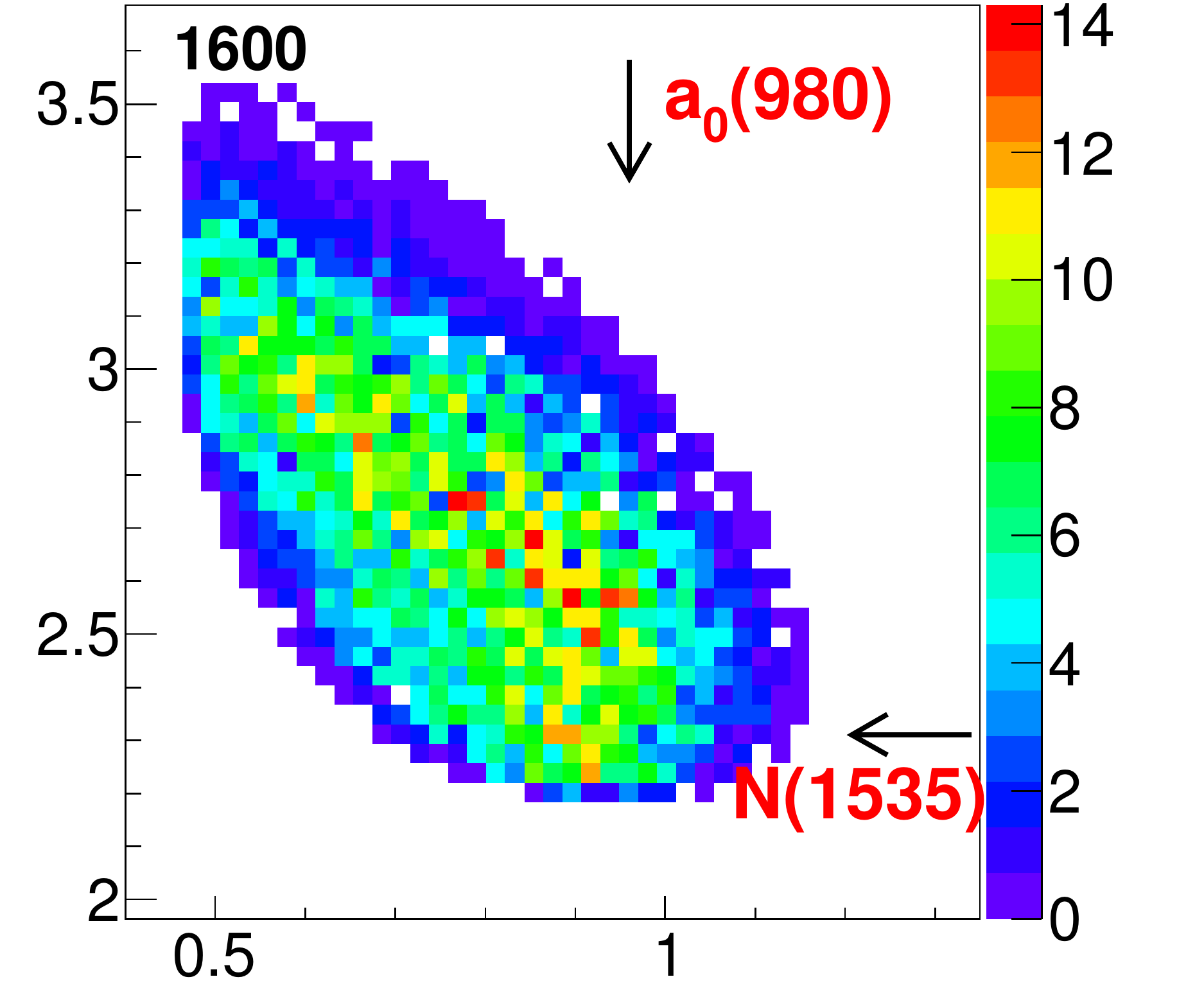}\\
\hspace{-1mm}\includegraphics[width=0.24\textwidth]{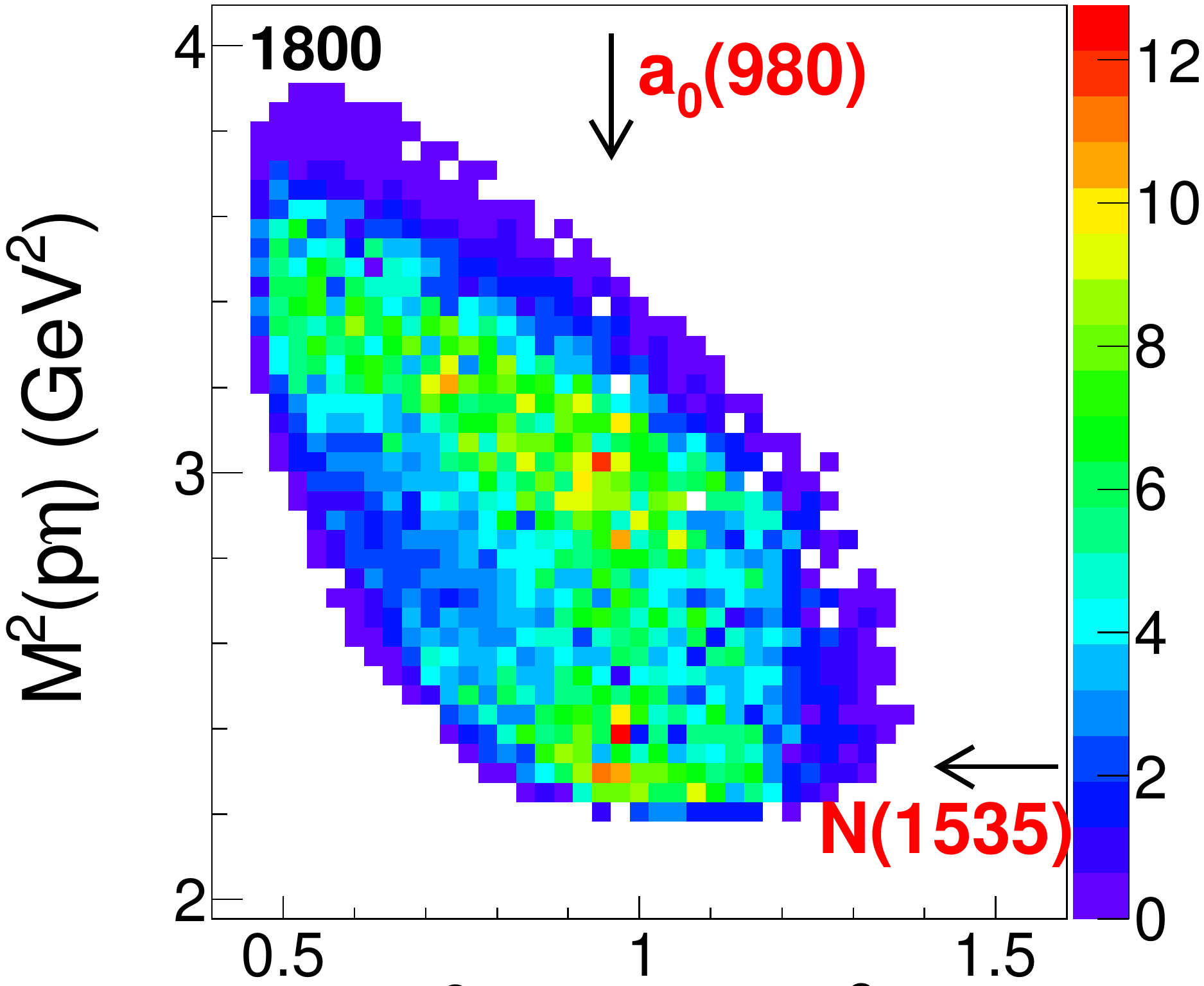}&
\hspace{-2mm}\includegraphics[width=0.24\textwidth]{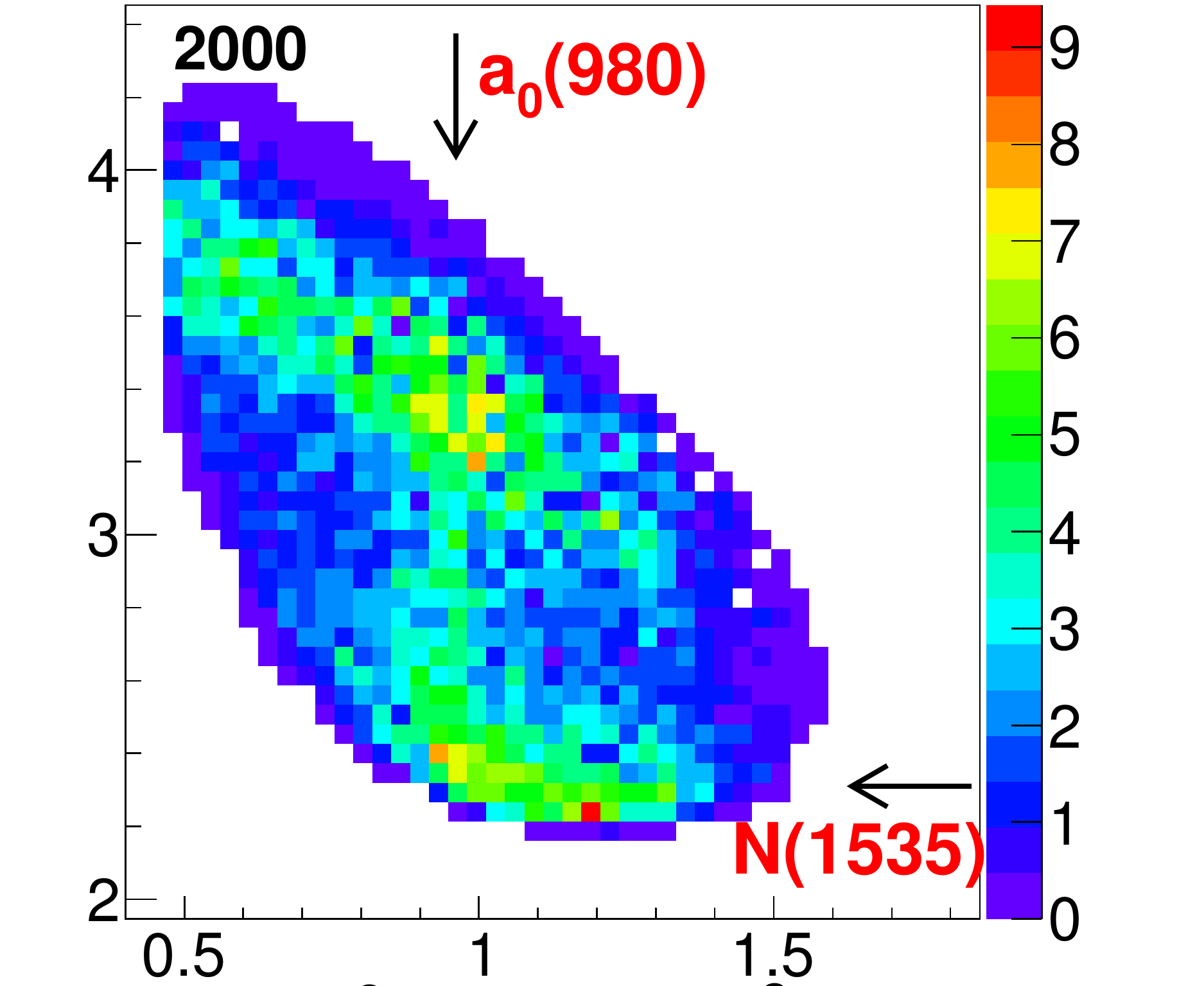}&
\hspace{-6mm}\includegraphics[width=0.24\textwidth]{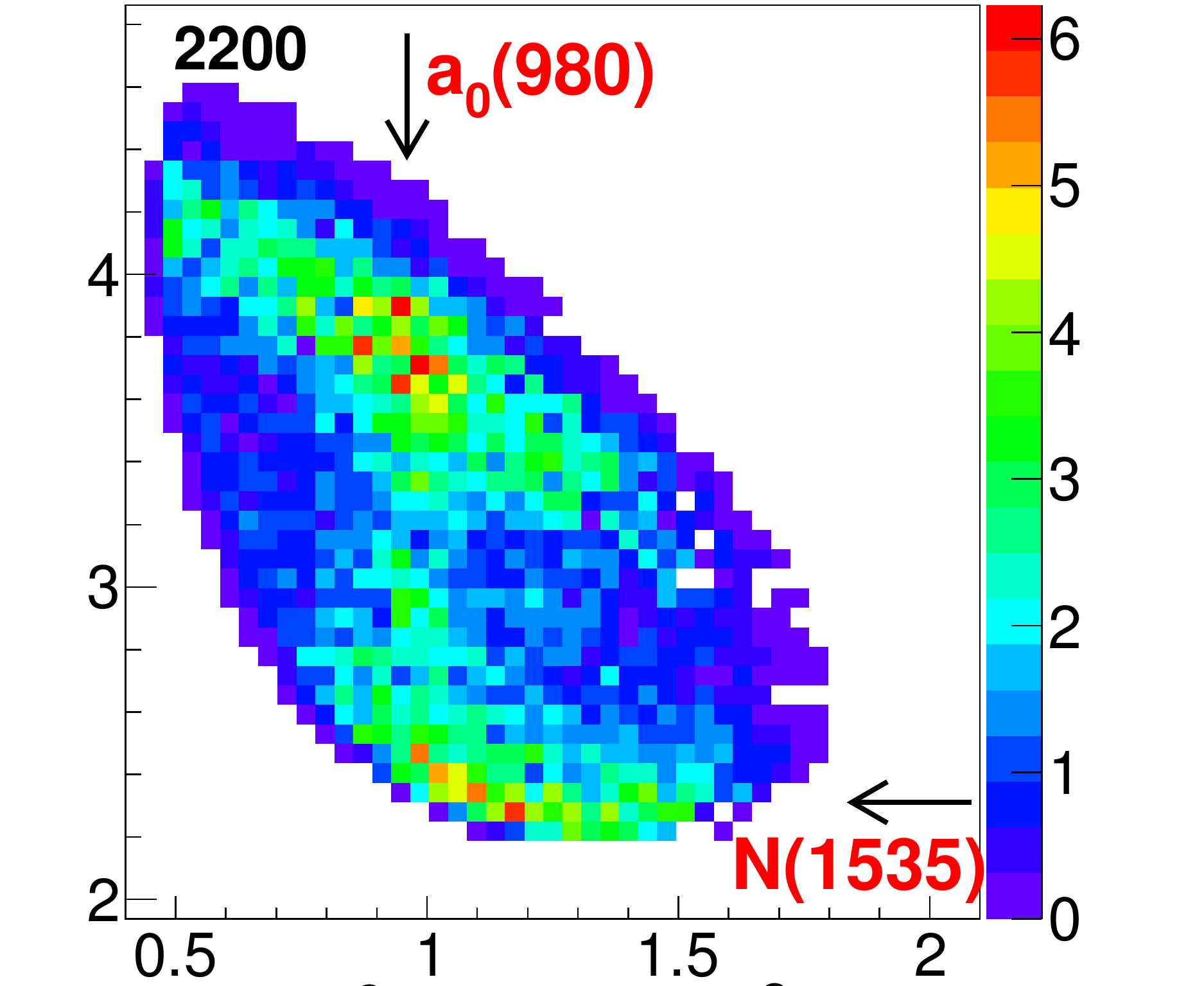}&
\hspace{-6mm}\includegraphics[width=0.24\textwidth]{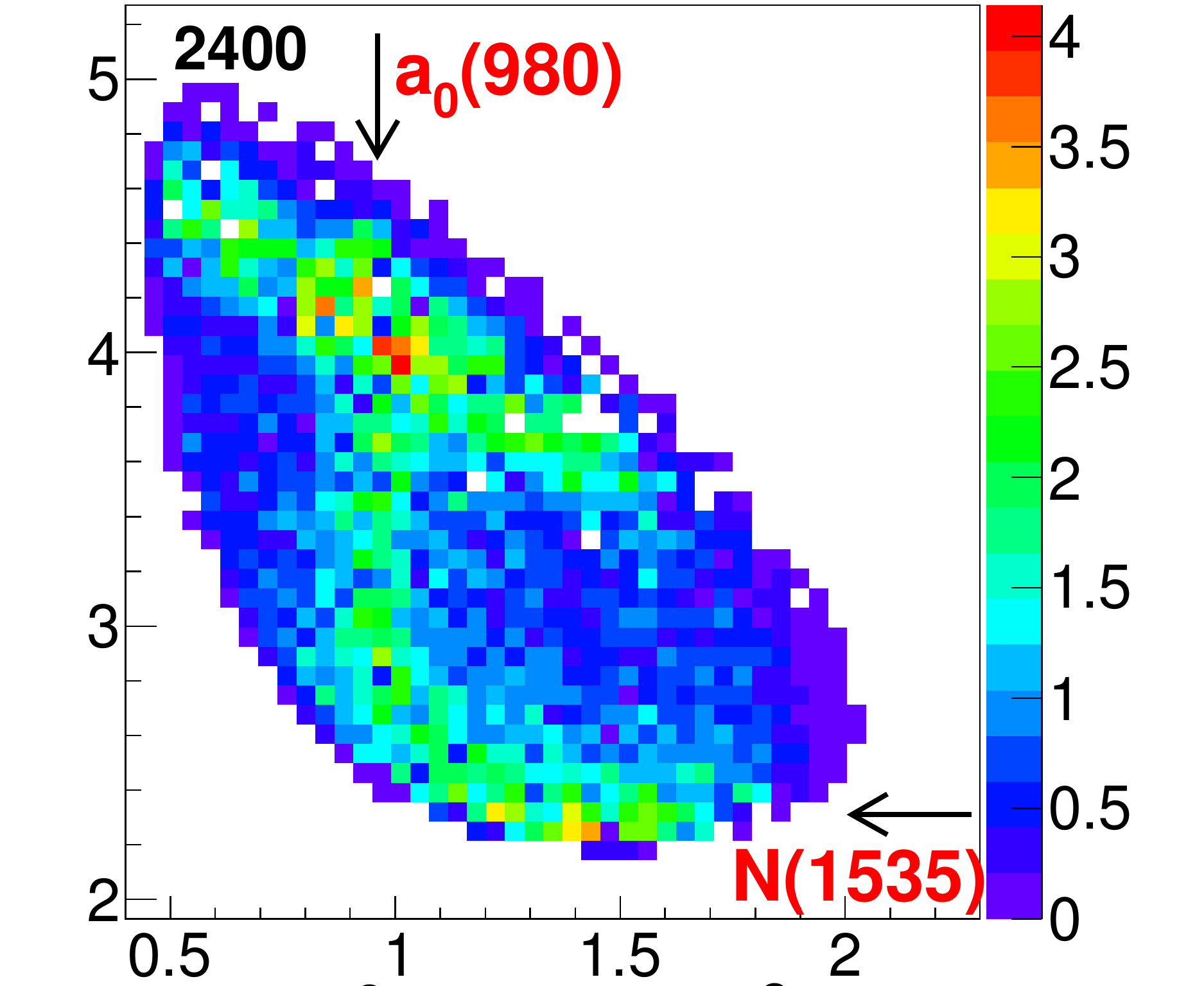}\\[2ex]
\end{tabular}
\caption{Dalitz plots $M^2(p\eta)$ versus $M^2(\pi^0\eta)$ for the
incoming photon energy ranges $1000\pm100$\,MeV to
$2400\pm100$\,MeV.\vspace{2mm}}
\label{fig:dalitz-b}
\end{figure*}

Fig.~\ref{fig:tot_wq} shows the total cross section for reaction
(\ref{ppi0eta}). The cross section is determined from a partial wave
analysis to the data described below. The partial wave analysis
allows us to generate a Monte Carlo event sample representing the
``true'' physics. For any distribution, the efficiency can then be
calculated as fraction of the reconstructed to the generated events.
The open circles in Fig.~\ref{fig:tot_wq} are given by the number of
events, normalized to the incoming photon flux, divided by the
efficiency. The red and blue open circles represent the two run
periods, CB/TAPS1 and CB/TAPS2, respectively. For CB/TAPS1, the $E_\gamma$ range 
covering the coherent peak (b) showed an ($\approx 10\%$) excess of the 
cross section compared to CB/TAPS2. The data are omitted from further analysis.  

The cross section as determined by us is smaller by about 15\%
compared to those measured at GRAAL \cite{Ajaka:2008zz} and at MAMI
\cite{Kashevarov:2009ww}. This is particularly intriguing since the 
cross section for the related reaction $\gamma p\to p\pi^0\pi^0$ 
\cite{Sokhoyan:2014} determined from the same data agrees very well with the GRAAL and
MAMI measurements. We checked all ingredients entering the 
normalization very carefully, and did not find any reason for this
discrepancy. However, the systematic error given in the latter two 
experiments is much smaller than our estimated error, hence we 
decided to normalize
our cross section using one scaling factor of 0.85 which is applied
to the photon flux in the full energy range. This factor is applied
in Fig.~\ref{fig:tot_wq}, in all subsequent figures, and in the
partial wave analysis. The error which originates from this uncertainty 
for the decay branching ratios is covered by the systematic error.

In the figure, we also show important isobar contributions as
derived in the partial wave analysis described below. The strongest
isobar $\Delta(1232)\eta$ defines up to 70\% of the total
cross section. $N(1535)\pi$ and $p\,a_0(980)$ provide a significant
contribution as well.

\begin{figure*}[pt]
\vspace{2mm}
\begin{tabular}{cc}
\hspace{-2mm}\includegraphics[width=0.50\textwidth,height=0.45\textwidth]{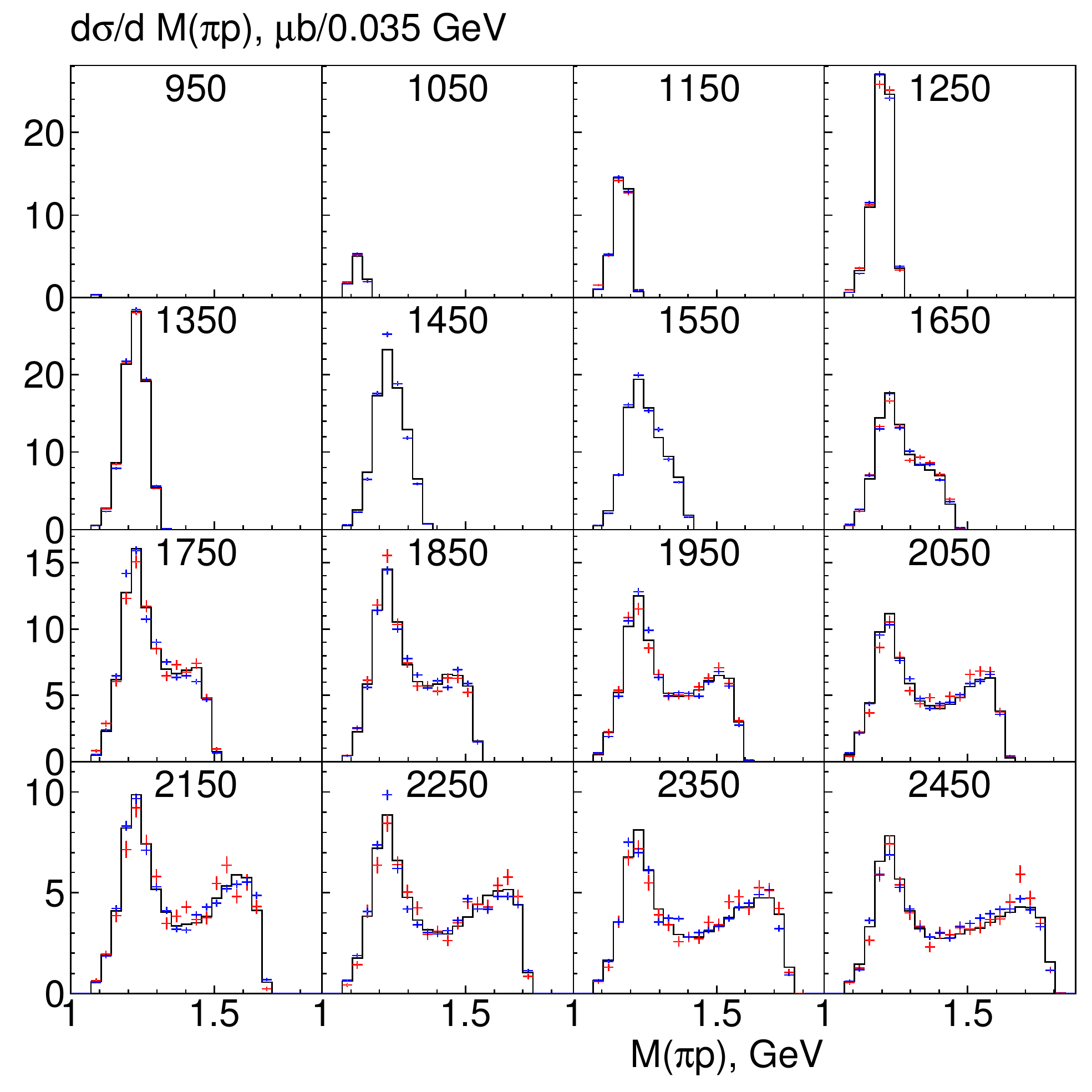}&
\hspace{-3mm}\includegraphics[width=0.50\textwidth,height=0.45\textwidth]{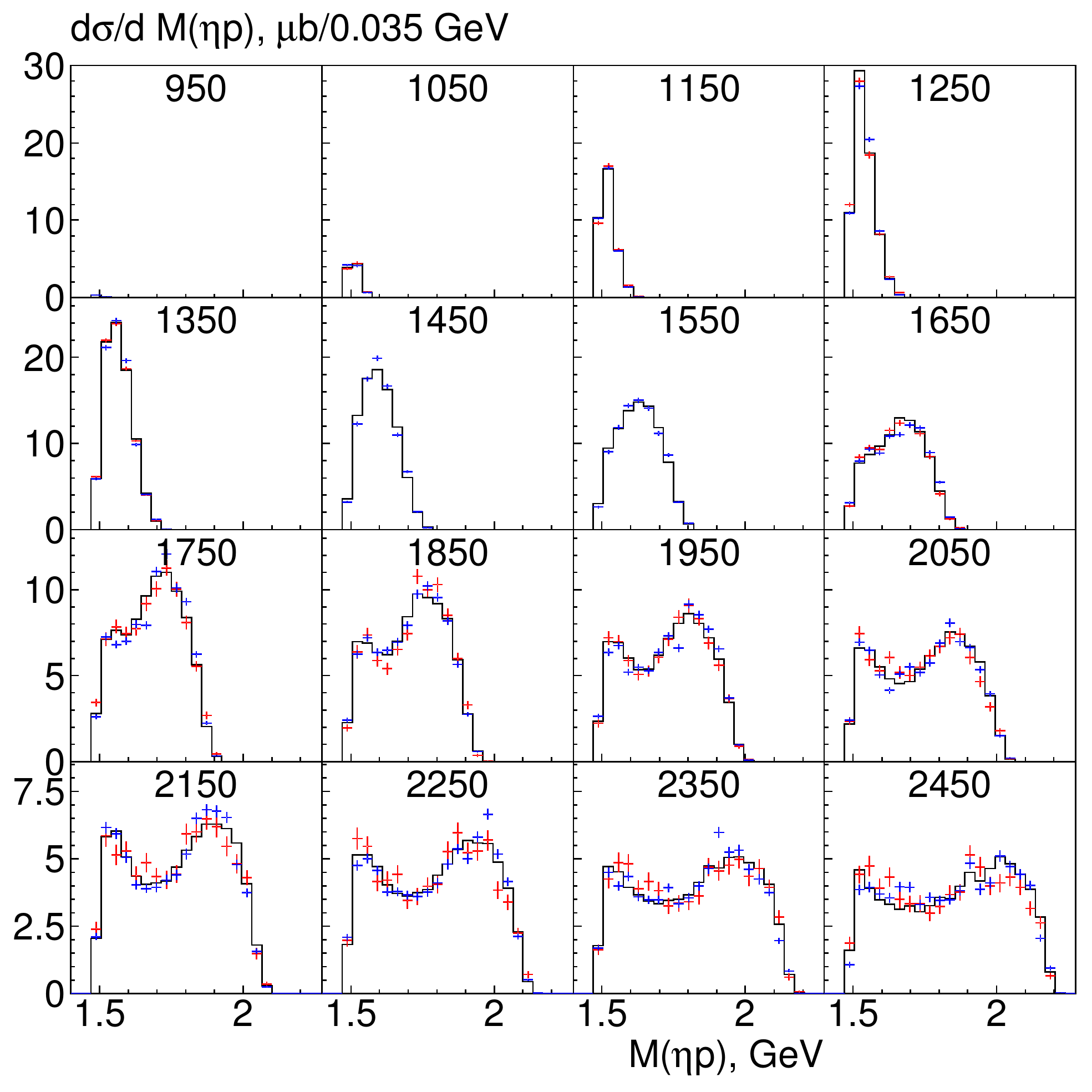}
\end{tabular}
\caption{Differential cross sections d$\sigma$/d$M_{\pi^0p}$ (left)
and d$\sigma$/d$M_{\eta p}$ (right). Red points: CB/TAPS1, blue
points: CB/TAPS2, histogram: BnGa2013 PWA fit. Numbers denote the
center of 100\,MeV wide bins in incoming photon energy.}
\label{fig:diff_m-a}
\end{figure*}
\subsection{Dalitz plots}
\label{sec:dalitz}

Figs.~\ref{fig:dalitz-c}-\ref{fig:dalitz-b} show three variants of
Dalitz plots for the reaction $\gamma p\to p\pi^0\eta$ for 100\,MeV
wide bins in incoming photon energy. Clear structures can be
observed in all three squared invariant masses. In
Fig.~\ref{fig:dalitz-c}, $M^2(p\pi^0)$ is shown versus $M^2(p\eta)$.
At low energy, an enhancement is observed which develops into two
separate active areas when higher energies are reached. The two
areas can be identified with the production of $N(1535)\pi$, with
$N(1535)$ decaying into $p\eta$, and of $\Delta(1232)\eta$, with
$\Delta$ decaying into $p\pi^0$. In the second diagonal, a faint
band is visible corresponding to the production of $a_0(980)$ mesons
recoiling against a proton. The $a_0(980)$ meson is directly visible
as a vertical band in Fig.~\ref{fig:dalitz-a} and
Fig.~\ref{fig:dalitz-b}. The former figure shows again clear
evidence for $\Delta(1232)\eta$ production, the latter one for
$N(1535)\pi$. The highest intensity is observed at the crossings of
the two bands. Evidently, interference will be important in the
partial wave analysis. These are particularly sensitive to the
phases of the amplitudes and help identify resonant behavior. The
findings observed in the visual inspection of the Dalitz plots are
confirmed in the mass distributions given below.

\subsection{Mass distributions}
\label{sec:diff}

\begin{figure}[pt]
\begin{tabular}{c}
\hspace{-3mm}\includegraphics[width=0.50\textwidth,height=0.45\textwidth]{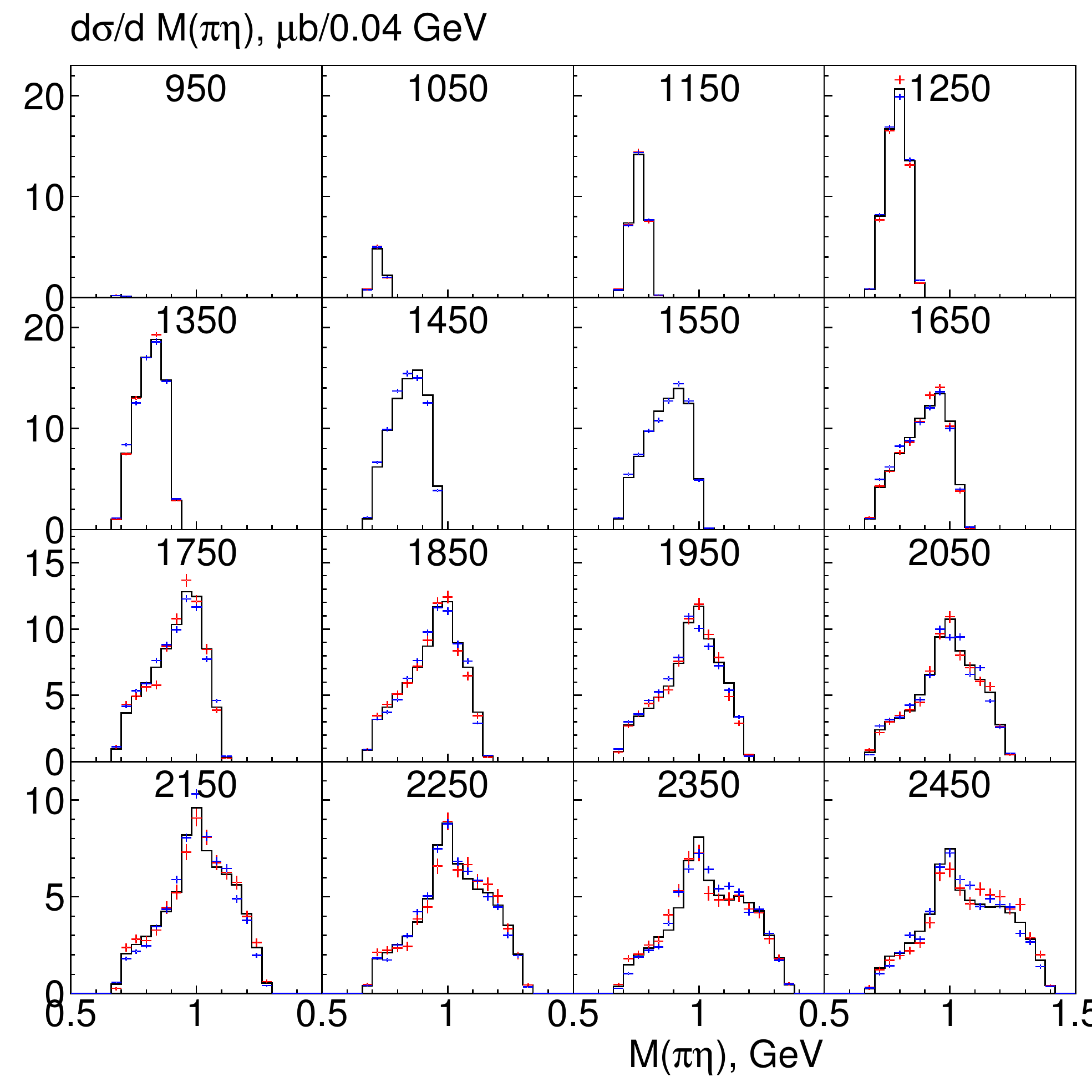}
\end{tabular}
\caption{Differential cross sections d$\sigma$/d$M_{\pi^0\eta}$. See
Fig.~\protect\ref{fig:diff_m-a} for further
explanations.}
\label{fig:diff_m-b}
\end{figure}

Figs.~\ref{fig:diff_m-a} and \ref{fig:diff_m-b} show mass
distributions (in $\mu$b per mass bin) for 100\,MeV wide bins in the
photon energy. From a summation over all mass bins in a photon
energy bin, the total cross section can be recalculated. At a photon
energy of 1650\,MeV, $\Delta(1232)$ production becomes visible above
a smooth background, see Fig.~\ref{fig:diff_m-a}~(left).
$\Delta(1232)$ production becomes increasingly important and
continues to make a significant contribution to the cross section up
to the highest energies. In Fig.~\ref{fig:diff_m-a} (right) an
identifiable threshold enhancement becomes visible at about
1700\,MeV photon energy; the clearest evidence for $N(1535)$
production is seen at around 2150\,MeV. Fig.~\ref{fig:diff_m-b}
exhibits a narrow peak at $\approx$980\,MeV which we identify with
$a_0(980)$. Its contribution to the cross section rises slowly with
photon energy and reaches a plateau at about 2000\,MeV.

\begin{figure}[pt]
\bc
\begin{tabular}{ccc}
\hspace{-6mm}\includegraphics[width=0.24\textwidth]{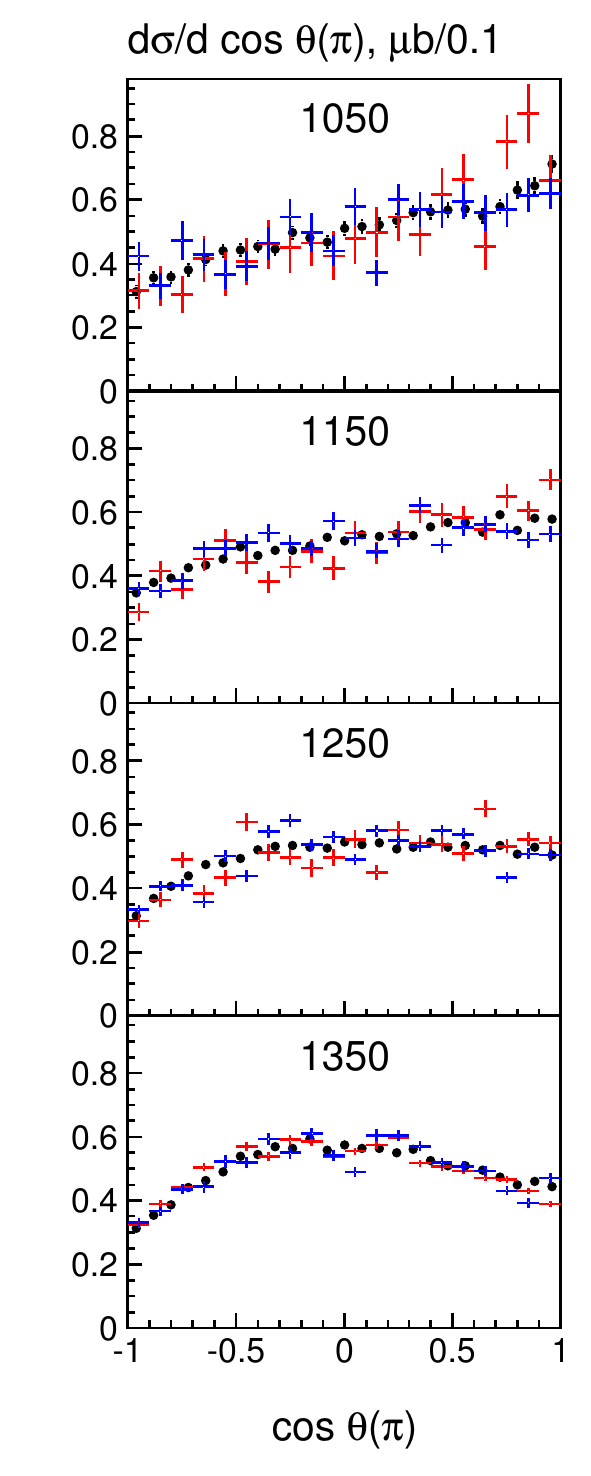}&
\hspace{-6mm}\includegraphics[width=0.24\textwidth]{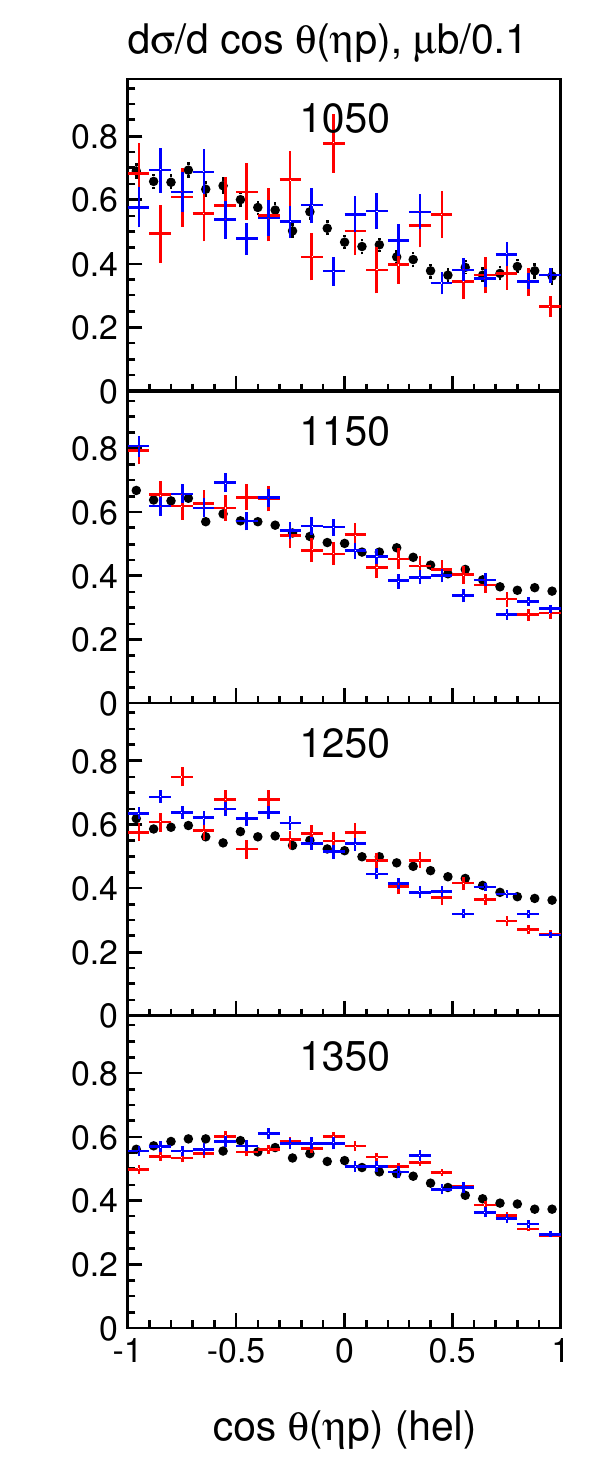}
\end{tabular}
\ec
\caption{Comparison of CBELSA/TAPS (red/blue) and MAMI (black)
Differential cross-sections in the cm (left) and helicity (right)
systems. Numbers denote the center of 100\,MeV wide bins in incoming
photon energy.}
\label{fig:comp}\end{figure}

\begin{figure*}[pt]
\begin{tabular}{ccc}
\hspace{-2mm}\includegraphics[width=0.33\textwidth]{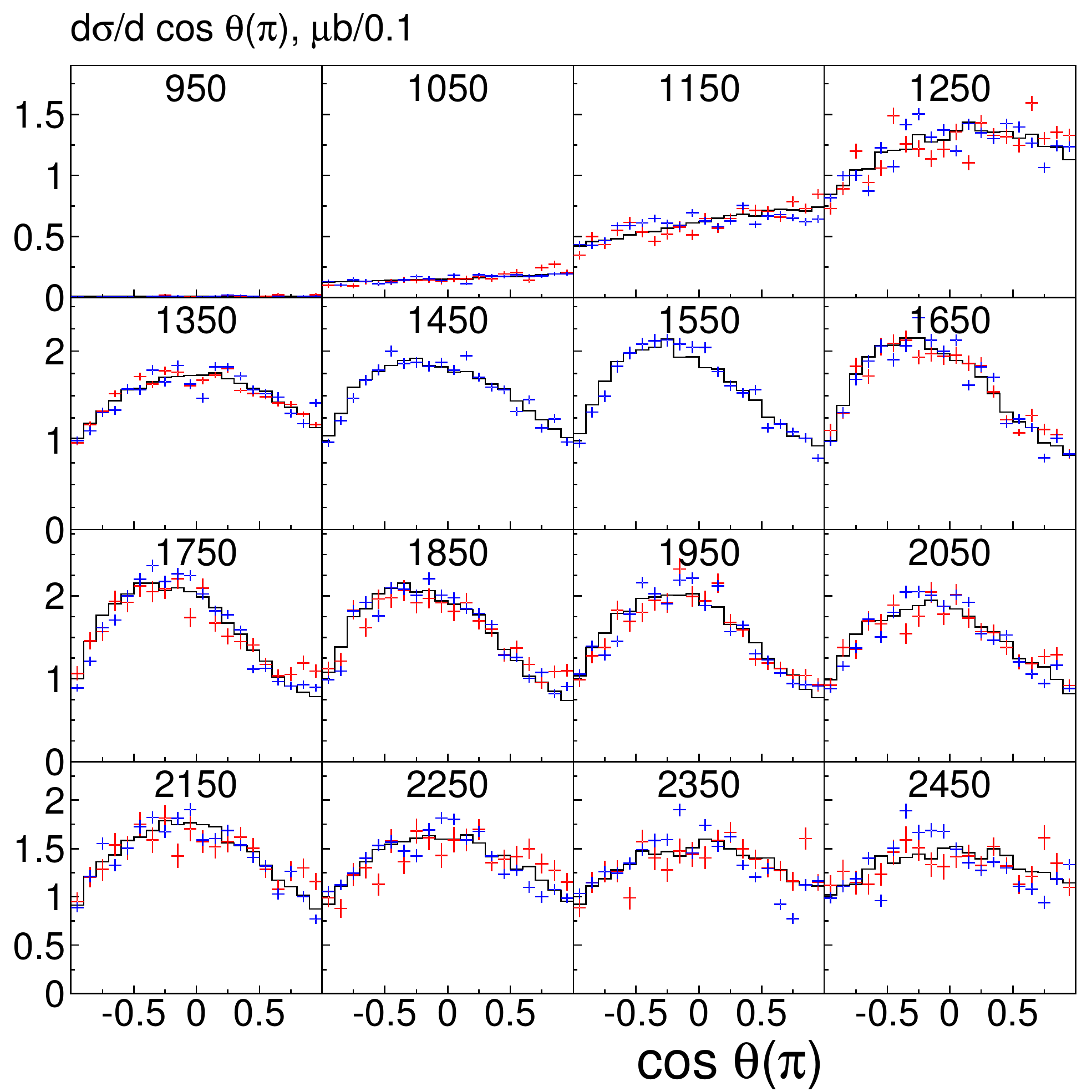}&
\hspace{-3mm}\includegraphics[width=0.33\textwidth]{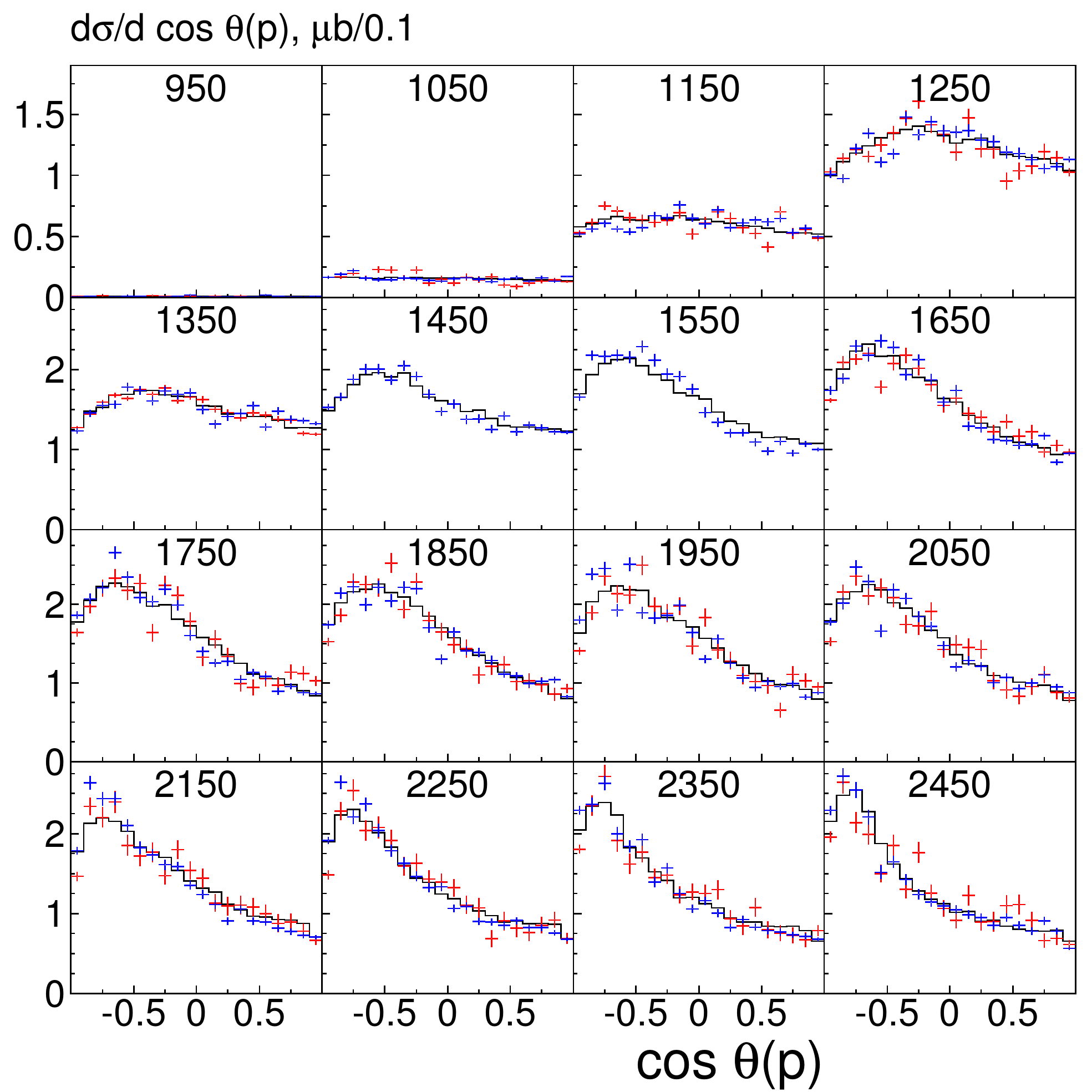}&
\hspace{-3mm}\includegraphics[width=0.33\textwidth]{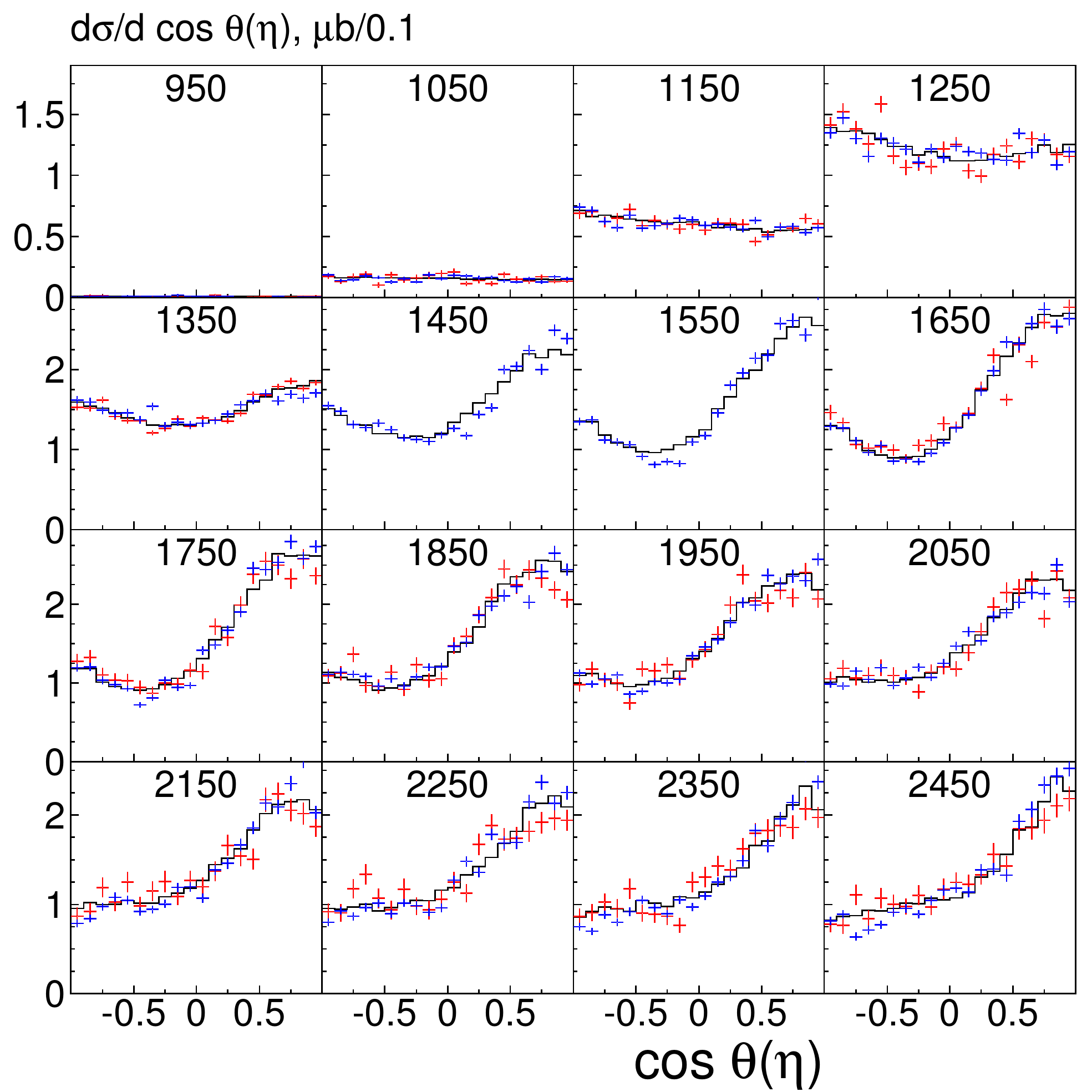}
\end{tabular}
\caption{Differential cross-sections in the center-of-mass system.
d$\sigma$/d$(\cos\pi)$ (left), d$\sigma$/d$(\cos p)$ (center),
d$\sigma$/d$(\cos\eta)$ (right.) Red points: CB/TAPS1, blue points:
CB/TAPS2, histogram: BnGa2013 PWA fit. Numbers denote the center of
100\,MeV wide bins in incoming photon energy. }
\label{fig:diff_cos}
\end{figure*}
\subsection{Comparison with other data}
\label{sec:comp}
In Fig.~\ref{fig:comp} our differential distributions are compared
with those measured at MAMI \cite{Kashevarov:2009ww}. The MAMI data
have a significant better statistics but are limited in energy. The
agreement between both measurements is very good, suggesting that
the acceptance is well understood in both experiments.\vspace{-2mm}

\subsection{Angular distributions}

The 3-body final state in the reaction $\gamma p\to p\pi^0\eta$ can
be characterized by several angular distributions. The $\pi^0\eta$
subsystem can be treated as an ordinary meson (and contains at least
the scalar $a_0(980)$ meson) and can be used to define differential
cross-sections in the center-of-mass system. In
Fig.~\ref{fig:diff_cos}, this differential cross-section is given as
a function of the cosine of the scattering angle of the recoiling
proton, $\cos\theta_p$. Similarly, $\cos\theta_\pi$ gives the
differential cross-section of the $\pi^0$ recoiling against the
$p\eta$ subsystem, and $\cos\theta_\eta$ for the $\eta$ recoiling
against $p\pi^0$. With a mass cut on $\Delta(1232)$ (and background
subtraction), $d\sigma/\cos\theta_\eta$ would be the cross-section
for the reaction $\gamma p\to \Delta(1232)\eta$.

\begin{figure*}[pt]
\bc
\begin{tabular}{ccc}
\hspace{-2mm}\includegraphics[width=0.32\textwidth]{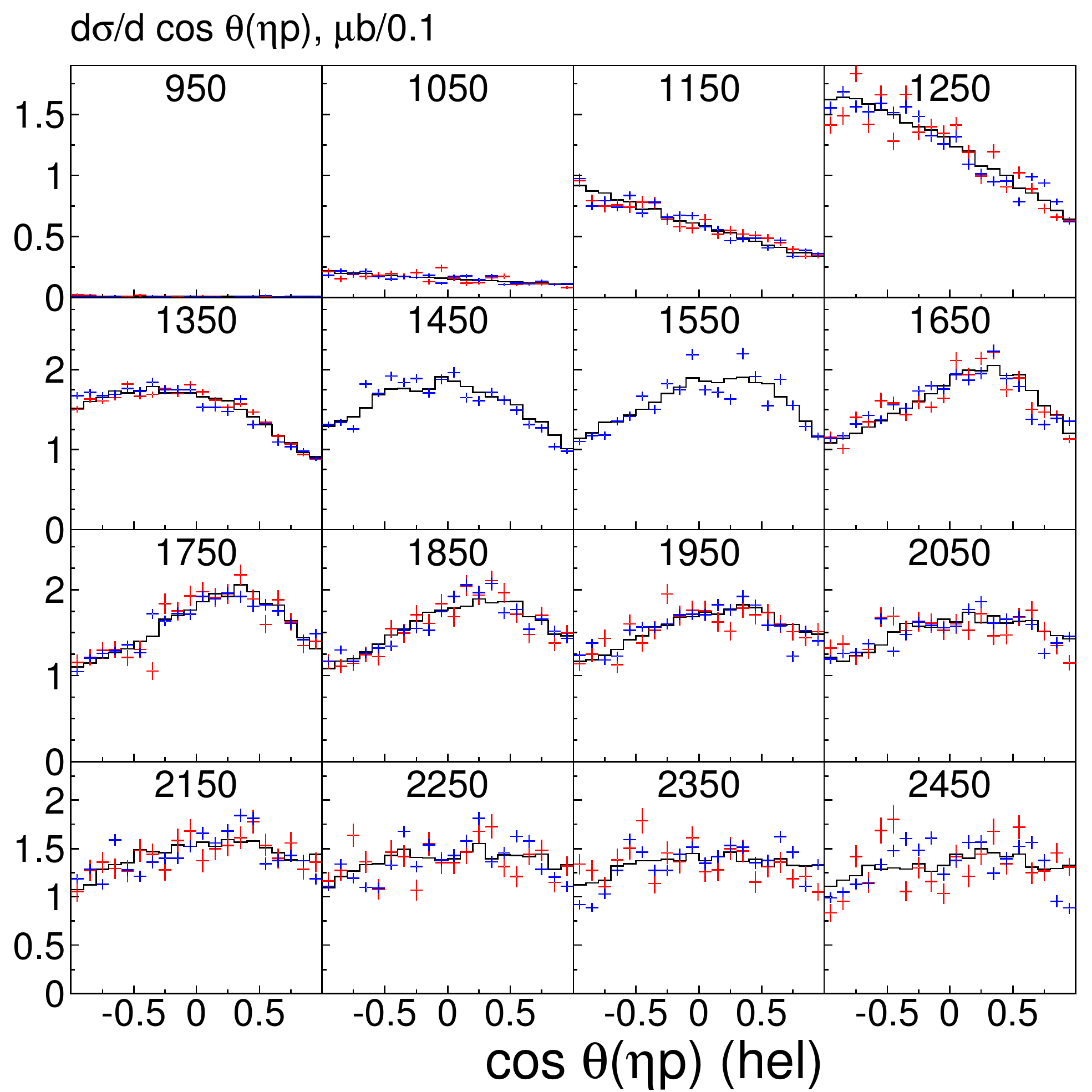}&
\hspace{-3mm}\includegraphics[width=0.32\textwidth]{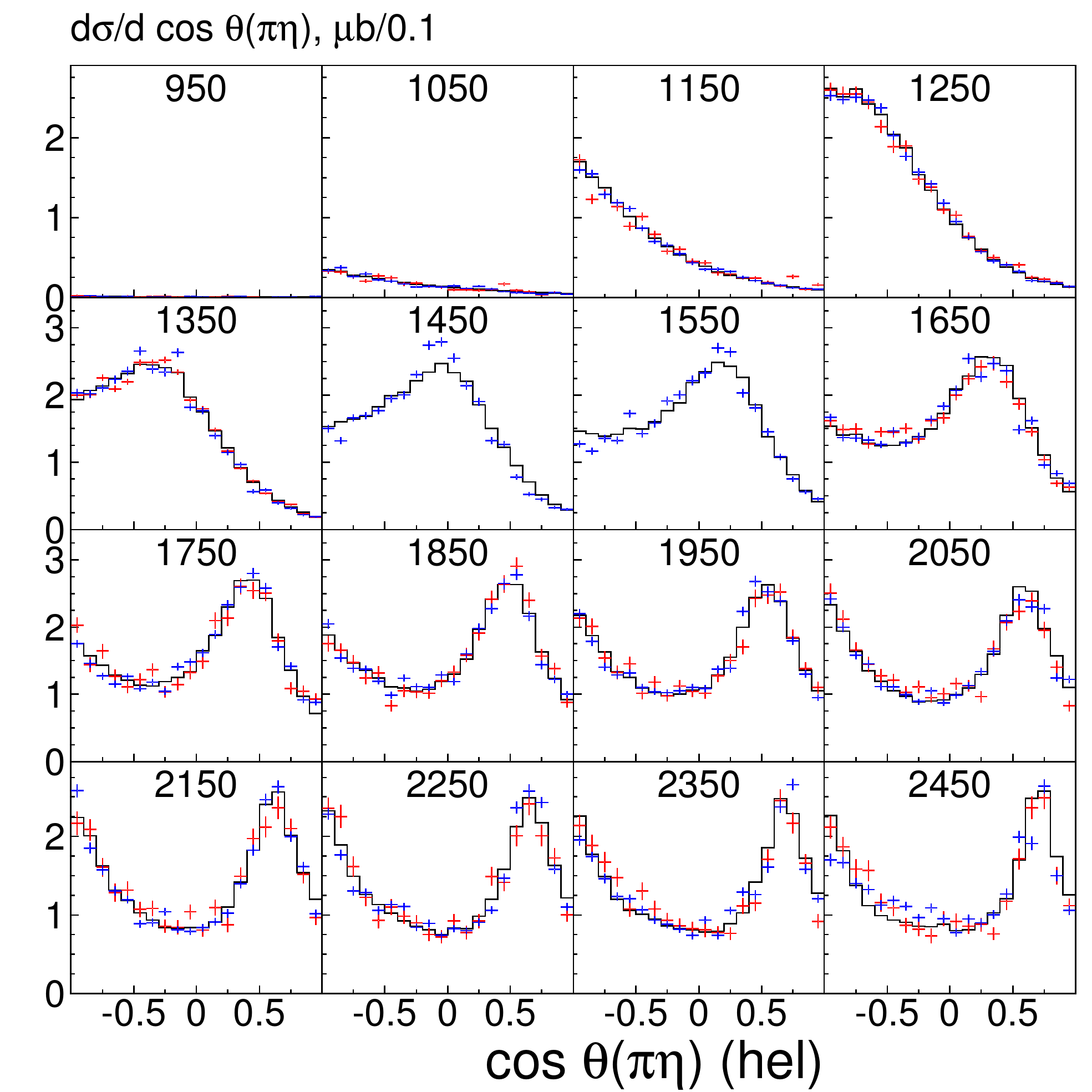}&
\hspace{-3mm}\includegraphics[width=0.32\textwidth]{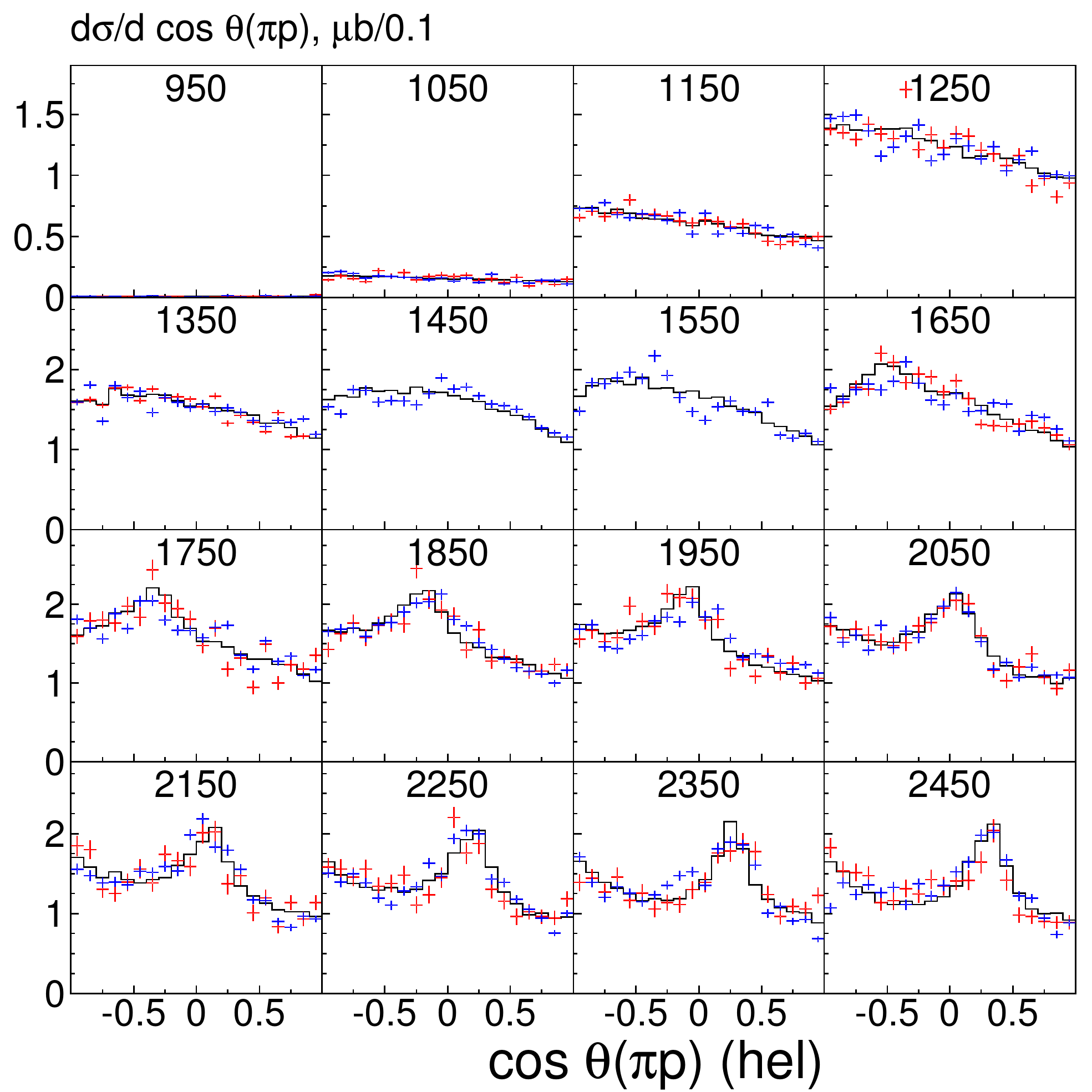}
\end{tabular}
\vspace{-2mm}\ec \caption{Differential cross-sections
d$\sigma$/d$\Omega$ in the helicity frame. See
Fig.~\protect\ref{fig:diff_cos} for further explanations.}
\label{fig:diff_he}
\bc
\begin{tabular}{ccc}
\hspace{-2mm}\includegraphics[width=0.32\textwidth]{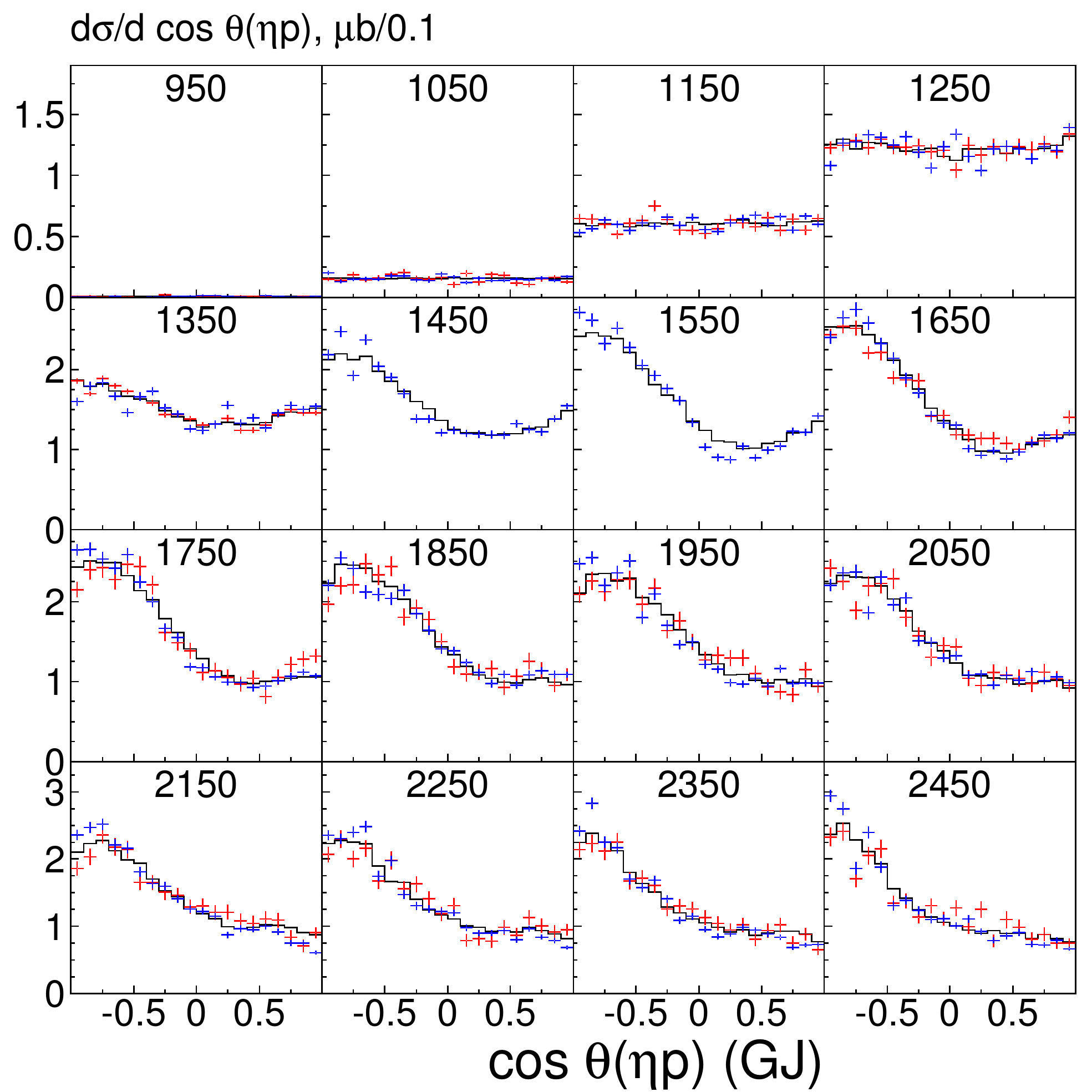}&
\hspace{-3mm}\includegraphics[width=0.32\textwidth]{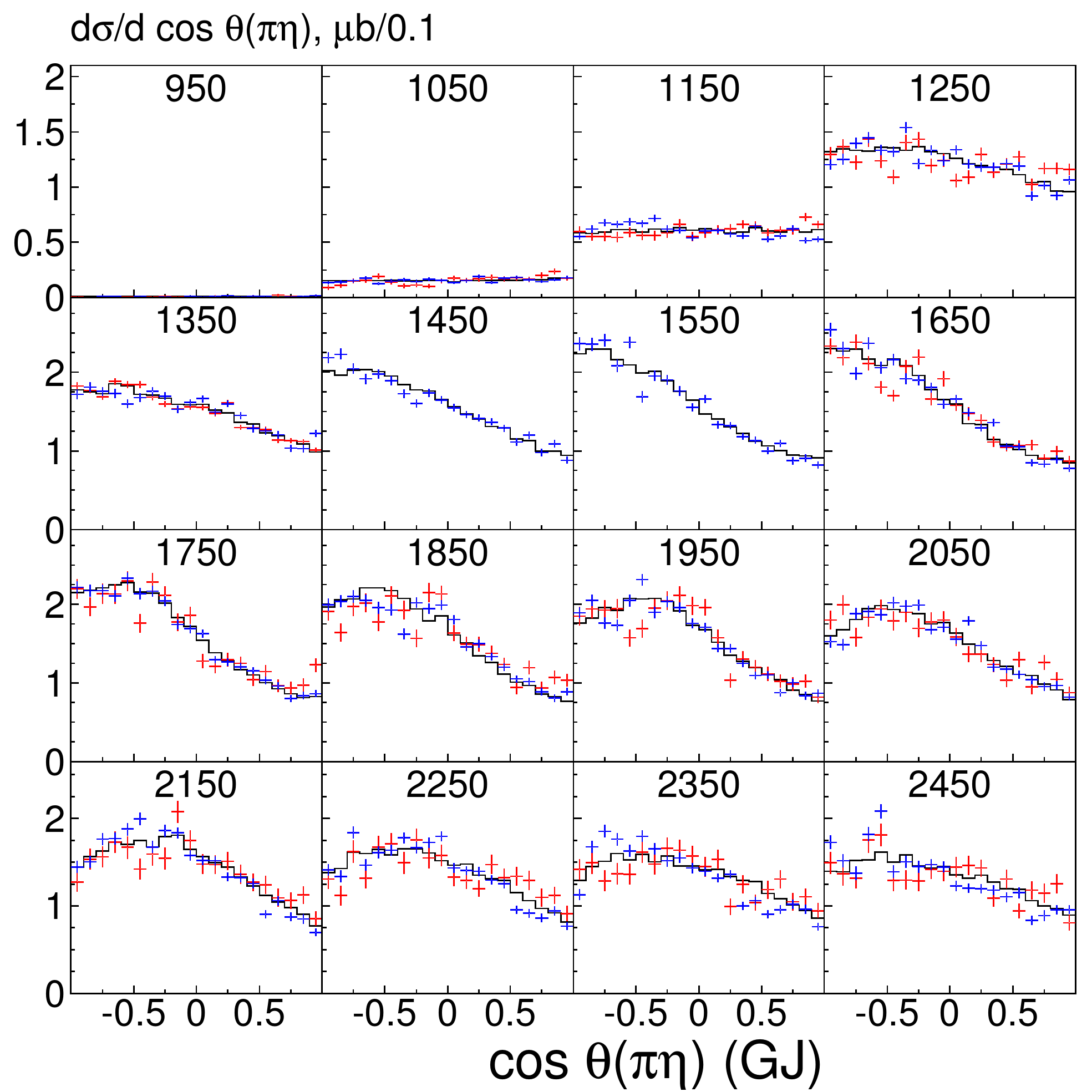}&
\hspace{-3mm}\includegraphics[width=0.32\textwidth]{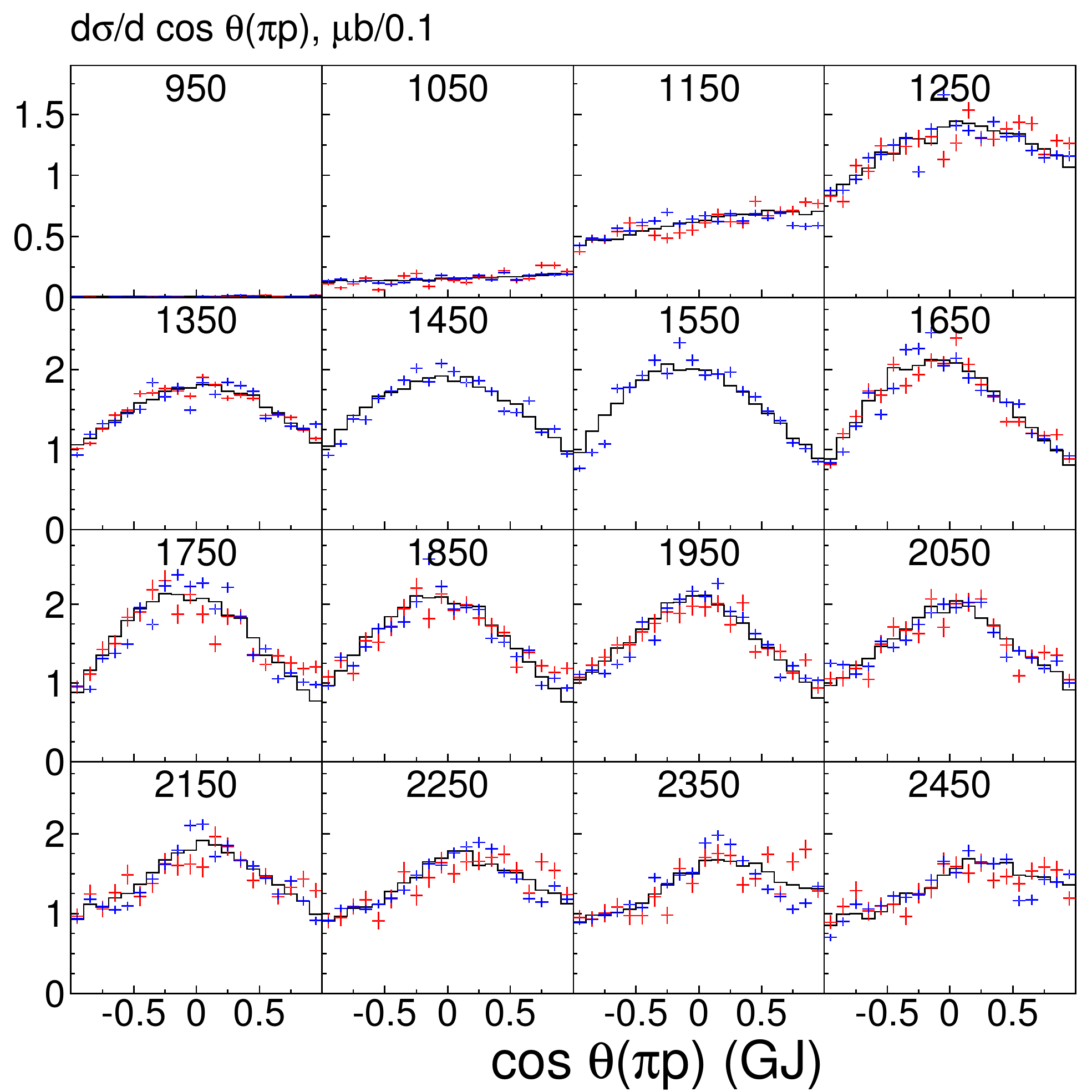}
\end{tabular}\vspace{-2mm}
\ec
\caption{Differential cross-sections d$\sigma$/d$\Omega$ in the
Gottfried-Jackson frame. See Fig.~\protect\ref{fig:diff_cos} for further
explanations.}
\label{fig:diff_gj}
\end{figure*}

Two additional angular distributions are often shown which are
defined in the rest frame of a two-particle subsystem, (e.g., of the
$\pi\eta$ subsystem). In both systems, the $y$-axis is defined by
the reaction-plane normal and the $x$- and $z$-axis lie in the
reaction plane. In the helicity frame (HEL), the $z$-axis is taken
in the direction of the outgoing two-particle system (e.g.
$\pi\eta$), while in the Gottfried-Jackson frame GJ the $z$-axis is
taken along the incoming photon. The distributions are shown in
Fig.~\ref{fig:diff_he} and~\ref{fig:diff_gj}.
\subsection{Polarization observables}
\label{sec:polobs}
A subset of the data presented above has been taken with a linearly
polarized photon-beam and an unpolarized target. It was analyzed
with respect to polarization observables.  Prior to our earlier
analysis \cite{Gutz:2009zh}, two-meson production
has been treated in a quasi two-body approach
\cite{Ajaka:2008zz,Gutz:2008}, resulting in the extraction of the
beam asymmetry $\Sigma$, known from single-meson photoproduction.
The beam asymmetries $\Sigma$ accessible in $\pi^{0}\eta$
photoproduction are presented in the next paragraphs. However, a
three-body final state like $p\pi^{0}\eta$ yields additional degrees
of freedom, reflected
\begin{figure*}[pt!]
\centering
\includegraphics[width=0.93\textwidth]{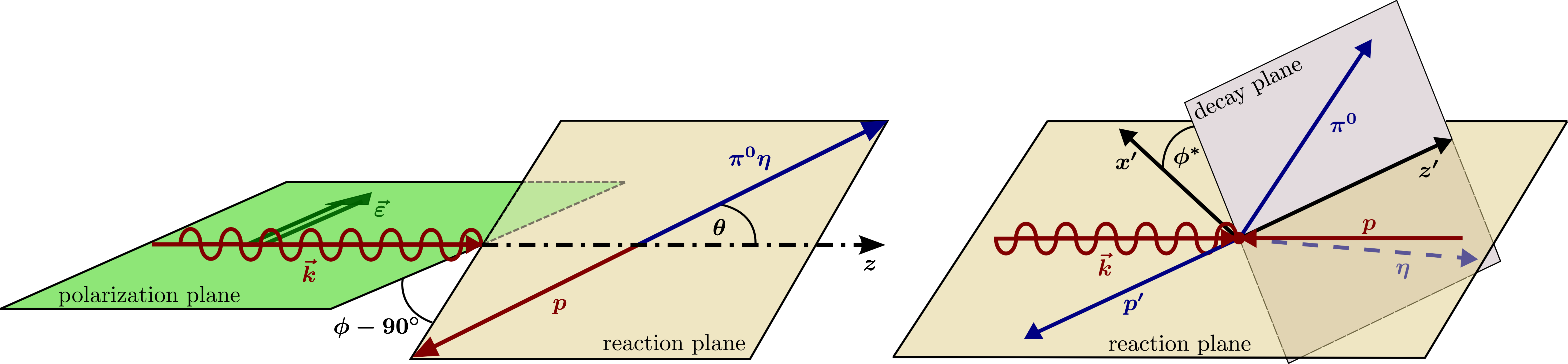}
\caption{Angle definitions for the extraction of beam asymmetries.
Left: Quasi two-body approach. Right: Additional degree of freedom
occurring in full three-body kinematics.}
\label{fig:angles}
\end{figure*}
in a different set of polarization observables
\cite{Roberts:2004mn}. These will be discussed in detail in the last
part of this section.

\begin{figure}[pb]
\centering
\includegraphics[width=0.49\textwidth]{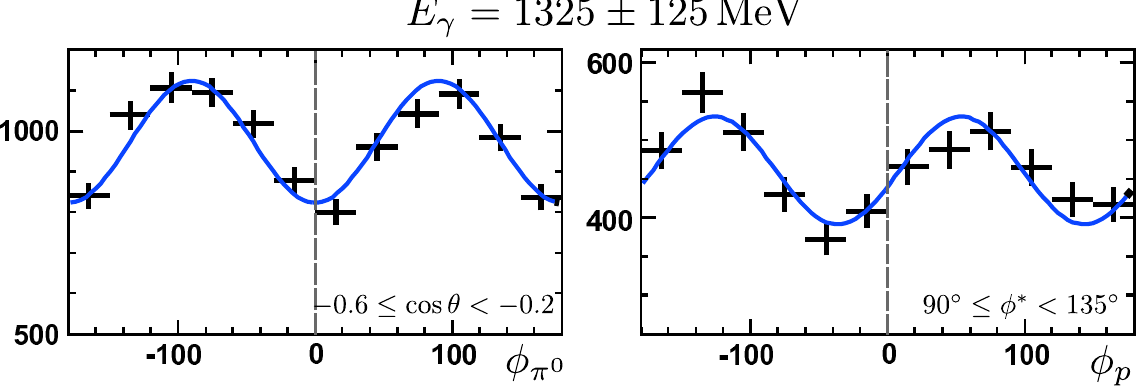}
\caption{Left: $\phi$-distribution of the final state $\pi^{0}$ in a
quasi two-body approach, binned in $\cos\theta_{\pi}$. Only the
$\cos2\phi$-modulation according to eq.~(\ref{eq:2body}) is visible.
Right: $\phi$-distribution of the final state $p$ in a full
three-body approach, binned in $\phi^{*}$. An additional
$\sin2\phi$-modulation, according to eq.~(\ref{eq:3body}) is
apparent.}
\label{fig:phidist}
\end{figure}

\paragraph{Quasi two-body approach}
In a first approach, one can apply the well-known techniques from
single-meson production to extract polarization observables for
two-meson final states. For the $p\pi^{0}\eta$ final state this
translates to the three quasi two-body reactions:
\begin{eqnarray*}
 \gamma p & \rightarrow & pX,\;\;\mathrm{with~} X\rightarrow \pi^{0}\eta,\\
 \gamma p & \rightarrow & \eta Y,\;\;\mathrm{with~} Y\rightarrow p\pi^{0},\\
 \gamma p & \rightarrow & \pi^{0} Z,\;\;\mathrm{with~} Z\rightarrow p\eta.
\end{eqnarray*}
The cross-section for such two-body final states can be written as
\cite{Worden:1972}
\begin{equation}
  \label{eq:2body}
\frac{\mathrm{d}\sigma}{\mathrm{d}\Omega} =
\left(\frac{\mathrm{d}\sigma}{\mathrm{d}\Omega}\right)_{0}\left(1+\delta_{l}\Sigma\cos2\phi\right),
\end{equation}
where $\left(\frac{\mathrm{d}\sigma}{\mathrm{d}\Omega}\right)_{0}$ is the
unpolarized cross-section and $\delta_{l}$ the degree of linear
polarization (see Fig.~\ref{fig:pol}). In this work, the angle $\phi$
is defined as depicted in Fig.~\ref{fig:angles} (left). The beam
asymmetry $\Sigma$ can then be extracted from the amplitude of the
$\cos2\phi$ modulation of the $\phi$-distributions of the individual
final state particles. An example is shown in
Fig.~\ref{fig:phidist}, left. Here, the $\phi$-distribution of the
final state $\pi^{0}$ has been fitted, according to
eq.~(\ref{eq:2body}), with the expression
\begin{equation}
f(\phi) = A + P\cdot B\cdot \cos 2\phi,
\end{equation}
where the degree of linear polarization, $P$, has been determined
event by event and was later averaged for each fitted bin. In this
ansatz, the ratio of the parameter $B/A$ translates to the beam
asymmetry $\Sigma$.

\begin{figure*}[!ht]
\begin{tabular}{cc}
\includegraphics[width=0.49\textwidth]{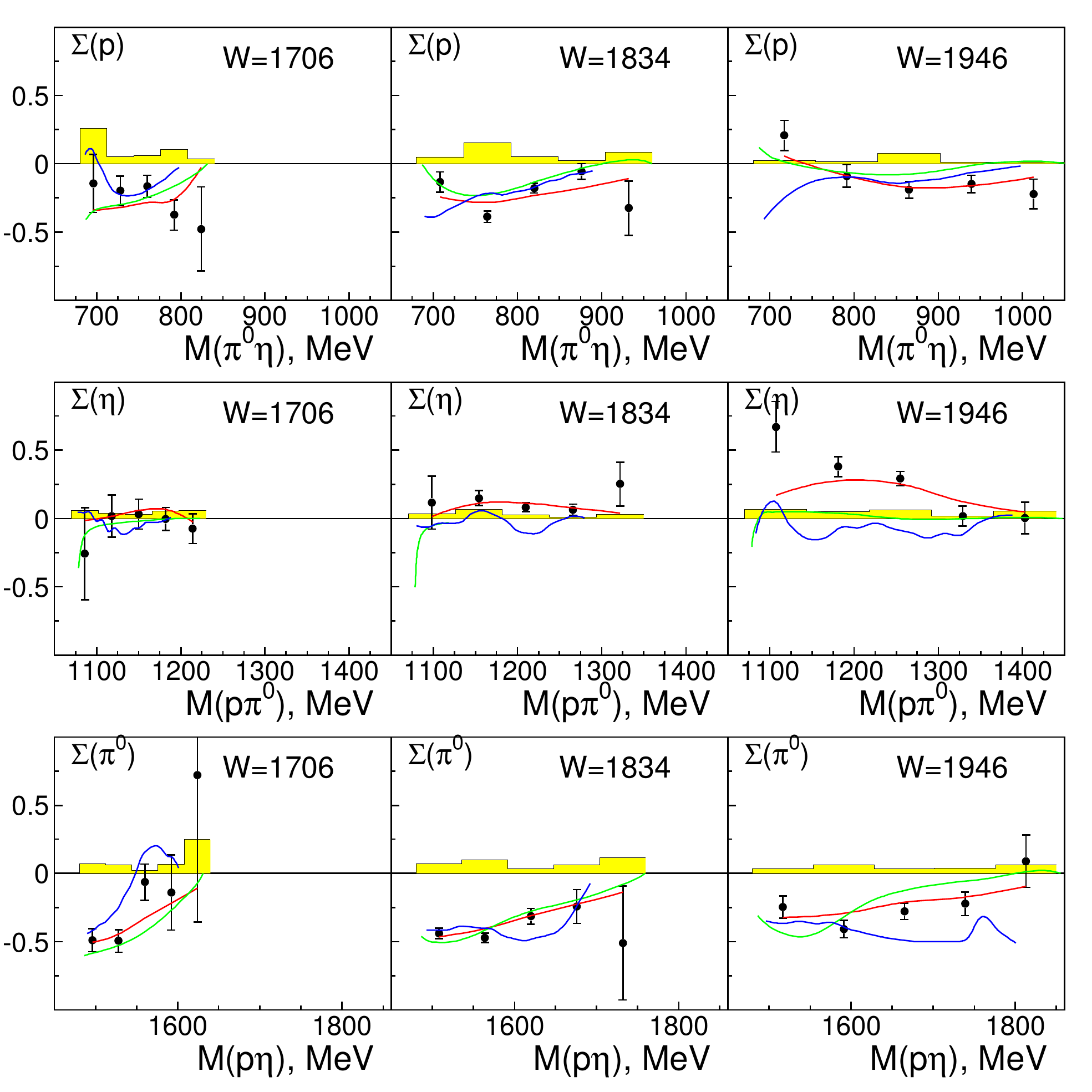}&
\hspace{-3mm}\includegraphics[width=0.49\textwidth]{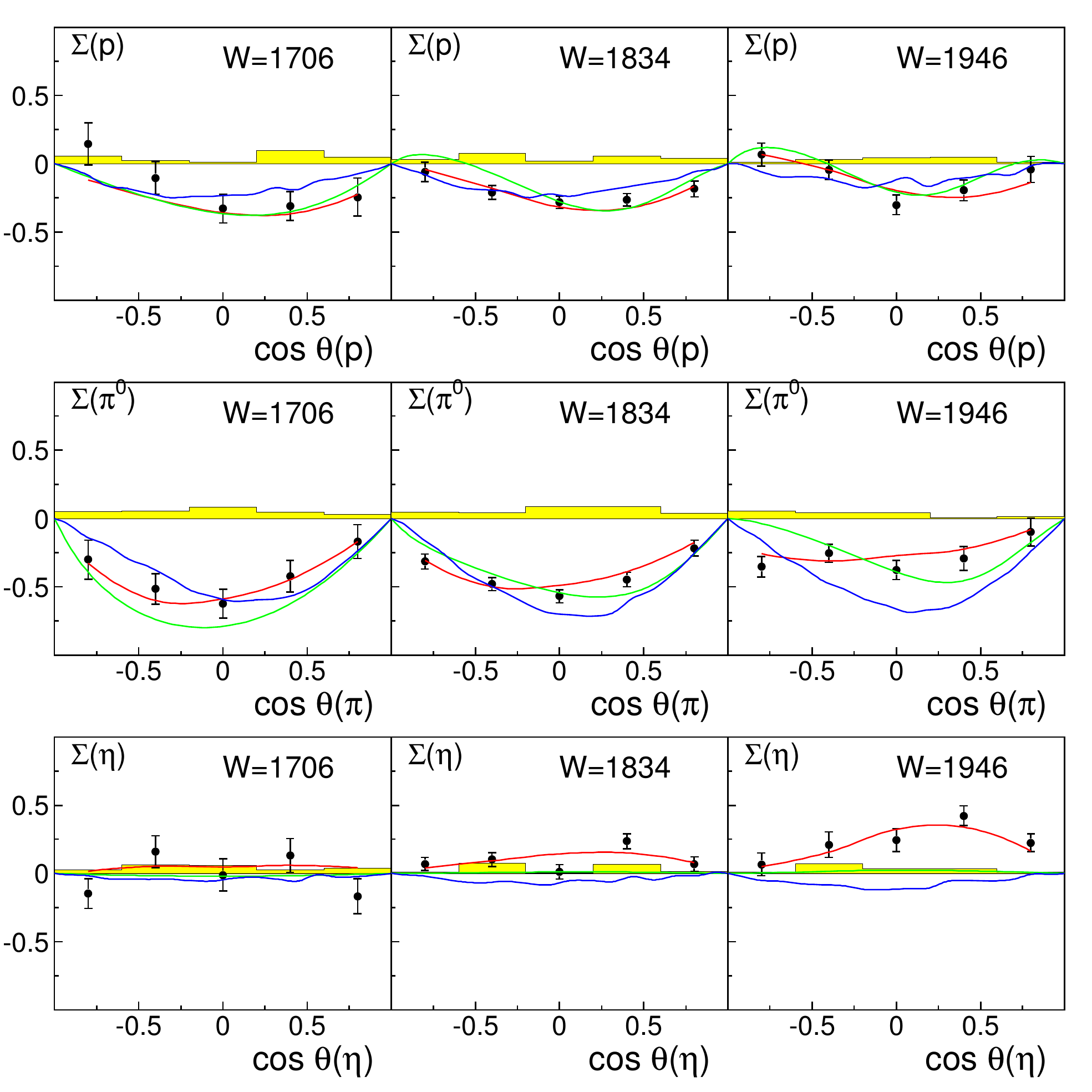}
\end{tabular}
\caption{Two-body beam asymmetry $\Sigma$ for the reaction $\gamma p
\rightarrow p \pi^{0}\eta$. Top to bottom: incoming photon energy
ranges $1085\pm115$\,MeV, $1325\pm125$\,MeV, $1550\pm100$\,MeV.
Left: Asymmetries obtained from the $\phi$-distributions of the
recoiling (left to right) $p$, $\eta$, $\pi^{0}$ as function of the
invariant mass of the other two particles  \cite{Gutz:2008}. Right:
The same as function of the $\cos\theta$ of the recoiling particle.
Systematic error estimate from acceptance studies (yellow). Curves:
BnGa-PWA (red), Fix et al. \cite{A1_Fix:2010} (green), D\"oring et
al. \cite{Doring:2010fw} (blue).\vspace{2mm}}
\label{fig:sigma}
\begin{tabular}{cc}
\includegraphics[width=0.49\textwidth]{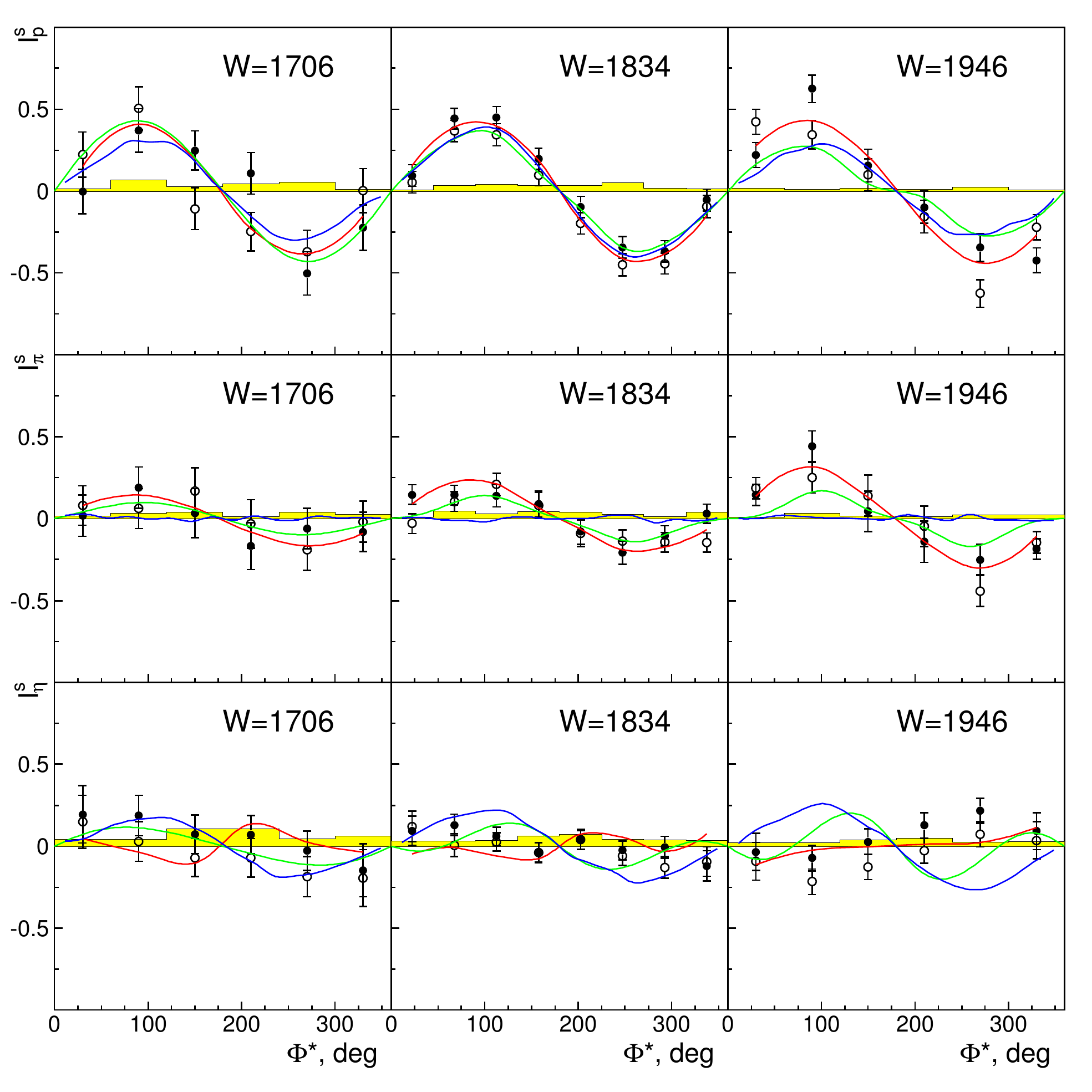} &
\hspace{-3mm}\includegraphics[width=0.49\textwidth]{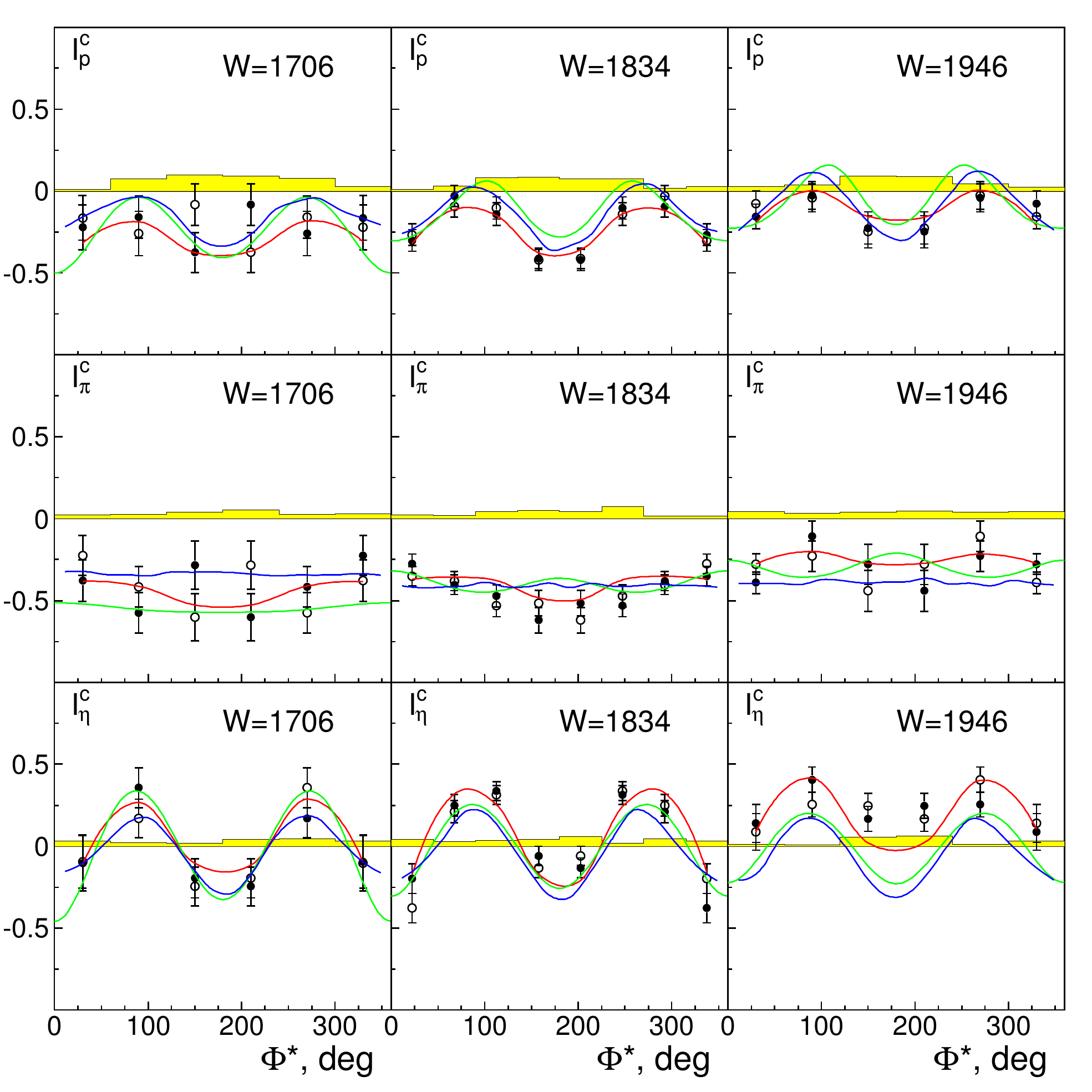}
\end{tabular}
\caption{Three-body beam asymmetries $I^{s}$ (left) and $I^{c}$
(right) \cite{Gutz:2009zh}. Closed symbols: $I^{s}(\phi^{*})$
($I^{c}(\phi^{*})$) as extracted from the data, open symbols:
($-I^{s}(2\pi-\phi^{*})$, $I^{c}(2\pi - \phi^{*})$ see
eqs.~(\ref{eq:ic_symm},~\ref{eq:is_symm}) Grey bars: Systematic
error estimate from acceptance studies. Curves: BnGa-PWA (red), Fix
et al. \cite{A1_Fix:2010} (green), D\"oring et al.
\cite{Doring:2010fw} (blue).}
\label{fig:isic}
\end{figure*}

The values for the observable are shown in Fig.~\ref{fig:sigma}
for the three energy ranges under consideration and extracted from
the $\phi$-distributions of all three final-state particles, as
function of the invariant masses of the respective other two
particles (left) and of the $\cos\theta$ of the recoiling particle
itself (right).

Using bin sizes in the 5-dimensional phase space sufficiently small
compared to variations of the acceptance, the asymmetries should not
be influenced by these factors since they cancel out. However, due
to geometrical limitations present in every experimental setup, the
phase space is not covered completely, leading to areas of vanishing
acceptance. Two different methods have been applied to the data to
estimate the influence of acceptance variations and other systematic
effects on the results. In a first step, the two-dimensional
acceptance has been determined from Monte Carlo simulations for the
three energy ranges separately as function of $\phi$ of the
recoiling particle and the variable intended for binning the
polarization observable. The data have been corrected for this
acceptance, determining the resulting change in the observables for
each bin. In a second step, the Bonn-Gatchina partial wave analysis
(BnGa-PWA, described below) has been used to determine the
five-dimensional acceptance for the reaction, accounting also for
the contributing physics amplitudes. Again, the differences in terms
of the beam asymmetry in the relevant binning have been calculated
using this correction. The larger absolute value derived from these
methods is given in Fig.~\ref{fig:sigma} as grey bars, corresponding
to an estimate of the systematic error.

\paragraph{Full three-body approach}
The cross-section for the production of pseudoscalar meson pairs can
be written in the form
\begin{equation}
  \label{eq:3body}
\frac{\mathrm{d}\sigma}{\mathrm{d}\Omega} =
\left(\frac{\mathrm{d}\sigma}{\mathrm{d}\Omega}\right)_{0}\left(1+\delta_{l}\left(I^{c}(\phi^{*})\cos2\phi+I^{s}(\phi^{*})\sin2\phi\right)\right).
\end{equation}
The two polarization observables $I^{s}$ and $I^{c}$ emerge as the
amplitudes of the respective modulations of the azimuthal
distributions of the final state particles (Fig.~\ref{fig:phidist},
right), once the acoplanar kinematics of the reaction are taken into
account (Fig.~\ref{fig:angles}, right). The angular dependence of
the observables
\begin{eqnarray}
  \label{eq:is_symm}
\hspace{-2mm}I^s(\phi^{*}) =
\sum\limits_{n=0}^{}a_{n}\sin(n\phi^{*}) & \rightarrow &
I^{s}(\phi^{*}) = - I^{s}(2\pi-\phi^{*})\hspace{3mm}
\end{eqnarray}
and
\begin{eqnarray}  \label{eq:ic_symm}
I^c(\phi^{*}) = \sum\limits_{n=0}^{}b_{n}\cos(n\phi^{*}) &
\rightarrow & I^{c}(\phi^{*}) = I^{c}(2\pi-\phi^{*})\hspace{3mm}
\end{eqnarray}
allows for a direct cross-check of the data with respect to
systematic effects \cite{Gutz:2009zh}. Additionally it is to be
noted that the two-body beam asymmetry $\Sigma$ occurs here as the
constant term in the expansion of $I^{c}$.

Fig. \ref{fig:isic} shows the obtained beam asymmetries as
function of the angle $\phi^{*}$ (solid symbols) for the three
photon-energy ranges 970-1200\,MeV, 1200-1450\,MeV and
1450-1650\,MeV, along with the data after application of
transformations (\ref{eq:is_symm}), (\ref{eq:ic_symm}) (open
symbols). In the absence of systematic effects, both sets of data
should coincide. Taking into account the statistical uncertainties,
this is well fulfilled, proving the quality of the data.

The comparison of the theoretical approaches used for the
interpretation of the data on the beam asymmetries shows the
complexity of the reaction $\gamma p\rightarrow p\pi^{0}\eta$, once
the incoming photon energy sufficiently exceeds the threshold energy
for the process. A good description at low energies is already
achieved by including just one resonant process, the excitation of
the $\Delta(1700)3/2^-$, dominating the threshold region, as first
introduced by the Valencia group in \cite{Doring:2005bx}. The
results are also consistent with the assumption that this resonance
is in fact dynamically generated by meson-baryon interactions
\cite{Doring:2010fw}. The continued dominance of the partial wave
with $(I)J^P=(3/2)3/2^-$, at higher energies populated by the
$\Delta(1940)3/2^-$, first claimed by the BnGa-PWA group in
\cite{Horn:2008qv}, is confirmed by the Fix-model
\cite{A1_Fix:2010}. Figs.~\ref{fig:sigma} and \ref{fig:isic} compare
our data with the fits from \cite{Doring:2010fw,A1_Fix:2010} and the
BnGa2013 fit.

\section{\label{Partial wave analysis}Partial wave analysis}
\subsection{Aims}
The partial wave analysis serves two purposes: it is required to
determine the acceptance and to extract the physical content of the
data. The use of the partial wave analysis for the acceptance
correction can best be understood in the case of a simple spectrum
like the angular distribution in $\cos\theta$. Often, the full
angular range is not covered experimentally, hence the distribution
has to be extrapolated into regions with no data. This can be done
using a polynomial extrapolation but, of course, an extrapolation
with a partial wave analysis is better. In the case of a three-body
final state, there are five independent variables, the photon energy
and, e.g., two invariant masses used to construct a Dalitz plot, and
two Euler angles describing the orientation of the Dalitz plot
relative to the direction of the incoming beam. Hence an
extrapolation into the four-dimensional phase space (for each bin in
$E_{\gamma}$) is required; this is not possible using a polynomial
expansion.

The second (and main) purpose of the partial wave analysis is to
determine properties of baryon resonances. In the presentation of
the data, evidence was seen for $\Delta(1232)\pi$,
$N(1535){1/2^-}\pi$, and $pa_0(980)$ as intermediate isobars. There
are immediate questions: how are they produced? Via $u$- or
$t$-channel exchange processes, as direct three-body production, or
via resonance formation and a subsequent cascade decay? If the
latter is the case, which primary resonances contribute? Are there
new resonances, so far {\it missing resonances} involved? Can
further intermediate isobars be identified?

Dynamical coupled-channels models based on effective chiral
Lagrangians provide a microscopical description of the background
\cite{Doring:2009bi,Doring:2009yv}. In some cases, resonances can
even be constructed from the iteration of background terms
\cite{Kaiser:1995cy,Meissner:1999vr,Nieves:2001wt}. We follow a more
phenomenological approach based on a K-matrix. The formalism is
described in detail in \cite{Anisovich:2004zz,Anisovich:2006bc}.
Here we give a short outline, mainly to introduce the definitions of
the quantities presented in the tables below.

\subsection{The formalism}
The transition amplitude for pion- and photo-produced reactions from
an initial state, e.g. $a=\pi N$ or $\gamma N$, to a final state,
e.g. $b=\Lambda K^+$, can be written in the form of a K-matrix as
\be
\label{kmatrix}
A_{ab}&=&K_{ac}(I-i\rho K)_{cb}^{-1}\,,
\ee
where $\rho$ is a diagonal matrix of effective phase volumes. For a
two body final state the effective phase volume in a partial wave
with a relative orbital angular momentum $L$ rises at threshold with
$q^{2L\!+\!1}$ where $q$ is the decay momentum. To describe the
high-energy behavior of the cross-section, form factors are used as
suggested by Blatt and Weisskopf \cite{Blatt:1952}:
\be
\rho_b(s)=\frac{2q}{\sqrt s}\frac{q^{2L}}{F(r^2,q^2,L)}.
\ee
Here $r$ is an effective interaction radius and the explicit form of
the Blatt-Weisskopf form factor can be found in Appendix C of
\cite{Anisovich:2004zz}. For three body final states we use the
dispersion integral (the explicit form is given in
\cite{Anisovich:2006bc}) which takes into account all corresponding
threshold singularities on the real axis and in the complex energy
plane.

The amplitude contains resonances in the form of K-matrix poles
(characterized by the pole position $M_\alpha$ and couplings
$g^\alpha_a$) and non-resonant terms for direct transitions $f_{ab}$
between different channels:
\be
K_{ab}=\sum\limits_\alpha \frac{g^{\alpha}_a
g^{\alpha}_b}{M_\alpha^2-s}+f_{ab}\,.
\ee
The background terms $f_{ab}$ can be arbitrary functions of $s$. In
general we introduce an $s$-dependence in the form
\be
f_{ab}=\frac{(a+b\sqrt{s})}{(s-s_0)}\,.
\label{backg}
\ee
In most cases, the form (\ref{backg}) did not lead to a noticeable
improvement of the fit but sometimes to a poor convergence. For the
majority of transitions (for all partial waves with orbital momentum
$L>1$ and for all transitions into three body final states) a
constant background term was sufficient to describe the data. In the
fits, amplitudes for reggeized exchanges in the $t$- and $u$-channel
were added.

In practice we neglect in our analysis contributions from $\gamma N$
loop diagrams in rescattering. Thus we do not take into account the
$\gamma N$ channel in the K-matrix equation (\ref{kmatrix}) and
describe photoproduction of mesons within the P-vector formalism
\cite{Chung:1995dx}:
\be
\label{p-vector}
A_{hb}&=&P_{hc}(I-i\rho K)_{cb}^{-1},\qquad P_{hc}=K_{hc},
\ee
where the $h$ denotes the helicity of the initial state and where
$b,c$ list the hadronic final states.

\subsection{Particle properties at the pole position}

We define the pole position by a zero of the amplitude denominator in
the complex plane
\be
\prod\limits_\alpha (M_\alpha^2-s)\,det(I-i\rho K)=0\,.
\ee
In the case of a one-pole K-matrix without non-resonant terms (which
corresponds to a relativistic Breit-Wigner amplitude) this equation
has a simple form:
\be
M^2-s-i\sum\limits_j g_j^2\rho_j(s)=0\,.
\ee
The residues for the transition amplitude from $a$ to $b$ can be
calculated by a contour integral of the amplitude around the pole
position in the energy $(\sqrt s)$ plane
\begin{eqnarray}
Res(a\to b)&=&\int\limits_{o}\frac{d\sqrt s}{2\pi
i}\sqrt{\rho_a}A_{ab}(s)\sqrt{\rho_b}\nonumber\label{transamp}\vspace{-3mm}\\
&=&\frac{1}{2M_p}\sqrt{\rho_a(M_p^2)}g^r_a\,g^r_b\sqrt{\rho_b(M_p^2)}\,.
\end{eqnarray}
Here $M_p$ is the position of the pole (complex number) and $g^r_a$
are pole couplings.
For elastic scattering of channel $a$ to $a$, e.g. for $\pi N\to \pi
N$, this gives the elastic residue
 \be
Res(\pi N\to N\pi)=\frac{1}{2M_p}
(g^r_{N\pi})^2\,\rho_{N\pi}(M_p^2)\,
\label{residuum}
 \ee

At the pole position the amplitude factorizes:
\be
Res^2(a\to b)=Res(a\to a)\times Res(b\to b)\,.
\ee
This relation can be used to define branching ratios at the pole
position as
\begin{eqnarray}
BR_{\rm pole}({\rm channel} \ b) = \frac{|Res (\pi N\to
b)|^2}{|Res(\pi N\to N\pi)|\cdot \left(\Gamma_{\rm pole}/2\right) }.
\end{eqnarray}

The helicity amplitudes for photoproduction of the final state $b$,
$\tilde A_{1/2}$ and $\tilde A_{3/2}$, calculated at the pole
position are given by
\begin{eqnarray}
\tilde A_{h}=\sqrt{\frac{\pi(2J+1)m_{\rm pole}}{k^2_{\rm pole}\,m_N
}}\, \frac{Res(h\to b)}{\sqrt{Res(b\to b)}}
\end{eqnarray}
where the residues ($Res$) are evaluated at the pole mass~($m_{\rm
pole}$), and $k_{\rm pole}^2=(m^2_{\rm pole}-m_N^2)^2/4m^2_{\rm
pole}$\cite{Workman:2013rca}. The transition amplitude from the
fixed helicity is correspondingly defined as
\be
(\gamma N)^h\!\to\!b=\sqrt{\frac{\pi(2J+1)m_{\rm pole}}{k^2_{\rm
pole}\,m_N }}\, Res(h\to b)\,.
\ee

In the Tables below, we give multipole transition amplitudes
$E_{L\pm}$ and $M_{L\pm}$ for production and decay of resonances.
The multipoles can be expressed in terms of the helicity amplitudes.
For states with $J=L\!+\!1/2$:
\be
\tilde E_{L+}&=&\frac{-1}{L\!+\!1}\left (\tilde A_{1/2}-\tilde
A_{3/2}\sqrt{\frac{L}{L\!+\!2}}\right )\qquad L\ge0,
\nonumber \\
\tilde M_{L+}&=&\frac{-1}{L\!+\!1}\left (\tilde A_{1/2}+\tilde
A_{3/2}\sqrt{\frac{L\!+\!2}{L}}\right )\qquad L\ge1\,,
\ee
where $J$ is the total momentum, $L$ is the orbital momentum in the
$\pi N$ channel. For states with $J=L\!-\!1/2$:
\be
\tilde E_{L-}&=&\frac{-1}{L}\left (\tilde A_{1/2}+\tilde
A_{3/2}\sqrt{\frac{L\!+\!1}{L\!-\!1}}\right )\qquad L\ge2,
\nonumber \\
\tilde M_{L-}&=&\frac{1}{L}\left (\tilde A_{1/2}-\tilde
A_{3/2}\sqrt{\frac{L\!-\!1}{L\!+\!1}}\right )\qquad L\ge1\,,
\ee
In Table~\ref{nucleon} below, the helicity and electric and magnetic
transition amplitudes at the pole position are abbreviated as $(\gamma p)^{h}\to b$,
($\gamma p\to b$; \ $E_{L\vec{\pm}}$), and ($\gamma p\to b$; \
$M_{L\vec{\pm}}$), respectively.

\subsection{Particle properties from Breit-Wigner representations}

A K-matrix fit returns an amplitude with poles in the complex plane,
Breit-Wigner parameters do not result from those fits. Here, the
Breit-Wigner parameters are calculated from the pole position of the
resonance. The procedure is described in \cite{Anisovich:2011fc}. We
shortly review it for convenience of the reader. We define a
Breit-Wigner amplitude as
\be
A_{ab}=\frac{f^2g^r_a g^r_b}{M_{BW}^2-s-if^2\sum\limits_a
|g^r_a|^2\rho_a(s)}
\label{bw}
\ee
where $g^r_a$ are coupling residues at the pole, $M_{BW}$ is the
Breit-Wigner mass and $f$ is a scaling factor. The Breit-Wigner mass
and scaling factor are adjusted to reproduce the pole position of
the resonance. In the case of very fast growing phase volumes, the
Breit-Wigner mass and width can shift from the pole position by a
large amount. In the 1600-1700\,MeV region, the large phase volume
leads to a very large Breit-Wigner width and an appreciable shift in
mass from the pole position (see for example \cite{Thoma:2007bm}).
The visible width, e.g. in the $N\pi$ invariant mass spectrum,
remains similar to the Breit-Wigner width. The large phase volume
effects are highly model dependent. We therefore extract the
Breit-Wigner parameters of resonances above the Roper resonance by
approximating the phase volumes for the three body channels in
eq.~(\ref{bw}) as $\pi N$ phase volume for the respective partial
wave. This procedure conserves the branching ratio between
three-particle final states and the $\pi N$ channel at the resonance
position.

The Breit-Wigner helicity amplitude is defined as
\be
(\gamma N)^h\!\to\!b=\frac{A^h_{BW}
fg^r_b}{M_{BW}^2-s-if^2\sum\limits_a |g^r_a|^2\rho_a(s)}\,,
\label{helibw}
\ee
where $A^h_{BW}$ is calculated to reproduce the pion
photo-production residues in the pole. In general, this quantity is a
complex number. However, for the majority of resonances its phase
deviates not significantly from 0 or 180 degrees and the sign can be
defined as sign of the real part. $A_{1/2}$ and $A_{3/2}$ are
normalized to satisfy
\be
\Gamma_{\gamma} = { {k_{\rm BW}^2}\over {\pi } } { {2 m_N}\over
{(2J+1)
  m_{\rm BW}} } \Bigl( | A_{1/2} |^2 + | A_{3/2} |^2 \Bigr) ~
\ee
when the pole position is replaced by the Breit-Wigner mass.

For Breit-Wigner amplitudes, $m_{\rm pole}=m_{\rm BW}$ and $\tilde
A_{h} = A_h$.

\subsection{\boldmath $t$- and $u$-channel exchange amplitudes}
The amplitudes for $t$-channel meson exchange in photoproduction are
described by reggeized trajectories:
\be
A=g_1(t)g_2(t)\frac{1+\xi \exp(-i\pi\alpha(t))}{\sin(\pi\alpha(t))}
\left (\frac{\nu}{\nu_0} \right )^{\alpha(t)} \;,
\ee
where $\nu=\frac 12 (s-u)$, ~$\alpha(t)$ is the function which
describes the trajectory, $\nu_0$ is a normalization factor  and
$\xi$ is the signature of the trajectory. In the case of $\pi^0\eta$
photoproduction, both $\rho$ and $\omega$ exchanges (which have
positive signature $\xi=+1$) can contribute to the fit. We do not
introduce these exchanges separately using an ``effective'' $\rho$
meson exchange. The structure of the upper and lower vertices is
given in detail in \cite{Anisovich:2004zz}. The upper and lower
vertices are parameterized with a form factor proportional to
$\exp(-\beta t)$, where $\beta$ is a fit parameter.

\begin{table}[pt]
\caption{Change in $\chi^2$ of the multichannel fit when the
coupling of a resonance to different decay modes is set to zero and
the data refitted. If contributions are found to be very small, they
are set to zero (to improve the fit stability). In this case, a - is
given. Mesonic $t$-channel exchange is represented by reggeized
$\rho/\omega$-exchange, $u$-channel exchange by proton and $\Delta$
exchange. }
\label{sign}\renewcommand{\arraystretch}{1.3}
\begin{tabular}{lccc}
\hline\hline\\[-2.8ex]
                  &$\Delta(1232)\eta$ &$N(1535){1/2^-}\pi$&$pa_0(980)$\\
\\[-2.8ex]\hline\\[-2ex]
 $N(1710){1/2^+}$ &           -       &      656           &      -    \\
 $N(1880){1/2^+}$ &           -       &       23           &    166    \\
 $N(1900){3/2^+}$ &           -       &      883           &    108    \\
 $N(2100){1/2^+}$ &           -       &      323           &    156    \\
 $N(2120){3/2^-}$ &           -       &      169           &      -    \\
 \\[-2.8ex]\hline\\[-2ex]
 $\Delta(1700){3/2^-}$  &  1333       &      263           &      -    \\
 $\Delta(1900){1/2^-}$  &   198       &        -           &      -    \\
 $\Delta(1905){5/2^+}$  &   328       &      337           &      -    \\
 $\Delta(1910){1/2^+}$  &  1195       &        -           &      -    \\
 $\Delta(1920){3/2^+}$  &   273       &        -           &    204    \\
 $\Delta(1940){3/2^-}$  &  1545       &      162           &      -    \\
 $\Delta(1950){7/2^+}$  &   476       &        -           &      -    \\
 \\[-2.8ex]\hline\\[-2ex]
 $\rho/\omega$ exchange         &   849       &        -           &    696    \\
 \boldmath$p$ exchange         &   299       &        -           &    189
 \\ \\[-2.8ex]
\hline\hline
\end{tabular}
\end{table}

The contributions from $u$-channel (proton) exchange are described
by a proton exchange amplitude which contains a proton propagator in
the $u$-channel and form factors at the vertices. The $t$- and
$u$-channel exchanges are introduced for the $\Delta(1232)\eta$ and
$a_0(980)p$ final states. For production of $N(1535)1/2^+\pi$ we
found very small contributions from $t$- and $u$-exchanges; hence
they were neglected in the final fit.

\subsection{Method}
The new data on $\gamma p\to p\pi^0\eta$, and the accompanying data
on $\gamma p\to p2\pi^0$, have been included in the large data set
used in the Bonn-Gatchina (BnGa) multichannel partial wave analysis.
The fitted data set contains results on $\pi N$ elastic and
inelastic ($\pi N\to N\eta, \Lambda K, \Sigma K$) reactions and the
very precise data from photoproduction off protons, including
information from polarization and double-polarization experiments. A
list of data, including references, is given elsewhere
\cite{Anisovich:2010an,Anisovich:2011ye}; the list is updated in
\cite{Anisovich:2011fc} and \cite{Anisovich:2013vpa}. The
partial wave analysis method used in this analysis is described in
detail in \cite{Anisovich:2004zz,Anisovich:2006bc}. Further
information is given in \cite{Anisovich:2010an}.

The fit minimizes the total log likelihood defined by
\begin{eqnarray}
-\ln {\cal L}_{\rm tot}= ( \frac 12\sum w_i\chi^2_i-\sum w_i\ln{\cal
L}_i ) \ \frac{\sum N_i}{\sum w_i N_i}
\label{likelihood}
\end{eqnarray}
where the summation over binned data contributes to the $\chi^2$
while unbinned data contribute to the likelihoods ${\cal L}_i$. The
data presented in this paper, and other data with three particles in
the final state, are fitted using an event-based maximum likelihood
method which takes all correlations between mass and angular
distributions into account. Differences in fit quality are given as
$\chi^2$ difference, $\Delta\chi^2=-2\Delta{\cal L}_{\rm tot}$. For
new data, the weight is increased from $w_i=1$ until a visually
acceptable fit is reached. Without weights, low-statistics data e.g.
on polarization variables may be reproduced unsatisfactorily without
significant deterioration of the total ${\cal L}_{\rm tot}$. The
total $\chi^2$ is normalized to avoid an artificial increase in
statistics. The $\chi^2$ values per data point are typically between
1 and 2 even though some data give a larger value.

\subsection{Significance}
The partial wave analysis requires a large number of resonances in
the intermediate state. In Table~\ref{sign}, a list of resonances is
given which are presently used in the analysis. Also given is the
statistical significance (as change in $\chi^2$) when a particular
decay mode is removed from the fit and the data refitted.

The strongest contributions come from the $\Delta(1232)\eta$ isobar
with $\Delta(1232)$ and $\eta$ in a relative $S$-wave, that is from
the $(I,J^P)=(3/2,3/2^-)$-wave. Strong couplings to
$\Delta(1232)\eta$ are known for the $\Delta(1700)3/2^-$
\cite{Horn:2008qv,Horn:2007pp,A1_Fix:2010}, and also
$\Delta(1940)3/2^-$ is known to make a significant contribution to
reaction (\ref{ppi0eta}) \cite{Horn:2008qv,Horn:2007pp}. A further
resonance with these quantum numbers seems to be required above
2\,GeV; it improves the fit but no clear minimum is found in a mass
scan. Tentatively, we call it $\Delta(2200)3/2^-$.

We believe decay modes leading to a $\chi^2$ change of more than
1000 to be rather certain (***), the evidence for decay modes with a
$\delta\chi^2\ge 500$ is estimated to be fair (**), and those with
$\delta\chi^2\ge 200$ to be poor (*).

It has be mentioned that a high significance is not necessarily
connected with a large branching ratio for the decay of a particular
resonance. The significance can be poor even in the case of a
resonance with a sizable branching ratio into a final state, e.g.
$N(1535)1/2^-\pi$, when the resulting pattern in the final state can
be described easily by the sum of two other resonances decaying into
the same final state.

At its nominal mass, $\Delta(1600){3/2^+}$ cannot contribute to the
three decay modes listed in Table~\ref{sign}. $\Delta(1600){3/2^+}$
is, however,  a wide resonance and it may contribute to these decays
via its tail. Mathematically, this is treated by an continuation of
the decay momentum (which enters, e.g., in the energy-dependent
width of the denominator of a Breit-Wigner amplitude)
\begin{equation}
q = \{(M^2 - (m_1 + m_2)^2) (M^2 - (m_1 - m_2)^2)\} ^{1/2} / 2M
\end{equation}
into the range of imaginary values. Due to this analytic
continuation of the amplitude, $\Delta(1600){3/2^+}$ contributes
significantly to the final states in Table~\ref{sign}. The branching
ratios are, however, defined by the coupling constants at the pole
position, and hence they vanish.

\subsection{Results}
Before presenting the full results of this analysis, we discuss a
possible interpretation of the branching ratios. It is the first
time that cascade processes of high-mass resonances into a resonance
with intrinsic orbital angular momentum (here $N(1535)1/2^-$) are
studied. The comparison of these decay modes with decays into $N\pi$
should help to identify mechanisms responsible for the decays of $N$
and $\Delta$ resonances.

\begin{table}[pt]
\caption{Branching ratios of nucleon and $\Delta$ resonances }
\label{brND}\renewcommand{\arraystretch}{1.3}
\bc
\begin{tabular}{cccc}
\hline\hline
Resonance & $\pi N$ & $N(1535)\pi$ & $\Delta(1232)\eta$ \\
\hline
 $N(1710){1/2^+}$ &      5\er 3\%          &     15\er 6\%              &    -    \\
 $N(1880){1/2^+}$ &      6\er 3\%          &      8\er 4\%              &    -    \\
 $N(1900){3/2^+}$ &      3\er 3\%          &      7\er 3\%              &    -    \\
 $N(2100){1/2^+}$ &      3\er 2\%         &     22\er 8\%              &    -    \\
 $N(2120){3/2^-}$ &      5\er 3\%          &     15\er 8\%              &    -    \\
 \\[-2.8ex]\hline\\[-2ex]
 $\Delta(1700){3/2^-}$  &   22\er 4\%      &     1\er 0.5\%             &     5\er 2\% \\
 $\Delta(1900){1/2^-}$  &   7\er 2\%       &       -                    &     1\er 1\% \\
 $\Delta(1905){5/2^+}$  &  13\er 2\%       &     $\leq 1$\%             &     4\er 2\% \\
 $\Delta(1910){1/2^+}$  &  12\er 3\%       &     5\er 3\%               &     9\er 4\% \\
 $\Delta(1920){3/2^+}$  &   8\er 4\%       &     $\leq 2$\%             &     11\er 6\%\\
 $\Delta(1940){3/2^-}$  &   2\er 1\%       &     8\er 6\%               &     10\er 6\%\\
 $\Delta(1950){7/2^+}$  &   46\er 2\%      &                            &   $\leq  1$\%\\
\hline\hline
\end{tabular}
\ec\end{table}

\begin{table*}[pt]
\caption{\label{nucleon} Nucleon and Delta resonances and their
properties from the BnGa multichannel partial wave analysis of $\pi
N$ elastic scattering data and from pion and photo-induced inelastic
reactions.  Along with the name of the resonance, the star rating of
the Particle Data Group \cite{Beringer:1900zz} is given. Helicity
couplings, abbreviated as $(\gamma p)^{h}$, are given in
GeV$^{-\frac 12}$. $\pi N\to \pi N$ stands for the elastic pole
residue, $2\,(\pi N\to X)/\Gamma$ for inelastic pole residues. They
are normalized by a factor $2/\Gamma$ with $\Gamma=\Gamma_{\rm
pole}$.  For $N(1880)1/2^+$ we find two solutions with distinct
photocouplings. Both are given below. \vspace{3mm}}
\renewcommand{\arraystretch}{1.2}\begin{scriptsize}
\begin{tabular}{cc}
\begin{tabular}{lrlr}
\multicolumn{4}{l}{\boldmath \fbox{\fbox{$N(1710){1/2^+}$}}\unboldmath\hfill***}\\[-0.7ex]
&&&\\\hline\hline\\[-1.5ex]
\multicolumn{4}{l}{$N(1710){1/2^+}$ pole parameters}\\[0.5ex]
$M_{pole}$ &  1690\er15 & $\Gamma_{pole}$& 170\er 20\\
$A^{1/2}$ & 0.052\er0.014& Phase: & (-10\er50)\oo\\
\multicolumn{3}{l}{$N(1710){1/2^+}$ transition residues }&\multicolumn{1}{c}{\hspace{6mm}phase}\\[0.2ex]
\multicolumn{2}{l}{$\pi N\to \pi N$} & 6\er 3 (MeV)& (120\er45)\oo \\
\multicolumn{2}{l}{$2\,(\pi N\to N(1535)\pi)/\Gamma$}  &10\er4\%& (140\er 40)\oo \\
\multicolumn{2}{l}{$(\gamma p)^{1/2}\to N(1535)\pi$\hfill$M_{1-}$}& 8.5\er3.5 $10^{-3}$ & (25\er 35)\oo \\
\hline\\[-1.5ex]
\multicolumn{4}{l}{$N(1710){1/2^+}$ Breit-Wigner parameters}\\[0.5ex]
$M_{BW}$ &  1715\er20 & $\Gamma_{BW}$& 175\er 15\\
Br($\pi N$)  & 5\er 3\% &Br($N(1535)\pi$)  & 15\er 6\% \\
$A^{1/2}_{BW}$ & 0.050\er0.010& & \\[1ex]
\hline\hline\\
\multicolumn{4}{l}{\boldmath \fbox{\fbox{$N(1880){1/2^+}$}}\unboldmath\hfill**}\\[-0.7ex]
&&&\\\hline\hline\\[-1.5ex]
\multicolumn{4}{l}{$N(1880){1/2^+}$ decay modes}\\[0.2ex]
$M_{pole}$ &  1870\er40 & $\Gamma_{pole}$& 220\er 50\\
$A^{1/2\,(1)}$ & 0.010\er0.05& Phase & -(170\er40)\oo\\
$A^{1/2\,(2)}$ & 0.038\er0.15& Phase & (80\er40)\oo\\
\multicolumn{3}{l}{$N(1880){1/2^+}$ transition residues}&\multicolumn{1}{c}{phase}\\[0.2ex]
\multicolumn{2}{l}{$\pi N\to \pi N$} & 6\er 4 (MeV)& (70\er60)\oo \\
\multicolumn{2}{l}{$2\,(\pi N\to N(1535)\pi)/\Gamma$}  &9\er5\% &(130\er 60)\oo \\
\multicolumn{2}{l}{$2\,(\pi N\to Na_0(980))/\Gamma$}  &4\er3\% & (40\er65)\oo\\[0.2ex]
\multicolumn{2}{l}{$\gamma p^{1/2\,(1)}\to N(1535)\pi$\hfill$M_{1-}$}  &1\er0.5  $10^{-3}$ &-(30\er50)\oo \\
\multicolumn{2}{l}{$\gamma p^{1/2\,(1)}\to Na_0(980)$\hfill$M_{1-}$}  &$<0.5$ $10^{-3}$ &not defined\\[0.2ex]
\multicolumn{2}{l}{$\gamma p^{1/2\,(2)}\to N(1535)\pi$\hfill$M_{1-}$}  &4.5\er2.5 $10^{-3}$&-(75\er 60)\oo \\
\multicolumn{2}{l}{$\gamma p^{1/2\,(2)}\to Na_0(980)$\hfill$M_{1-}$}  &1\er0.5 $10^{-3}$& -(40\er50)\oo\\[0.2ex]
\hline
\multicolumn{4}{l}{$N(1880){1/2^+}$ Breit-Wigner parameters}\\[0.2ex]
$M_{BW}$ &  1875\er40 & $\Gamma_{BW}$& 230\er 50\\
Br($\pi N$)  & 6\er 3\% & Br($N(1535)\pi$)  & 8\er 4\% \\
Br($N\,a_0(980)$)  & 3\er 2\% \\
$A^{1/2\,(1)}_{BW}$ & -0.010\er0.005&
$|A^{1/2\,(2)}_{BW}|$ &0.038\er0.015\\
\hline\hline\\
\end{tabular}&
\begin{tabular}{lrlr}
\multicolumn{4}{l}{\boldmath \fbox{\fbox{$N(1900){3/2^+}$}}\unboldmath\hfill***}\\[-0.7ex]
&&&\\\hline\hline\\[-1.5ex]
\multicolumn{4}{l}{$N(1900){3/2^+}$ decay modes}\\[0.2ex]
$M_{pole}$ &  1910\er30 & $\Gamma_{pole}$& 280\er 50\\
$A^{1/2}$ & 0.026\er0.014& Phase & (60\er35)\oo\\
$A^{3/2}$ & -(0.070\er0.030)& Phase & (30\er 50)\oo\\
\multicolumn{3}{l}{$N(1900){3/2^+}$ transition residues}&\multicolumn{1}{c}{phase}\\[0.2ex]
\multicolumn{2}{l}{$\pi N\to \pi N$} & 4\er 2 (MeV)& -(10\er40)\oo \\
\multicolumn{2}{l}{$2\,(\pi N\to N(1535)\pi)/\Gamma$}  & 4\er1\% & (170\er 30)\oo \\
\multicolumn{2}{l}{$(\gamma p)^{1/2}\to N(1535)\pi$}  & 3\er1.5 $10^{-3}$&-(100\er 45)\oo \\
\multicolumn{2}{l}{$(\gamma p)^{3/2}\to N(1535)\pi$}  & 7\er3.5 $10^{-3}$&(90\er 45)\oo \\
\multicolumn{2}{l}{$\gamma p\to N(1535)\pi$\hfill$E_{2+}$}  & 3.5\er2 $10^{-3}$&(90\er 45)\oo \\
\multicolumn{2}{l}{$\gamma p\to N(1535)\pi$\hfill$M_{2+}$}  & 5\er3 ~\, $10^{-3}$&-(75\er 40)\oo \\
\hline
\multicolumn{4}{l}{$N_{3/2^+}(1900)$ Breit-Wigner parameters}\\[0.2ex]
$M_{BW}$ &  1910\er30 & $\Gamma_{BW}$& 270\er 50\\
Br($\pi N$)  & 3\er 2\% & Br($N(1535)\pi$)  & 7\er 3\% \\
$A^{1/2}_{BW}$ & 0.024\er0.014&
$A^{3/2}_{BW}$ &-0.067\er0.030\\
\hline\hline\\
\multicolumn{4}{l}{\boldmath \fbox{\fbox{$N(2120){3/2^-}$}}\unboldmath\hfill**}\\[-0.7ex]
&&&\\\hline\hline\\[-1.5ex]
\multicolumn{4}{l}{$N(2120){3/2^-}$ decay modes}\\[0.2ex]
$M_{pole}$ &  2115\er40 & $\Gamma_{pole}$& 345\er 35\\
$A^{1/2}$ & 0.130\er0.045& Phase & -(50\er20)\oo\\
$A^{3/2}$ & 0.160\er0.060& Phase & -(30\er 15)\oo\\
\multicolumn{3}{l}{$N(2120){3/2^-}$ transition residues}&\multicolumn{1}{c}{phase}\\[0.2ex]
\multicolumn{2}{l}{$\pi N\to \pi N$} & 11\er 6 (MeV)& -(30\er20)\oo \\
\multicolumn{2}{l}{2\,$(\pi N\to N(1535)\pi)/\Gamma$}  &15\er8\% & -(90\er 40)\oo \\
\multicolumn{2}{l}{$(\gamma p)^{1/2}\to N(1535)\pi$}  &55\er35 $10^{-3}$& (180\er 60)\oo \\
\multicolumn{2}{l}{$(\gamma p)^{3/2}\to N(1535)\pi$}  &60\er35 $10^{-3}$&-(160\er 60)\oo \\
\multicolumn{2}{l}{$\gamma p\to N(1535)\pi$\hfill$E_{2-}$}  &50\er30 $10^{-3}$&(10\er 60)\oo \\
\multicolumn{2}{l}{$\gamma p\to N(1535)\pi$\hfill$M_{2-}$}  &18\er10 $10^{-3}$& (150\er 60)\oo \\
\hline
\multicolumn{4}{l}{$N(2120){3/2^-}$ Breit-Wigner parameters}\\[0.2ex]
$M_{BW}$ &  2120\er35 & $\Gamma_{BW}$& 340\er 35\\
Br($\pi N$)  & 5\er 3\% & Br($N(1535)\pi$)  & 15\er 8\% \\
$A^{1/2}_{BW}$ & 0.130\er0.050&
$A^{3/2}_{BW}$ & 0.160\er0.065\\
\hline\hline\\
\end{tabular}
\end{tabular}
\end{scriptsize}
\renewcommand{\arraystretch}{1.2}\begin{scriptsize}
\begin{tabular}{cc}
\begin{tabular}{lrlr}
\multicolumn{4}{l}{\boldmath
\fbox{\fbox{$\Delta(1900){1/2^-}$}}\unboldmath
\hfill**}\\[-0.7ex]
&&&\\\hline\hline\\[-1.5ex]
\multicolumn{4}{l}{$\Delta(1900){1/2^-}$ decay modes}\\[0.2ex]
$M_{pole}$ &  1845\er20 & $\Gamma_{pole}$& 295\er 35\\
$A^{1/2}$ & 0.059\er0.015& Phase & (60\er20)\oo\\
\multicolumn{3}{l}{$\Delta(1900){1/2^-}$ transition residues}&\multicolumn{1}{c}{phase}\\[0.2ex]
\multicolumn{2}{l}{$\pi N\to \pi N$} & 11\er2 (MeV)& -(115\er20)\oo \\
\multicolumn{2}{l}{$2\,(\pi N\to \Delta(1232)\eta)/\Gamma$}  &1.3\er0.6\% &not defined \\
\multicolumn{2}{l}{$(\gamma p)^{1/2}\to \Delta(1232)\eta$\hfill$E_{0+}$}  &1.1\er0.6 $10^{-3}$ &-(110\er60)\oo \\
\hline
\multicolumn{4}{l}{$\Delta(1900){1/2^-}$ Breit-Wigner parameters}\\[0.2ex]
$M_{BW}$ &  1840\er20 & $\Gamma_{BW}$& 295\er30\\
Br($\pi N$)  & 7\er 2\% & BR($\Delta(1232)\eta$) & 1\er1\%\\
$A^{1/2}_{BW}$ & 0.057\er 0.014& &\\
\hline\hline\\
\end{tabular}&
\hspace{10mm}\begin{tabular}{lrlr} \multicolumn{4}{l}{\boldmath
\fbox{\fbox{$\Delta(1910){1/2^+}$}}\unboldmath
\hfill****}\\[-0.7ex]
&&&\\\hline\hline\\[-1.5ex]
\multicolumn{4}{l}{$\Delta(1910){1/2^+}$ decay modes}\\[0.2ex]
$M_{pole}$ &  1840\er40 & $\Gamma_{pole}$& 370\er 60\\
$A^{1/2}$ & 0.027\er0.009& Phase & -(30\er60)\oo\\
\multicolumn{3}{l}{$\Delta(1910){1/2^+}$ transition residues}&\multicolumn{1}{c}{phase}\\[0.2ex]
\multicolumn{2}{l}{$\pi N\to \pi N$} & 25\er 6 (MeV)& -(155\er30)\oo \\
\multicolumn{2}{l}{$2\,(\pi N\to \Delta(1232)\eta)/\Gamma$}  &11\er4\%&-(150\er50)\oo \\
\multicolumn{2}{l}{$(\gamma p)^{1/2}\to \Delta(1232)\eta$\hfill$M_{1-}$}  &4\er3 $10^{-3}$ &-(80\er??)\oo \\
\hline
\multicolumn{4}{l}{$\Delta(1910){1/2^+}$ Breit-Wigner parameters}\\[0.2ex]
$M_{BW}$ &  1845\er40 & $\Gamma_{BW}$& 360\er 60\\
Br($\pi N$)  & 12\er 3\% & BR($\Delta(1232)\eta$) & 9\er4\%\\
$A^{1/2}_{BW}$ & 0.026\er 0.008& &\\
\hline\hline\\
\end{tabular}
\end{tabular}
\end{scriptsize}
\end{table*}
\begin{table*}[pt]
{\bf Table~6 continued.}\vspace{3mm}\\
\renewcommand{\arraystretch}{1.2}\begin{scriptsize}
\begin{tabular}{cc}
\begin{tabular}{lrlr}
\multicolumn{4}{l}{\boldmath
\fbox{\fbox{$\Delta(1700){3/2^-}$}}\unboldmath
\hfill****}\\[-0.7ex]
&&&\\\hline\hline\\[-1.5ex]
\multicolumn{4}{l}{$\Delta(1700){3/2^-}$ decay modes}\\[0.2ex]
$M_{pole}$ &  1685\er10 & $\Gamma_{pole}$& 300\er 15\\
$A^{1/2}$ & 0.175\er0.020& Phase & (50\er10)\oo\\
$A^{3/2}$ & 0.180\er0.020& Phase & (45\er 10)\oo\\
\multicolumn{3}{l}{$\Delta(1700){3/2^-}$ transition residues}&\multicolumn{1}{c}{phase}\\[0.2ex]
\multicolumn{2}{l}{$\pi N\to \pi N$} & 40\er 6 (MeV)& -(1\er10)\oo \\
\multicolumn{2}{l}{2\,$(\pi N\to \Delta(1232)\eta)/\Gamma$}  &12\er2\% &-(60\er12)\oo \\
\multicolumn{2}{l}{2\,$(\pi N\to N(1535)\pi)/\Gamma$}  &3.5\er1.5\%&-(75\er 30)\oo \\
\multicolumn{2}{l}{$(\gamma p)^{1/2}\to \Delta(1232)\eta$}  &16\er3 $10^{-3}$ &-(10\er15)\oo \\
\multicolumn{2}{l}{$(\gamma p)^{3/2}\to \Delta(1232)\eta$}  &16\er3 $10^{-3}$ &-(10\er14)\oo \\
\multicolumn{2}{l}{$(\gamma p)^{1/2}\to N(1535)\pi$}  &4.5\er2 $10^{-3}$ &-(25\er 30)\oo \\
\multicolumn{2}{l}{$(\gamma p)^{3/2}\to N(1535)\pi$}  &4.5\er2 $10^{-3}$ &-(30\er 30)\oo \\
\multicolumn{2}{l}{$\gamma p\to \Delta(1232)\eta$\hfill$E_{2-}$}  &22\er4 $10^{-3}$ & (170\er15)\oo \\
\multicolumn{2}{l}{$\gamma p\to \Delta(1232)\eta$\hfill$M_{2-}$}  &3.5\er0.7 $10^{-3}$&-(5\er20)\oo \\
\multicolumn{2}{l}{$\gamma p\to N(1535)\pi$\hfill$E_{2-}$}  &6\er3 $10^{-3}$&(150\er 30)\oo \\
\multicolumn{2}{l}{$\gamma p\to N(1535)\pi$\hfill$M_{2-}$}  &1\er0.5 $10^{-3}$&-(20\er 35)\oo \\
\hline
\multicolumn{4}{l}{$\Delta(1700){3/2^-}$ Breit-Wigner parameters}\\[0.2ex]
$M_{BW}$ &  1715\er20 & $\Gamma_{BW}$& 300\er 25\\
Br($\pi N$)  & 22\er 4\% & BR($\Delta(1232)\eta$) & 5\er2\%\\
Br($N(1535)\pi$)  & 1\er 0.5\% & &\\
$A^{1/2}_{BW}$ & 0.165\er0.020&
$A^{3/2}_{BW}$ & 0.170\er0.025\\
\hline\hline\\
\multicolumn{4}{l}{\boldmath
\fbox{\fbox{$\Delta(1920){3/2^+}$}}\unboldmath
\hfill****}\\[-0.7ex]
&&&\\\hline\hline\\[-1.5ex]
\multicolumn{4}{l}{$\Delta(1920){3/2^+}$ decay modes}\\[0.2ex]
$M_{pole}$ &  1875\er30 & $\Gamma_{pole}$& 300\er 40\\
$A^{1/2}$ & 0.110\er0.030& Phase & -(50\er20)\oo\\
$A^{3/2}$ & 0.100\er0.040& Phase & -(180\er 20)\oo\\
\multicolumn{3}{l}{$\Delta(1920){3/2^+}$ transition residues}&\multicolumn{1}{c}{phase}\\[0.2ex]
\multicolumn{2}{l}{$\pi N\to \pi N$} & 16\er 6 (MeV)& -(50\er25)\oo \\
\multicolumn{2}{l}{$2\,(\pi N\to \Delta(1232)\eta)/\Gamma$}  &15\er 4\% &(70\er20)\oo \\
\multicolumn{2}{l}{$2\,(\pi N\to N(1535)\pi)/\Gamma$}  &3\er 2\% & (35\er45)\oo \\
\multicolumn{2}{l}{$2\,(\pi N\to Na_0(980))/\Gamma$}  & 3\er2\%& -(85\er 45)\oo\\[0.2ex]
\multicolumn{2}{l}{$(\gamma p)^{1/2}\to \Delta(1232)\eta$}  &20\er 7 &(55\er30)\oo \\
\multicolumn{2}{l}{$(\gamma p)^{3/2}\to \Delta(1232)\eta$}  &17\er 5 &-(75\er25)\oo \\
\multicolumn{2}{l}{$(\gamma p)^{1/2}\to N(1535)\pi$}  &5\er3 $10^{-3}$ & (10\er 35)\oo \\
\multicolumn{2}{l}{$(\gamma p)^{3/2}\to N(1535)\pi$}  &4\er3 $10^{-3}$ &-(110\er 40)\oo \\
\multicolumn{2}{l}{$(\gamma p)^{1/2}\to Na_0(980)$}  &5\er3 $10^{-3}$ &-(100\er 50)\oo \\
\multicolumn{2}{l}{$(\gamma p)^{3/2}\to Na_0(980)$}  &4\er2 $10^{-3}$ & (130\er 50)\oo \\
\multicolumn{2}{l}{$\gamma p\to \Delta(1232)\eta$\hfill$E_{1+}$}  &18\er 6 $10^{-3}$&-(100\er30)\oo \\
\multicolumn{2}{l}{$\gamma p\to \Delta(1232)\eta$\hfill$M_{1+}$}  &11\er 5 $10^{-3}$& (150\er30)\oo \\
\multicolumn{2}{l}{$\gamma p\to N(1535)\pi$\hfill$E_{1+}$}  &5\er3 $10^{-3}$ &-(150\er 35)\oo \\
\multicolumn{2}{l}{$\gamma p\to N(1535)\pi$\hfill$M_{1+}$}  &5\er3 $10^{-3}$ & (110\er 40)\oo \\
\multicolumn{2}{l}{$\gamma p\to Na_0(980)$\hfill$E_{1+}$}  &4\er2 $10^{-3}$ &(100\er 50)\oo \\
\multicolumn{2}{l}{$\gamma p\to Na_0(980)$\hfill$M_{1+}$}  &4\er2 $10^{-3}$ & -(10\er 50)\oo \\
\hline
\multicolumn{4}{l}{$\Delta(1920){3/2^+}$ Breit-Wigner parameters}\\[0.2ex]
$M_{BW}$ &  1880\er30 & $\Gamma_{BW}$& 300\er 40\\
Br($\pi N$)  & 8\er 4\% & BR($\Delta(1232)\eta$) &11\er6\%\\
Br($N(1535)\pi$)  & $<2$\% & & \\
$A^{1/2}_{BW}$ & 0.110\er0.030&
$A^{3/2}_{BW}$ &0.105\er0.035\\
\hline\hline\\
\multicolumn{4}{l}{\boldmath
\fbox{\fbox{$\Delta(1950){7/2^+}$}}\unboldmath
\hfill****}\\[-0.7ex]
&&&\\\hline\hline\\[-1.5ex]
\multicolumn{4}{l}{$\Delta(1950){7/2^+}$ decay modes}\\[0.2ex]
$M_{pole}$ &  1888\er4 & $\Gamma_{pole}$& 245\er 8\\
$A^{1/2}$ & -0.067\er0.004& Phase & -(10\er5)\oo\\
$A^{3/2}$ & -0.095\er0.004& Phase & -(10\er5)\oo\vspace{1mm}\\
\end{tabular}&
\begin{tabular}{lrlr}
\multicolumn{4}{l}{\boldmath
\fbox{\fbox{$\Delta(1940){3/2^-}$}}\unboldmath
\hfill**}\\[-0.7ex]
&&&\\\hline\hline\\[-1.5ex]
\multicolumn{4}{l}{$\Delta(1940){3/2^-}$ decay modes}\\[0.2ex]
$M_{pole}$ &  2040\er50 & $\Gamma_{pole}$& 450\er 90\\
$A^{1/2}$ & $0.170^{+0.120}_{-0.080}$& Phase &-(10\er30)\oo\\
$A^{3/2}$ & 0.150\er0.080& Phase & -(10\er 30)\oo\\
\multicolumn{3}{l}{$\Delta(1940){3/2^-}$ transition residues}&\multicolumn{1}{l}{phase}\\[0.2ex]
\multicolumn{2}{l}{$\pi N\to \pi N$} & 4\er 3 (MeV)& -(50\er35)\oo \\
\multicolumn{2}{l}{$2\,(\pi N\to \Delta(1232)\eta)/\Gamma$}  &$<1$\%&not drfined \\
\multicolumn{2}{l}{$2\,(\pi N\to N(1535))\pi/\Gamma$}  &$<3$\% &not defined \\
\multicolumn{2}{l}{$(\gamma p)^{1/2}\to \Delta(1232)\eta$}  &6.5\er3 $10^{-3}$ &-(110\er55)\oo \\
\multicolumn{2}{l}{$(\gamma p)^{3/2}\to \Delta(1232)\eta$}  &4\er2 $10^{-3}$ &-(110\er45)\oo \\
\multicolumn{2}{l}{$(\gamma p)^{1/2}\to N(1535)\pi$}  &16\er6 $10^{-3}$ &-(30\er 20)\oo \\
\multicolumn{2}{l}{$(\gamma p)^{3/2}\to N(1535)\pi$}  &11\er4 $10^{-3}$ &-(30\er 20)\oo \\
\multicolumn{2}{l}{$\gamma p\to \Delta(1232)\eta$\hfill$E_{2-}$}  &6.7\er3 $10^{-3}$ & (65\er55)\oo \\
\multicolumn{2}{l}{$\gamma p\to \Delta(1232)\eta$\hfill$M_{2-}$}  &2\er1 $10^{-3}$ &-(110\er55)\oo \\
\multicolumn{2}{l}{$\gamma p\to N(1535)\pi$\hfill$E_{2-}$}  &18\er6 $10^{-3}$ & (150\er 15)\oo \\
\multicolumn{2}{l}{$\gamma p\to N(1535)\pi$\hfill$M_{2-}$}  &6\er3 $10^{-3}$ &-(30\er 15)\oo \\
\hline
\multicolumn{4}{l}{$\Delta(1940){3/2^-}$ Breit-Wigner parameters}\\[0.2ex]
$M_{BW}$ &  2050\er40 & $\Gamma_{BW}$& 450\er 70\\
Br($\pi N$)  & 2\er 1\% & BR($\Delta(1232)\eta$) &10\er6\%\\
Br($N(1535)\pi$)  & 8\er 6\% & \\
$A^{1/2}_{BW}$ & $0.170^{+0.110}_{-0.080}$&
$A^{3/2}_{BW}$ & 0.150\er0.080\\
\hline\hline\\
\multicolumn{4}{l}{\boldmath
\fbox{\fbox{$\Delta(1905){5/2^+}$}}\unboldmath
\hfill****}\\[-0.7ex]
&&&\\\hline\hline\\[-1.5ex]
\multicolumn{4}{l}{$\Delta(1905){5/2^+}$ decay modes}\\[0.2ex]
$M_{pole}$ &  1800\er6 & $\Gamma_{pole}$& 290\er 15\\
$A^{1/2}$ & 0.025\er0.005& Phase & -(28\er12)\oo\\
$A^{3/2}$ & -0.050\er0.004& Phase &  (5\er 10)\oo\\
\multicolumn{3}{l}{$\Delta(1905){5/2^+}$ transition residues}&\multicolumn{1}{c}{phase}\\[0.2ex]
\multicolumn{2}{l}{$\pi N\to \pi N$} & 19\er 2 (MeV)& -(45\er4)\oo \\
\multicolumn{2}{l}{$2\,(\pi N\to \Delta(1232)\eta)/\Gamma$}  &7\er2\% &(40\er20)\oo \\
\multicolumn{2}{l}{$2\,(\pi N\to N(1535)\pi)/\Gamma$}  &2.5\er1\% &(130\er 35)\oo \\
\multicolumn{2}{l}{$(\gamma p)^{1/2}\to \Delta(1232)\eta$}  &1.8\er0.4 $10^{-3}$ & (23\er20)\oo \\
\multicolumn{2}{l}{$(\gamma p)^{3/2}\to \Delta(1232)\eta$}  &4.0\er1 $10^{-3}$ & (-135\er20)\oo \\
\multicolumn{2}{l}{$(\gamma p)^{1/2}\to N(1535)\pi$}  &0.6\er0.2 $10^{-3}$ & (120\er 40)\oo \\
\multicolumn{2}{l}{$(\gamma p)^{3/2}\to N(1535)\pi$}  &1.3\er0.5 $10^{-3}$ & -(40\er35)\oo \\
\multicolumn{2}{l}{$\gamma p\to \Delta(1232)\eta$\hfill$E_{3-}$}  &1.4\er0.4 $10^{-3}$ & (60\er20)\oo \\
\multicolumn{2}{l}{$\gamma p\to \Delta(1232)\eta$\hfill$M_{3-}$}  &1.5\er0.3 $10^{-3}$ & (40\er20)\oo \\
\multicolumn{2}{l}{$\gamma p\to N(1535)\pi$\hfill$E_{3-}$}  &0.4\er0.2 $10^{-3}$ & (150\er 35)\oo \\
\multicolumn{2}{l}{$\gamma p\to N(1535)\pi$\hfill$M_{3-}$}  &0.5\er0.1 $10^{-3}$ & (130\er35)\oo \\
\hline
\multicolumn{4}{l}{$\Delta(1905){5/2^+}$ Breit-Wigner parameters}\\[0.2ex]
$M_{BW}$ &  1856\er6 & $\Gamma_{BW}$& 325\er 15\\
Br($\pi N$)  & 13\er 2\% & BR($\Delta(1232)\eta$) &4\er2\%\\
Br($N(1535)\pi$)  & $<$1\% & & \\
$A^{1/2}_{BW}$ & 0.025\er0.005&
$A^{3/2}_{BW}$ &-0.050\er0.005\\
\hline\hline\\[-2ex]
\boldmath$\Delta(1950){7/2^+}$\unboldmath continued\\[1ex]
\hline\hline
\multicolumn{3}{l}{$\Delta(1950){7/2^+}$ transition residues}&\multicolumn{1}{c}{phase}\\[0.2ex]
\multicolumn{2}{l}{$\pi N\to \pi N$} & 58\er 2 (MeV)&  -(24\er3)\oo \\
\multicolumn{2}{l}{$2\,(\pi N\to \Delta(1232)\eta)/\Gamma$}  &3.5\er0.5\% &(90\er25)\oo \\
\multicolumn{2}{l}{$(\gamma p)^{1/2}\to \Delta(1232)\eta$}  &1.1\er0.2 $10^{-3}$ &-(80\er25)\oo \\
\multicolumn{2}{l}{$(\gamma p)^{3/2}\to \Delta(1232)\eta$}  &1.6\er0.3 $10^{-3}$ &-(80\er25)\oo \\
\multicolumn{2}{l}{$\gamma p\to \Delta(1232)\eta$\hfill$E_{3+}$}  &0.04\er0.02 $10^{-3}$ & not defined \\
\multicolumn{2}{l}{$\gamma p\to \Delta(1232)\eta$\hfill$M_{3+}$}  &0.75\er0.07 $10^{-3}$ & (100\er25)\oo \\
\hline
\multicolumn{4}{l}{$\Delta(1905){5/2^+}$ Breit-Wigner parameters}\\[0.2ex]
$M_{BW}$ &  1917\er4 & $\Gamma_{BW}$& 251\er 8\\
Br($\pi N$)  & 46\er 2\% & BR($\Delta(1232)\eta$) &$<$1\%\\
$A^{1/2}_{BW}$ &-0.067\er0.005&
$A^{3/2}_{BW}$ &-0.094\er0.004\\
\hline\hline
\end{tabular}
\end{tabular}
\end{scriptsize}
\end{table*}

In Table~\ref{brND}, the branching ratios for decays into $N\pi$,
$N(1535)\pi$, and $\Delta(1232)\eta$ are listed while
table~\ref{nucleon} collects the full information on decay modes of
nucleon and $\Delta(1232)$ resonances into $\Delta(1232)\eta$ (for
$\Delta$ resonances only), into $N(1535){1/2^-}\pi$, and into
$pa_0(980)$. The values are the central values derived from 12
representative fits to the data. The fits differ in the number of
resonances in a given partial wave and in the weight given to
individual data sets, see eq.~(\ref{likelihood}). The errors are
defined from the variance of results. The results on masses, widths,
and photo-couplings are consistent with previously published values
\cite{Beringer:1900zz}, and we refrain from a more detailed
discussion.

We return to a discussion of the branching ratios. In the decays
listed in Table~\ref{brND} the phase space is rather different: the
decay momentum for $\Delta(1950){7/2^+}\to N\pi$ is 730\,MeV/c, the
angular momentum between $N$ and $\pi$ is $L=3$. The branching
fraction is nevertheless very large. $\Delta(1920){3/2^+}$ decays
into $N\pi$ with about the same decay momentum but an angular
momentum between $N$ and $\pi$ of $L=1$ is sufficient. The angular
momentum does not seem to be the decisive quantity which governs the
decay. For $\Delta(1920){3/2^+}\to \Delta(1232)\eta$, the decay
momentum is 335\,MeV/c; this decay fraction is nevertheless of the
same order of magnitude as that for its decay into $N\pi$. Also the
linear momentum has not a highly significant impact on the decay
branching ratios. The decay into $\Delta(1232)\eta$ requires an
$S$-wave for $\Delta(1940){3/2^-}$, a $P$-wave for
$\Delta(1910){1/2^+}$, $\Delta(1920){3/2^+}$, and
$\Delta(1905){5/2^+}$. The latter three decay branching ratios are
of similar magnitudes as the former ones. On average, the different
branching ratios are of similar size, at least, there seems to be no
large suppression of decay modes due to phase space or angular
momentum barrier factors. We notice, however, that some resonances
prefer decays into $\pi N$, in other cases, cascade decays are
equally important. For the resonances
\begin{eqnarray}
\Delta(1700){3/2^-} & \Delta(1910){1/2^+}\label{diq}
\end{eqnarray}\begin{displaymath} \Delta(1920){3/2^+} \quad\Delta(1905){5/2^+}
\quad\Delta(1950){7/2^+}
\end{displaymath}
the decay branching ratios to $\pi N$ is larger than the branching
ratios for decays into $N(1535)1/2^-\pi$. For all nucleon resonances
and for $\Delta(1940){3/2^-}$ the reverse is true; they prefer
cascade decays to ground state transitions. The $N(1535)1/2^-\pi$
decay mode is not observed for the resonance $\Delta(1900){1/2^-}$;
we predict that it has a large coupling to $N(1520)3/2^-\pi$.

A similar observation has been made in $p\bar p$ annihilation:
$p\bar p$ annihilation prefers to produce high mass and not high
momentum \cite{Vandermeulen:1988hh}; there is little or no
dependence of the decay branching ratios for $p\bar p$ annihilation
into two mesons on the linear or orbital angular momentum between
the two mesons produced in the process \cite{Klempt:2005pp}. It may
be illustrative to remind the reader that not all decay processes
show a preference for high momenta. A well known example is the
ejection of an electron of an excited atom: the Auger-Meitner
process \cite{Auger:1925,Meitner:1922} preferentially ejects
electrons of low kinetic energies.

The puzzling results on decay modes require detailed theoretical
studies and may be a key to shed light on the structure of excited
nucleon and delta resonances. As in atomic and nuclear spectroscopy
the transition rates may be much more sensitive to the intrinsic
structure of excited states. In quark models, the resonances listed
in (\ref{diq}) have a comparably simple spatial wave function
\cite{Klempt:2012fy}. In the harmonic oscillator approximation, the
excited states are represented by excitations of two oscillators
$\lambda$ and $\rho$. Two excitations are possible, orbital ($l_i$,
$i=\rho,\lambda$) and radial ($n_i$) excitations. The resonances
listed in (\ref{diq}) have spatial wave functions which contain only
components in which either $\lambda$ or $\rho$ carry one kind of
excitation. In $\Delta(1900){1/2^-}$ and $\Delta(1940){3/2^-}$ both
$l$ and $n$ are excited, in $\Delta(1700){3/2^-}$ only $l$. In the
four positive-parity $\Delta(1232)$ states, either $\rho$ or
$\lambda$ carries two units of excitation, while $N(1900){3/2^+}$
has, very likely \cite{Anisovich:2011su}, a wave function in which
both oscillators $l_\rho$ and $l_\lambda$ carry one unit of
excitation. Possibly, baryon resonances in a complicated excitation
mode prefer to cascade down by de-excitation of one oscillator with
subsequent de-excitation to the ground state. Clearly, more examples
are needed to support or reject this conjecture.

\section{Summary}
\label{Summary}
We have reported a study of the reaction $\gamma p\to p\pi^0\eta$
for photon energies ranging from the threshold up to 2.5\,GeV. After
a detailed presentation of the experiment and of the selection of
the final state, the data are presented in the form of Dalitz plots.
The Dalitz plots reveal several contributing isobars: traces from
$\Delta(1232)\eta$, $N(1535)1/2^-\pi$, and $p\,a_{0}(980)$ can be
identified clearly. Angular distributions are presented in the
center-of-mass system, in the Gottfried-Jackson and in the helicity
frame. In the low-energy region the data are compared with the data
from Mainz on the same reaction. Very good consistency was achieved.

Part of the data were taken with linearly polarized photons. This
allowed us to extract several polarization variables: the
conventional beam asymmetry $\Sigma$; in a three-body final state
$\Sigma$ can be determined with respect to the three final-state
particles. Exploiting the full three-body kinematics, two further
variables can be deduced, $I^s$ and $I^c$, which characterize the
spin-alignment of the intermediate state when a linearly polarized
photon is absorbed.

The new data are included in the database of the BnGa partial wave
analysis. As three-body final state, they enter the program
event-by-event in a likelihood fit. The fit returned masses, widths,
and photo-couplings of four nucleon and seven $\Delta$
resonances compatible with previous results, and it returned decay
modes of these resonances into $\Delta(1232)\eta$,
$N(1535)1/2^-\pi$, $p\,a_{0}(980)$, and $N\pi$. It is conjectured
that strong cascade decays like $N(1900)3/2^+\to N(1535)1/2^-\pi\to
N\pi\eta$ (and weak $N\pi$ decays) signal a non-trivial structure of
the wave function of a decaying baryon resonance.

\subsection*{Acknowledgements}
We thank the technical staff at ELSA and at all the participating
institutions for their invaluable contributions to the success of
the experiment. We acknowledge support from the Deutsche
Forschungsgemeinschaft (within the SFB/TR16), the U.S. National
Science Foundation (NSF), and from the Schweizerische Nationalfonds. 
The collaboration with St. Petersburg
received funds from DFG and the Russian Foundation for Basic
Research.  This work comprises part of the PhD
thesis of E. Gutz. \phantom{Dummy}


\end{document}